\documentclass[twocolumn,aps,prd,showpacs,superscriptaddress,floats,floatfix,showkeys,notitlepage]{revtex4-1}

\usepackage{lipsum}
\usepackage{graphicx}
\usepackage{subfigure}
\usepackage{palatino}
\usepackage{changes}
\usepackage{hyperref}
\hypersetup{colorlinks=true,linkcolor=blue,urlcolor=blue,citecolor=blue}
\usepackage[toc,page]{appendix}
\usepackage[normalem]{ulem}
\usepackage{adjustbox}
\usepackage{latexsym}
\usepackage{amsmath}
\usepackage{amssymb}
\usepackage{amsfonts}
\usepackage{times}
\usepackage{dcolumn}
\usepackage{bm}
\usepackage{tikz}
\usepackage{bigints}
\usepackage{array,tabularx,multirow}
\usepackage[tracking=true]{microtype}
\SetTracking{}{500}
\SetTracking{encoding={*}, shape=sc}{40}
\UseRawInputEncoding 
\allowdisplaybreaks

\usepackage{braket}
\usepackage{tensor}
\usepackage{slashed}
\usepackage{booktabs} 
\usepackage{float}

\begin{document} \sloppy

\title{Nonlinearly charged black holes: Shadow and Thin-accretion disk}

\author{Akhil Uniyal}
\email{akhil\_uniyal@iitg.ac.in}
\affiliation{Department of Physics, Indian Institute of Technology, Guwahati 781039, India}

\author{Sayan Chakrabarti}
\email{sayan.chakrabarti@iitg.ac.in}
\affiliation{Department of Physics, Indian Institute of Technology, Guwahati 781039, India}

\author{Reggie C. Pantig}
\email{rcpantig@mapua.edu.ph}
\affiliation{Physics Department, Map\'ua University, 658 Muralla St., Intramuros, Manila 1002, Philippines}

\author{Ali \"Ovg\"un}
\email{ali.ovgun@emu.edu.tr}
\affiliation{Physics Department, Eastern Mediterranean University, Famagusta, 99628 North Cyprus via Mersin 10, Turkey}

\date{\today}
\begin{abstract}
In this paper, we explore the effect of non-linear electrodynamics (NLED) parameters and magnetic charges on various aspects of black holes, like how they bend light, how they emit radiation, and how they appear as a shadow by considering a thin accretion disk model. We initially examine the overall behavior of the photonsphere and the corresponding shadow silhouette under the effects of these parameters. Using the EHT data for Sgr. A* and M87*, we aim to find constraints for $q_m$. Our results indicate that M87* gives better constraints, and as the value of $\beta$ is varied to increase, the constrained range for $q_m$ widens. At lower values of $q_m$, we find that the shadow radius is close to the observed value. Then, we study different things like how much energy the black hole emits, the temperature of the disk around it, and the kind of light it gives off. We also look at how the black hole shadow appears in different situations. We also study how matter falls onto the black hole from all directions. Finally, we investigate how the magnetic charge affects all these things when we take into account a theory called  NLED along with gravity. This study helps us understand the complex relationship between magnetic charge and black holes.
\end{abstract}

\maketitle

\date{\today}

\date{\today}


\section{Introduction}\label{sec:intro}
Understanding black hole (BH) physics has always been an interesting topic, both for theoreticians as well as for observational astrophysicists as it can cast light into the predictions from the theory of general relativity, and thereby a large part of the gravitating Universe itself could be explained by understanding BH physics \cite{LIGOScientific:2016aoc}. Compact objects, particularly BHs have received a lot of attention within the premises of  Einstein’s general relativity from the perspective of observations. While the existence of BHs is now unquestionable, their detection still remains an extremely difficult task. In this direction, the concept of the shadow of a BH plays a significant role, since the very idea of imaging a black hole leads one to study the extremely strong gravity regime near the black hole horizon. One of the first works towards this direction for a static spherically symmetric black hole was performed by Synge \cite{Synge:1966okc} and Luminet \cite{Luminet:1979nyg}, and later Bardeen \cite{Bardeen:1973} extended the idea for studying BH shadow in the rotating Kerr geometry. In recent times, the Event Horizon Telescope (EHT) has published the black hole shadow image in the center of the M87$^*$ galaxy \cite{EventHorizonTelescope:2019dse} and the shadow of Sagittarius A$^*$ (Sgr A$^*$) black hole in the center of Milky Way galaxy \cite{EventHorizonTelescope:2022xnr}. Recent observations by EHT and Stratospheric Observatory for Infrared Astronomy (SOFIA), have shown that \cite{EventHorizonTelescope:2021srq, lopez2021extragalactic} the polarised appearance of the black holes, these studies is a significant milestone since the polarisation of light provides valuable information that can help us comprehend the physics underlying the image.

Immediately after the release of these images a number of works flooded the literature, many of which focused towards understanding the alternate theories of gravity around the black hole (for a detailed discussion and references, see \cite{Vagnozzi:2022moj,Cunha:2018gql,Takahashi:2004xh,Okyay:2021nnh,Allahyari:2019jqz,Chen:2022nbb,Roy:2021uye,Khodadi:2020jij,Wang:2018prk,Cunha:2019hzj,Pantig:2022toh,Pantig:2022whj,Pantig:2022sjb,Pantig:2022ely,Pantig:2022gih,Lobos:2022jsz,Zuluaga:2021vjc,Li:2021ypw,Rahaman:2021kge,Stashko:2021lad,Liu:2021yev,Guerrero:2021ues,Gan:2021xdl,Heydari-Fard:2020iiu,Heydari-Fard:2021ljh,Kazempour:2022asl,He:2022yse,Bisnovatyi-Kogan:2022ujt,Bauer:2021atk,Li:2021riw,Ovgun:2020gjz,Ovgun:2019jdo,Ovgun:2018tua,Ovgun:2021ttv,Ling:2021vgk,Belhaj:2020okh,Belhaj:2020rdb,Abdikamalov:2019ztb,Abdujabbarov:2016efm,Atamurotov:2015nra,Papnoi:2014aaa,Abdujabbarov:2012bn,Atamurotov:2013sca,Cunha:2018acu,Perlick:2015vta,Nedkova:2013msa,Li:2013jra,Cunha:2016wzk,Johannsen:2015hib,Johannsen:2015mdd,Shaikh:2019fpu,Yumoto:2012kz,Cunha:2016bpi,Moffat:2015kva,Giddings:2016btb,Cunha:2016bjh,Zakharov:2014lqa,Tsukamoto:2017fxq,Hennigar:2018hza,Kumar:2020hgm,Li2020,Cimdiker:2021cpz,Hu:2020usx,Zhong:2021mty,Kumaran:2022soh,Rayimbaev:2022hca,Kuang:2022xjp,Atamurotov:2022knb,Mustafa:2022xod,Uniyal:2022vdu,Ovgun:2023ego, Pantig:2023yer}). The work in \cite{Vagnozzi:2022moj} studies a wide range
of well-motivated modifications to classical General Relativistic black hole solutions and constrains them using the EHT observations of Sgr A$^*$. In the process of studying so, they connect the size of the bright ring of emission to that of the underlying BH shadow using the Sgr A$^*$'s mass-to-distance ratio. Different scenarios including regular BHs, string-inspired BH space-times, alternative models of the theory of gravity, and BH mimickers including wormhole and naked singular space-times were studied in the above paper to constrain the fundamental parameters of those modified theories and exotic objects, it was demonstrated that the EHT image of Sgr A$^∗$ puts very stringent constraints on the models and predicts a shadow size larger than that of a Schwarzschild BH of a given mass. On another front, it is to be mentioned here that since one can only have information about the diameter of the shadow size from observations made so far, one can easily get into the trouble of shadow-degeneracy, i.e. two black holes from two different theories of gravity casting similar shadow \cite{Lima:2021las}  and in the recent work that comparison is also done with the rotating Kerr black hole and non-rotating dilaton black hole \cite{Mizuno:2018lxz}. Therefore, in future, we expect to overcome such degeneracies by observing the black hole with more precision.

One of the alternative theories to general relativity (GR) is known as the non-linear electrodynamic theory (NLE). It was the first outcome of the motivation to remove the central singularity of the point charge and the related energy divergence by generalizing Maxwell’s theory in the 1930s by Born and Infeld \cite{Born:1934gh}. After that Plebanski extended the work in the special relativity framework by including an arbitrary function of the electromagnetic field invariants in the theory \cite{plebanski1966non}. After this, a lot of work has been done in this field having the motivation that this theory can produce regular black holes,  some kinds of NLE theories are limiting cases of certain types of models in the string theory and also solving singularity problems in cosmology \cite{Seiberg:1999vs, Tseytlin:1999dj,Novello:2003kh,Novello:2006ng,Vollick:2008dx}. Among all of them, the spherically symmetric solutions are the simplistic ones in the sense that we only have radial electric and magnetic charges in the theory \cite{Pellicer:1969cf}. Therefore, in this work, we will be interested in understanding to look into one of the purely magnetic charge spherical symmetric solutions of NLE \cite{kruglov2022nonlinearly} with the motivation to first constrain the magnetic charge with the help of known EHT results and then look into the accretion disk and shadow behaviour for different ideal intensity profiles.

To learn more about black holes, scientists study the accretion disk that surrounds them. Scientists have been studying black holes for many years, and in 1973 Cunningham and Bardeen \cite{1973Apx} were the first to study what a black hole would look like if a star was orbiting around it. In 1979, Luminet \cite{Luminet:1979nyg} theorized what the accretion disk around a black hole would look like using a thin disk model. The disk's structure was described by other scientists, including Shakura and Sunyaev in 1973 \cite{1973A}, Novikov and Thorne in 1973 \cite{NovikovThorne}, and Page and Thorne in 1974 \cite{1974ApJ...191..499P}. The disk is thin and opaque, meaning that as gas moves within it radiate energy due to the strong gravity of the black hole \cite{Tucker_2018, Corral-Santana:2015fud}. The radiation emitted from the accretion disk depends on how particles move within the gas and the structure of space-time around the black hole. By studying the radiation emitted from accretion disks, scientists can test the theory of gravity around black holes. Therefore, this paper aims in the same direction to understand the effect of a magnetically charged NLED black hole on the particles moving along the geodesics path around the black hole.

The main motivation of this work is to understand the effect of magnetically charged NLED black holes on the particles' geodesics, radiation emitted from the thin accretion disk, and shadow of the black hole. Moreover, we first find the constraints to the NLED coupling parameter $(\beta)$ and magnetic charge $(q_m)$ of the black hole by using the data released by EHT for M87* and Sagittarius A*. Then, we look at things like the energy flux $(F)$, temperature $(T)$, luminosity $(dL_{\infty})$, and spectrum of the disk $(\nu \mathcal{L}_{\nu,\infty})$ coming form the thin accretion disk, and how they change when we include NLED in our calculations. We also look at how things change with $\beta$ and $q_m$ when we consider different emission profiles emitted from the thin accretion disk and observed by the observer at infinity including with an infalling spherical accretion in the equatorial plane of the accretion disk. We use a specific unit system $G=M=c=1$ and metric signature $(-,+,+,+)$ in our paper.

\section{Black Hole metric}\label{sec:formal}
\begin{figure*}[htbp]
    \centering
    \includegraphics[width=0.48\textwidth]{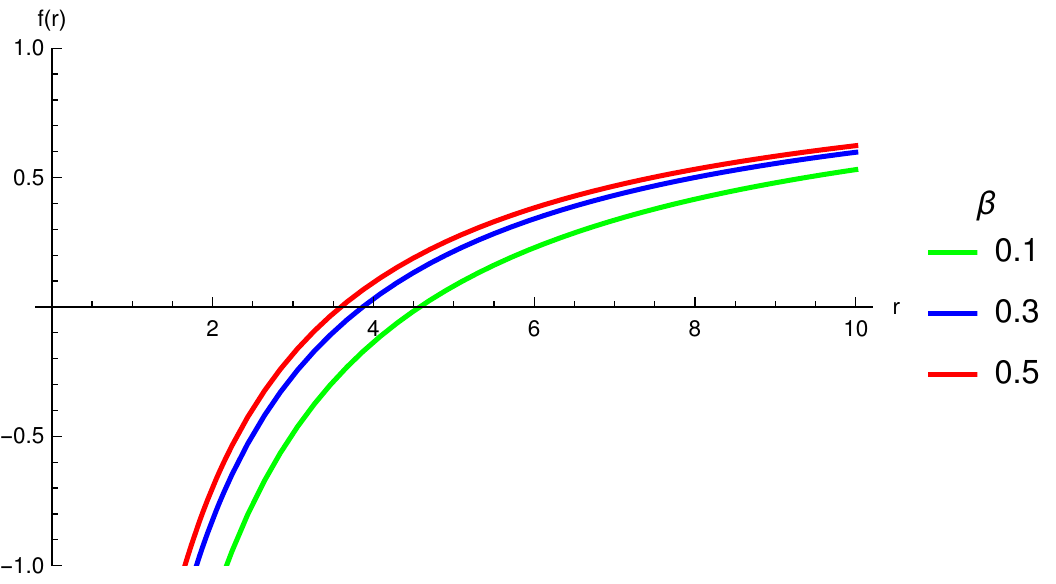}
    \includegraphics[width=0.48\textwidth]{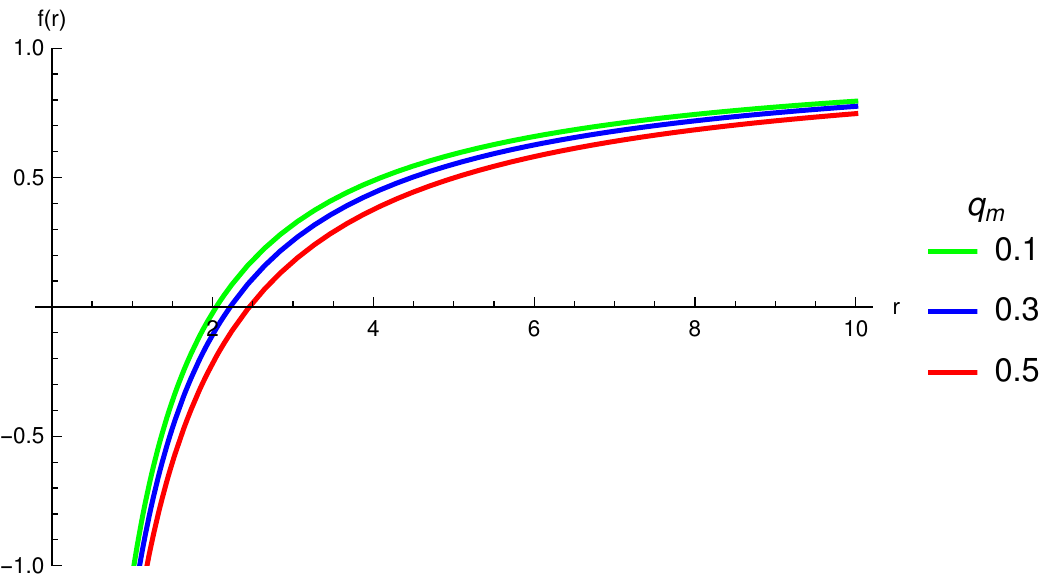}
    \caption{Metric function defined by Eq. \eqref{12} for different values of NLE parameter $\beta=0.1 (\text{green})$, $\beta=0.3 (\text{blue})$, and $\beta=0.5 (\text{red})$ when $q_m=0.3$ in the left side plot while for magnetic charge $q_m=0.1 (\text{green})$, $q_m=0.3 (\text{blue})$ and $q_m=0.5 (\text{red})$ when $\beta=0.3$ in the right.} \label{fig:met}
\end{figure*}

The definition of the action for the NED theory in general relativity is provided as \cite{Kruglov:2022lnc}
\begin{equation}
I=\int d^{4}x\sqrt{-g}\left(\frac{R}{16\pi G_N}+\mathcal{L}(\mathcal{F}) \right),
\label{1}
\end{equation}
 the Newton constant $G_N$ is also included. The NED Lagrangian utilized in this work is suggested in \cite{Kruglov:2017xmb} (see also \cite{Kruglov:2021stm})
\begin{equation}
{\cal L}(\mathcal{F}) =-\frac{{\cal F}}{1+\sqrt{2|\beta{\cal F}|}}.
\label{2}
\end{equation}
where we have the Lorentz invariant ${\cal F}=F^{\mu\nu}F_{\mu\nu}/4=(B^2-E^2)/2$. The electric and magnetic induction fields are denoted as $E$ and $B$ respectively. When $\beta$ is set to zero in Equations \eqref{1} and \eqref{2}, we get the action of black holes with Maxwell electrodynamics. By making variations to the action \eqref{1}, we can obtain the equations for the gravitational and electromagnetic fields given below
\begin{equation}
R_{\mu\nu}-\frac{1}{2}g_{\mu \nu}R =8\pi G_N T_{\mu \nu},
\label{3}
 \end{equation}
\begin{equation}
\partial _{\mu }\left( \sqrt{-g}\mathcal{L}_{\mathcal{F}}F^{\mu \nu}\right)=0,
\label{4}
\end{equation}
where$\mathcal{L}_{\mathcal{F}}=\partial \mathcal{L}( \mathcal{F})/\partial \mathcal{F}$.
The stress tensor of electromagnetic fields is given by
\begin{equation} \label{5}
 T_{\mu\nu }=F_{\mu\rho }F_{\nu }^{~\rho }\mathcal{L}_{\mathcal{F}}+g_{\mu \nu }\mathcal{L}\left( \mathcal{F}\right).
\end{equation}
Now, let us consider a space-time with spherical symmetry as follows,
\begin{equation} \label{6}
ds^{2}=-f(r)dt^{2}+\frac{1}{f(r)}dr^{2}+r^{2}\left( d\theta^{2}+\sin^{2}\theta d\phi ^{2}\right).
\end{equation}
Then the tensor $F_{\mu \nu}$ possesses the radial electric field $F_{01}=-F_{10}$ and radial magnetic field $F_{23}=-F_{32}=q_m \sin(\theta)$ with the magnetic charge $q_m$. The stress tensor is diagonal, $T_{0}^{~0}=T_{r}^{~r}$ and $T_{\theta}^{~\theta}=T_{\phi}^{~\phi}$. The metric function in Eq. \eqref{6} can be written as \cite{Bronnikov:2000vy},
\begin{equation}
f(r)=1 - \frac{2m(r)G_N}{r},
\label{7}
\end{equation}
and the mass function is given by,
\begin{equation}
m(r)=m_0+ \int_{0}^{r} \rho (r)r^{2}dr.
\label{8}
\end{equation}
In equation \eqref{8}, the integration constant $m_0$ represents the Schwarzschild mass, and $\rho(r)$ denotes the energy density. It should be noted that models involving electrically charged black holes that conform to Maxwell's weak-field limit have singularities \cite{Bronnikov:2000vy}. Therefore, in this study, we only consider magnetic black holes with $\mathcal{F}=q_m^2/(2r^4)$.
When we substitute $q_e=0$ in equation \eqref{5}, we obtain the energy density:
\begin{equation}
\rho=\frac{q_m^2}{2r^2(r^2+q_m\sqrt{\beta})}.
\label{9}
\end{equation}
 Making use of Eqs. \eqref{8} and \eqref{9} to find out the mass function as,
\begin{equation}
m(r)=m_0+\frac{q_m^{3/2}}{2\beta^{1/4}}\arctan\left(\frac{r}{\sqrt{q_m}\beta^{1/4}}\right).
\label{10}
\end{equation}
So, we are creating a definition for the magnetic mass of the black hole
\begin{equation}
m_M=\int_0^\infty \frac{q_m^2}{2(r^2+ q_m\sqrt{\beta})}dr=\frac{\pi q_m^{3/2}}{4\beta^{1/4}}.
\label{11}
\end{equation}
To explain, when $\beta=0$, Eq. \eqref{11} shows that the magnetic energy becomes infinite at the Maxwell limit. This is because a magnetic monopole (as well as point-like charges) has an infinite total magnetic energy. Hence, the coupling constant $\beta$ helps smooth out the NLED model's singularities under consideration. A similar situation can be observed in Born-Infeld electrodynamics. By using Eqs. \eqref{7} and \eqref{10}, we can calculate the metric function as presented in \cite{Kruglov:2022lnc}
\begin{equation}
f(r)=1-\frac{2m_0G_N}{r}-\frac{q_m^{3/2}G_N}{\beta^{1/4}r}\arctan\left(\frac{r}{\sqrt{q_m}\beta^{1/4}}\right).
\label{12}
\end{equation}
As we increase the value of the coupling constant $\beta$ to a very large value, $\beta\rightarrow \infty$, the metric function \eqref{12} takes the form of the Schwarzschild black hole that is already known. From Eq. \eqref{12}, we can find the metric function as $r$ tends to infinity \cite{Kruglov:2022lnc},
\begin{equation}
f(r)=1-\frac{2(m_0+m_M)G_N}{r}+\frac{q_m^2G_N}{r^2}-\frac{q_m^3\sqrt{\beta}G_N}{3r^4}+\mathcal{O}(r^{-6})
\label{13}
\end{equation}
Equation \eqref{13} indicates that there are adjustments to the familiar Reissner--Nordstrom solution. Therefore, we can associate the ADM mass of the black hole with the total mass $M\equiv m_0+m_M$, which is the combination of the Schwarzschild mass $m_0$ and the magnetic mass $m_M$, according to Eq. \eqref{13}. When $m_0=0$, the asymptotic metric as $r\rightarrow \infty$ can be obtained by using Eq. \eqref{12} \cite{Kruglov:2022lnc}
\begin{equation}
f(r)=1-\frac{q_mG_N}{\sqrt{\beta}}+\frac{G_Nr^2}{3\beta}+\mathcal{O}(r^{4})~~~\mbox{as}~r\rightarrow 0,
\label{14}
\end{equation}
the metric function given in equation \eqref{14} corresponds to a black hole with mass and charge-like terms. As per equation \eqref{14}, the metric function remains finite at the origin $r=0$, hence it is regular at $m_0=0$. Figure \ref{fig:met} shows the plot of the metric function \eqref{12} with $G_N=1$ for the different values of NLE parameter $\beta$ in left and magnetic charge $q$ in right with fixing other one to unity as given in \cite{Kruglov:2022lnc}.


According to the figure shown in Fig.~\ref{fig:met}, it can be observed that the black hole has only one horizon for both the parameters $\beta$ and $q_m$. It can be observed in the Figure that the event horizon decreases with increasing the value of $\beta$ (left) while increases with increasing the value of $q_m$ (right). Therefore, one can understand that in this space-time the effect of the $\beta$ and $q_m$ are opposite to each other.

\section{Photonsphere and the Shadow of Nonlinearly charged black hole}
The concept of the black hole shadow is of great importance in the field of astrophysics because it provides a rare opportunity to test the predictions of Einstein's theory of general relativity. The shadow's size and shape are determined by the curvature of spacetime around the black hole, which is a fundamental prediction of general relativity. By observing and scrutinizing the shadow, scientists can test the theory's predictions and search for any deviations that could lead to a better understanding of the fundamental laws of physics. In this section, we study the deviation caused by a black hole under the effect of the NLED parameter and magnetic charge.

\begin{figure}
  \centering
    \includegraphics[width=0.48\textwidth]{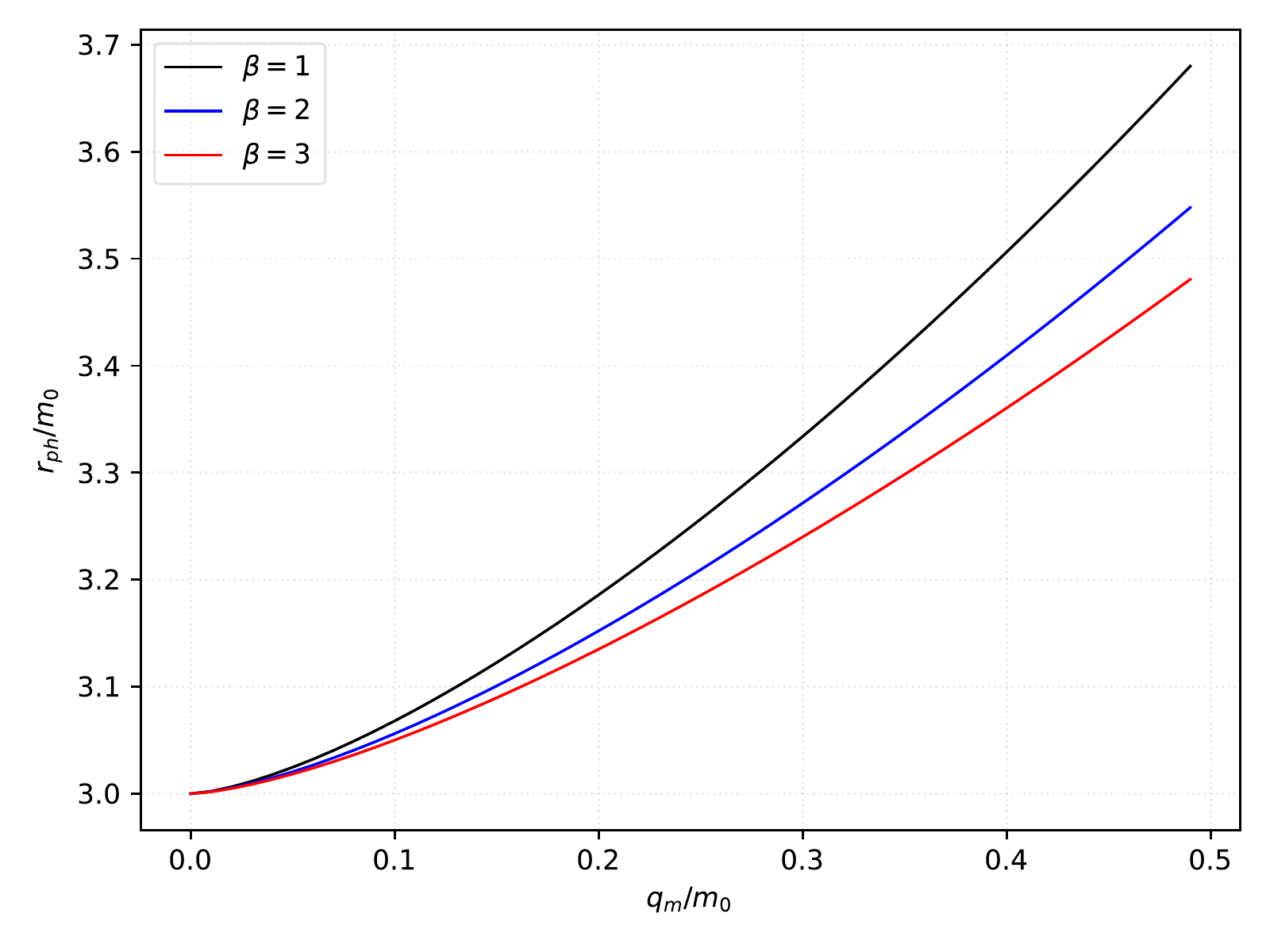}
    \caption{Behavior of the photonsphere as $q_m$ and $\beta$ varies. In this plot, we used $m_0 = 1$.}
    \label{fig:rph}
\end{figure}

To carefully study the shadow, we must initially investigate the deviations in the photonsphere. Using Eqs. \eqref{6} and \eqref{12}, and following the standard prescription in Ref. \cite{Perlick:2015vta}, the location of the photonsphere can be determined by solving the equation numerically.
\begin{align}
\frac{3 q_m^{3/2} \tan ^{-1}\left(\frac{r}{\sqrt[4]{\beta } \sqrt{q_m}}\right)}{\sqrt[4]{\beta }}+6 m_0=r \left(\frac{q_m^2}{\sqrt{\beta } q_m+r^2}+2\right)
\end{align}
We plot the results in Fig. \ref{fig:rph}, where we explored various effects of $q_m$ and $\beta$. Clearly, the photonsphere radius becomes larger as $q_m$ increases for a given value of $\beta$. The effect of increasing $\beta$, however, is to suppress the increase in the radius due to $q_m$.

Now that we know how to locate the photonsphere, we are now in the position to find constraints to $q_m$ using the EHT's results for M87* and Sgr. A*. We follow the method in Ref. \cite{Perlick:2018}, where we can find the expression for the angular radius as
\begin{equation}
    \tan(\alpha_{\text{sh}}) = \lim_{\Delta x \to 0}\frac{\Delta y}{\Delta x} = \left(r^2 f(r)\right)^{1/2} \frac{d\phi}{dr} \bigg|_{r=r_o},
\end{equation}
which can be simplified to
\begin{equation} \label{e17}
    \sin^{2}(\alpha_\text{sh}) = \frac{b_\text{crit}^{2}}{h(r_o)^{2}}
\end{equation}
with the help of the orbit equation
\begin{equation}
    \left(\frac{dr}{d\phi}\right)^2 =r^2 f(r) \left(\frac{h(r)^2}{b^2}-1\right).
\end{equation}
Here, the function $h(r)$ is defined as \cite{Perlick:2015vta}
\begin{equation} \label{e19}
    h(r)^2 = \frac{r^2}{f(r)}.
\end{equation}
Note the use of the critical impact parameter in Eq. \eqref{e17} and $H(r)$ evaluated at the observer position $r_o$. Using the prescription in Refs. \cite{Pantig:2022ely,Pantig:2022sjb},
\begin{equation} \label{ebcrit}
    b_\text{crit}^2 = \frac{4r_\text{ph}^2}{r f'(r) |_{r=r_\text{ph}} + 2f(r_\text{ph})},
\end{equation}
by which we find
\begin{align}   
    b_\text{crit}^2 &= 4 r_\text{ph}^{3} \beta^{\frac{1}{4}} \left(q_m \sqrt{\beta}+r_\text{ph}^{2}\right) \biggl[ 2 q_m \left(m_0 -r_\text{ph} \right) \beta^{\frac{3}{4}} \nonumber \\  
    & + \left(r_\text{ph}^{2} q_m^{\frac{3}{2}}+q_m^{\frac{5}{2}} \sqrt{\beta}\right) \arctan \! \left(\frac{r_\text{ph}}{\sqrt{q_m}\, \beta^{\frac{1}{4}}}\right) \nonumber \\ 
    & + 2 r_\text{ph} \left(m_0 r_\text{ph} +\frac{1}{2} q_m^{2}-r_\text{ph}^{2}\right) \beta^{\frac{1}{4}} \biggr]^{-1}.
\end{align}
Finally, with the help of Eqs. \eqref{e17} and \eqref{e19}, the shadow radius can be obtained as
\begin{equation}
    R_\text{sh} = b_\text{crit}.
\end{equation}

\begin{figure*}
  \centering
    \includegraphics[width=0.48\textwidth]{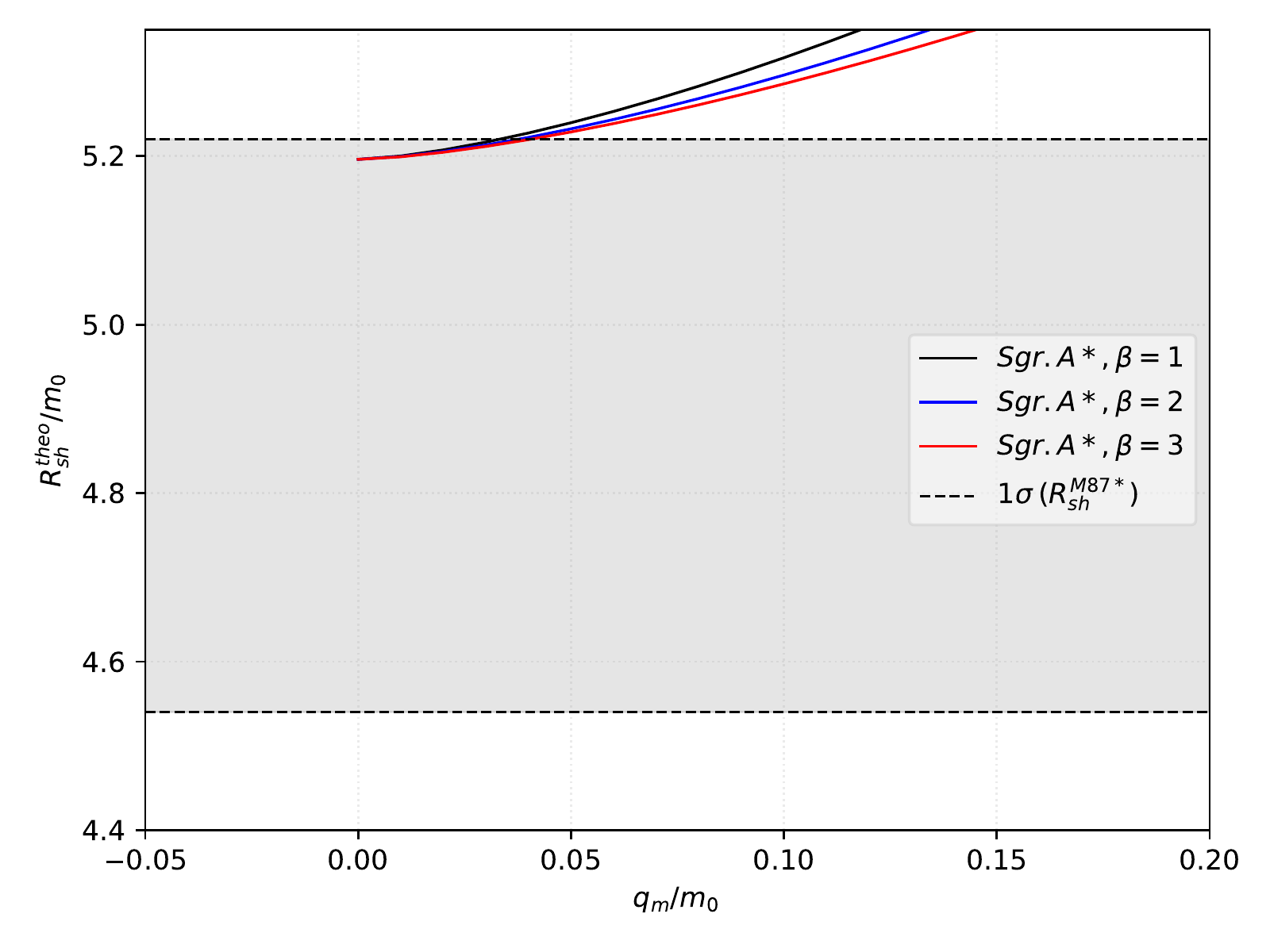}
    \includegraphics[width=0.48\textwidth]{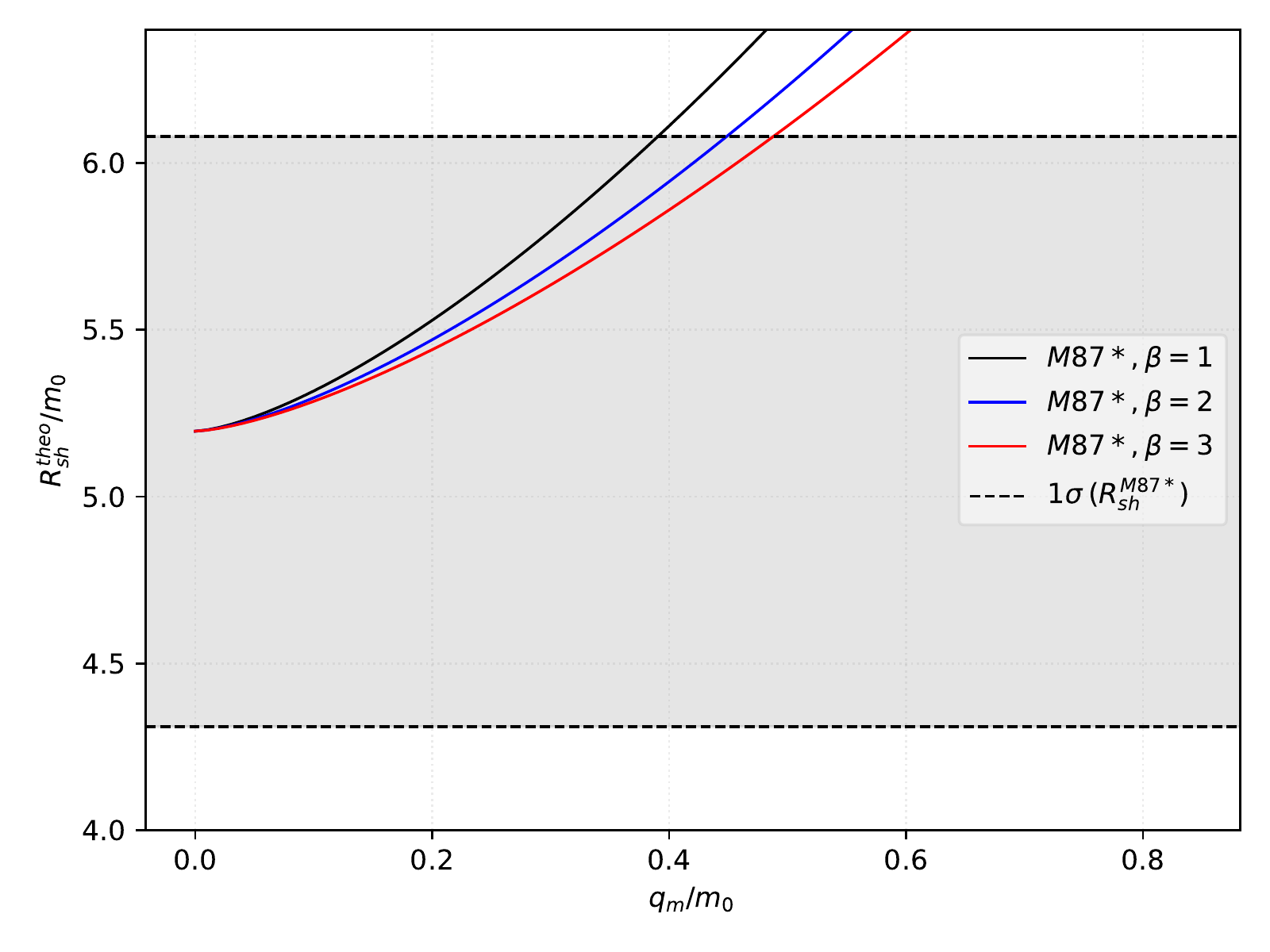}
    \caption{Constraints in $q_m$ for Sgr. A* (left), and M87* (right) showing $1-\sigma$ level.}
    \label{fig:shacons}
\end{figure*}

The shadow contour is expected to be a perfect circle since this is a non-rotating black hole. However, we could see the dependence of the shadow radius $R_\text{sh}$ on the $r_\text{ph}$, and in the metric function Eq. \eqref{12} evaluated at the observer position. Relative to Sgr. A*, the observer position are $8277$ pc, while from M87*, $16.8$ kpc. In terms of $M_\odot$, the masses of the central black holes Sgr. A* and M87* are $4.3 \pm 0.013$x$10^6$ (VLTI \cite{EventHorizonTelescope:2022xnr,EventHorizonTelescope:2021dqv,Vagnozzi:2022moj}) and $6.5 \pm 0.90$x$10^9$, respectively. Furthermore, we used the allowable bounds for Sgr. A* and M87*, set by the EHT, which uses the reported Schwarzschild deviation at $68\%$ confidence level. These are, $ 4.55M \leq R_{sh} \leq 5.22M$, and $ 4.31M \leq R_{sh} \leq 6.08M$, respectively \cite{EventHorizonTelescope:2019dse,EventHorizonTelescope:2022xnr,EventHorizonTelescope:2021dqv,Vagnozzi:2022moj}. With these data, we can now find constraints to $q_m$, where the results are plotted in Fig. \ref{fig:shacons}. Note that M87* allows a wider range of $q_m$ values as compared to Sgr. A*. Moreover, there are some values of $q_m$ that coincide with the reported shadow radius for M87*, which makes the SMBH a better candidate for constraining NLED BH. 
As a final remark, the plot also indicates how the shadow radius behaves when $q_m$ varies. It turns out that the $R_\text{sh}$ behaves in a similar manner as in $r_\text{ph}$.

\section{Thin Accretion Disk Model}
First, we will examine how particles move around the black hole by investigating the thin accretion disk in the equatorial plane using the Novikov-Thorne model \cite{NovikovThorne}. This model is an extension of the Shakura-Sunyaev thin disk model \cite{1973A}.

\subsection{Novikov-Thorne accretion model}
\begin{figure*}
    \centering
    \includegraphics[width=0.48\textwidth]{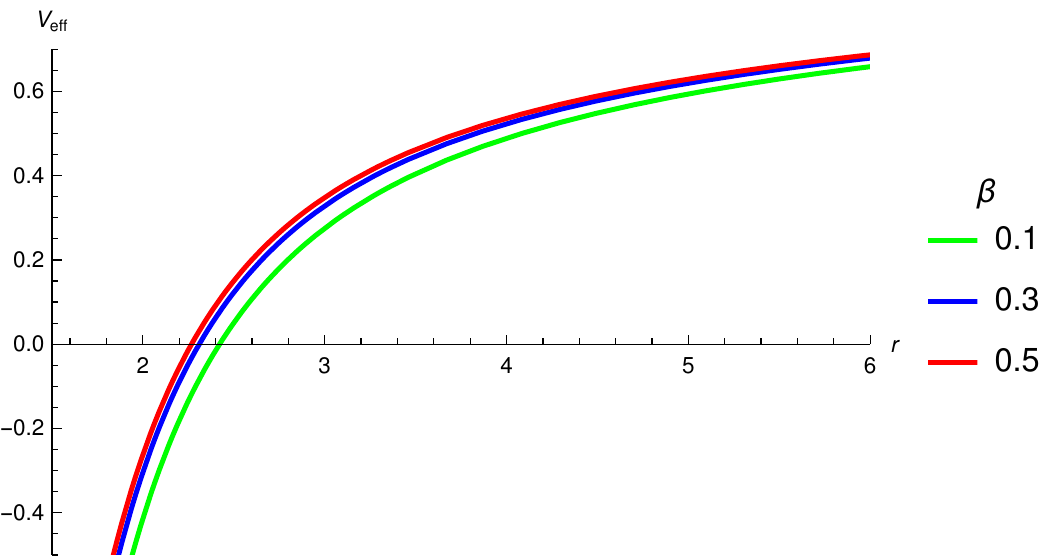}
    \includegraphics[width=0.48\textwidth]{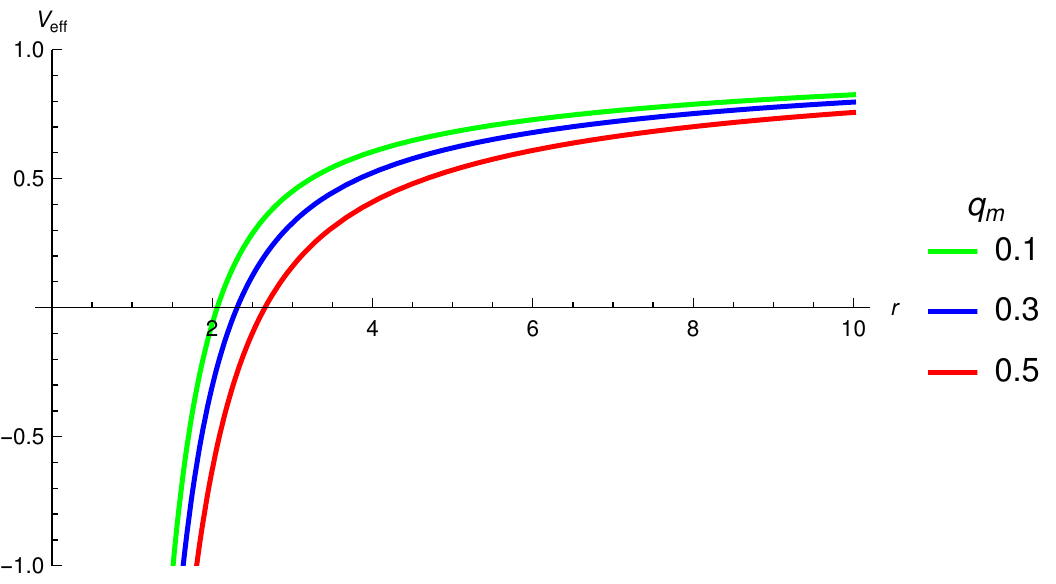}
    \caption{Effective potential for different values of NLE parameter $\beta=0.1 (\text{green})$, $\beta=0.3 (\text{blue})$, and $\beta=0.5 (\text{red})$ when $q_m=0.3$ in the left side plot while for magnetic charge $q_m=0.1 (\text{green})$, $q_m=0.3 (\text{blue})$ and $q_m=0.5 (\text{red})$ when $\beta=0.3$ in the right. } \label{fig:e}
\end{figure*}

To obtain the spectrum of the disk, we start by examining the Lagrangian of a test particle that is orbiting around a compact object \cite{Heydari-Fard:2020iiu}
\begin{equation}
    \mathcal{L}=\frac{1}{2}g_{\mu \nu}\dot{x}^\mu \dot{x}^{\nu}.
\end{equation}
To obtain the solution for a general space-time, we begin by considering the following form of a static,  where $g_{\mu\nu}$ is the metric of the space-time and a dot on the right-hand side indicates differentiation with respect to the affine parameter
\begin{equation}
    ds^2=g_{tt}dt^2+g_{rr}dr^2+g_{\theta \theta}d\theta^2+g_{\phi \phi}d\phi^2,
\end{equation}
we assume that the metric components $g_{tt}$ and $g_{rr}$ only vary with the radial coordinate $r$. Then, by applying the Euler-Lagrange equations in the equatorial plane of the accretion disk (where $\theta=\pi/2$), we obtain the following equations:
\begin{equation}
    \dot{t}=-\frac{E}{g_{tt}}, \label{18}
\end{equation}
\begin{equation}
    \dot{\phi}=\frac{L}{g_{\phi \phi}}, \label{19}
\end{equation}
Let's say there's a particle moving around a compact object in the equatorial plane, and we use the terms "specific energy $E$" and "specific angular momentum $L$ " to describe its movement. If we assume that a certain equation (which we call "Eq. \eqref{18} ") holds true, and we also assume that the value of "2L" is equal to negative one $2\mathcal{L}=-1$ for this particular particle, we can use another equation (which we call "Eq. \eqref{19} ") to come up with a new equation that looks like this: 
\begin{equation}
    -g_{tt}g_{rr}\dot{r}^2+V_\text{eff}=E^2,
\end{equation}
where the $V_\text{eff}$ is known as the effective potential for the particle and defined as,
\begin{equation}
    V_\text{eff}=-g_{tt}\left(1+\frac{L^2}{g_{\phi \phi}} \right).
\end{equation}
Figure \ref{fig:e} displays the effective potential for $\beta$ (left) and $q_m$ (right). As discussed earlier that the effect of these parameters is completely opposite to each other, the similar behaviour we have seen in this case also. It was observed that the effect of $q_m$ is larger and more prominent with small values than the effect of $\beta$. However, we tried to look at the effect of both the parameters in this paper for the thin accretion disk study and later on focused only on $q_m$ with fixing $\beta=1$ for the shadow part. It can be seen in the Figure that as we increase the magnetic charge $q_m$ or NLE parameter $\beta$, the effective potential starts showing maxima close to the black hole which acts as a centrifugal barrier to the test particle falling into the black hole due to its high gravity. Therefore, we study the effect of the parameters on the massive particle trajectory around the black hole in the equatorial plane of the disk.


To achieve circular orbits that are stable, certain conditions must be met, including having an effective potential of zero (written as $V_\text{eff}=0$) and a radial effective potential of zero (written as $V_{\text{eff},r}=0$). Additionally, we need to determine the specific energy $(E)$, specific angular momentum $(L)$, and angular velocity $(\Omega)$ of the test particle that's moving in the equatorial plane of the accretion disk around the compact object. These values can be obtained in the following manner:
\begin{equation}
    E=-\frac{g_{tt}}{\sqrt{-g_{tt}-g_{\phi \phi}\Omega^2}}
\end{equation}
\begin{equation}
    L=-\frac{g_{\phi \phi}\Omega}{\sqrt{-g_{tt}-g_{\phi \phi}\Omega^2}}
\end{equation}
\begin{equation}
    \Omega=\frac{d\phi}{dt}=\sqrt{\frac{-g_{tt,r}}{g_{\phi \phi,r}}}
\end{equation}

In this paper, it is considered that the inner edge of the accretion disk is the innermost stable circular orbit $r_\text{ISCO}$ position therefore to obtain the $r_\text{ISCO}$ radius for any general spherically symmetric black hole space-time, we use the following required condition, $V_{\text{eff},rr}=0$, and get the following equation,
\begin{equation}
    E^2g_{\phi \phi,rr}+L^2g_{tt,rr}+(g_{tt}g_{\phi \phi})_{rr}=0
\end{equation}
We looked at a black hole space-time that's described by the metric function (Eq. \eqref{12} ), and we presented a table (Table~\ref{Tab:I}) that shows the positions of the horizon, marginally bound orbit, and ISCO (innermost stable circular orbit) for different magnetic charge values ($q=0.1,0.3,0.5$) and similarly for the $\beta$ in (Table~\ref{Tab:I1}). It was observed that the positions of the horizon, marginally stable orbit and ISCO increase with increasing the magnetic charge $q_m$ while decreasing with increasing the NLED parameter $\beta$. The magnitude of this change is notable when compared to the Schwarzschild case. Therefore, one can study the properties of accretion disks and the images of black hole shadows to understand the impact of magnetically charged NLED black holes on observational features.

\begin{table*}
\centering
\begin{tabular}{ c  c  c  c }
\hline
\hline
 Parameter & $q_m=0.1$ & $q_m=0.3$ & $q_m=0.5$ \\
\hline
Horizon ($r_h$)   & $2.06229$    & $2.31019$ &   $2.65752$ \\
 \\ marginally bound orbit ($r_\text{mb}$) &   $4.12618$  & $4.63310$   & $5.34596$ \\
\\ Innermost stable circular orbit ($r_\text{ISCO}$) & $6.18684$ &  $6.93044$ & $7.97331$\\
\hline
\end{tabular}
\caption{The position of the horizon, marginally bound orbit, and ISCO for different coupling parameter value $q_m$ when $\beta=0.3$.}
\label{Tab:I}
\end{table*}

\begin{table*}
\centering
\begin{tabular}{ c  c  c  c }
\hline
\hline
 Parameter & $\beta=0.1$ & $\beta=0.3$ & $\beta=0.5$ \\
\hline
Horizon ($r_h$)   & $2.42203$    & $2.31019$ &   $2.26779$ \\
 \\ marginally bound orbit ($r_\text{mb}$) &   $4.85638$  & $4.6331$0   & $4.54840$ \\
\\ Innermost stable circular orbit ($r_\text{ISCO}$) & $7.26623$ &  $6.93044$ & $6.80303$\\
\hline
\end{tabular}
\caption{The position of the horizon, marginally bound orbit, and ISCO for different coupling parameter value $\beta$ when $q_m=0.3$.}
\label{Tab:I1}
\end{table*}


In this study, we focus on a thin accretion disk and assume that its vertical height ($H$) is much smaller than its radius $(r)$. We also assume that the disk is in a state of local hydrodynamic equilibrium, which means that the pressure and vertical entropy gradients are insignificant. The disk is able to maintain its vertical structure because it has efficient cooling, which prevents it from accumulating heat generated by the dynamic friction in the disk. The disk is also in a steady state, which means that its mass accretion rate ($\dot{M}$) remains constant over time. The inner edge of the disk is fixed at the ISCO position, while the matter that's far from the black hole follows Keplerian motion. We can represent the stress energy-momentum tensor for the matter that's accreting around the compact object in a specific form \cite{NovikovThorne,1974ApJ...191..499P}
\begin{equation}
    T^{\mu \nu}=\rho_0 u^\mu u^\nu+2u^{(\mu}q^{\nu)}+t^{\mu \nu}.
\end{equation}
The conditions $u_\mu q^\mu=0$ and $u_\mu t^{\mu \nu}=0$ apply here. In these equations, $u^\mu$ represents the four-velocity of the orbiting particles, while $\rho_0$, $q^\mu$, and $t^{\mu \nu}$ represent the rest mass density, energy flow vector, and stress tensor of the accreting matter, respectively. We can use the conservation of rest mass, represented by $\triangledown_\mu (\rho_0 u^\mu)=0$, to calculate the time-averaged mass accretion rate $(\dot{M})$. This accretion rate is independent of the radius of the accretion disk
\begin{equation}
    \dot{M}=-2\pi \sqrt{-g}\Sigma u^r=\text{const}.
\end{equation}
The variable $\Sigma$ represents the time-averaged surface density and is defined as such:
\begin{equation}
    \Sigma= \int_{-h}^{+h} <\rho_0>dz.
\end{equation}
In this context, $z$ is being used as a coordinate in the cylindrical coordinate system. We have utilized the conservation laws for both energy, represented by $\triangledown_\mu E^\mu=0$, and angular momentum $(L)$, represented by $\triangledown_\mu J^\mu=0$, to obtain the following equations:
\begin{equation} \label{e77}
    [\dot{M}E-2\pi \sqrt{-g} \Omega W^r_\phi]_{,r}=4\pi rF(r)E,
\end{equation}
and
\begin{equation} \label{e78}
    [\dot{M}L-2\pi \sqrt{-g} \Omega W^r_\phi]_{,r}=4\pi rF(r)L,
\end{equation}
where $W^r_\phi$ is called the averaged torque and given by
\begin{equation}
    W^r_\phi= \int_{-h}^{h} <t^r_\phi> dz.
\end{equation}

The value $<t^r_\phi>$ is the $(\phi,r)$ component of the stress-energy tensor which is determined by averaging it over a time interval $\delta t$ and an angle $\delta \phi=2\phi$. Equation \eqref{e77} represents a balanced-energy equation where the rest mass energy of the disk ($\dot{M} E$) and the energy caused by the torque present in the disk ($2 \pi \sqrt{-g} \Omega W^r_\phi$) are in equilibrium with the energy radiated ($4 \pi r F(r) E$) from the surface of the disk.

The equation \eqref{e78} depicts the balance of angular momentum, in which the transferred angular momentum by the rest mass of the disk ($\dot{M} L$) and that due to the present torque in the disk ($2\pi \sqrt{-g} \Omega W^r_\phi$) is equally counterbalanced by the transferred angular momentum from the surface of the disk through the outgoing radiation ($4 \pi r F(r) L$).

\begin{figure*}
    \centering
    \includegraphics[width=0.48\textwidth]{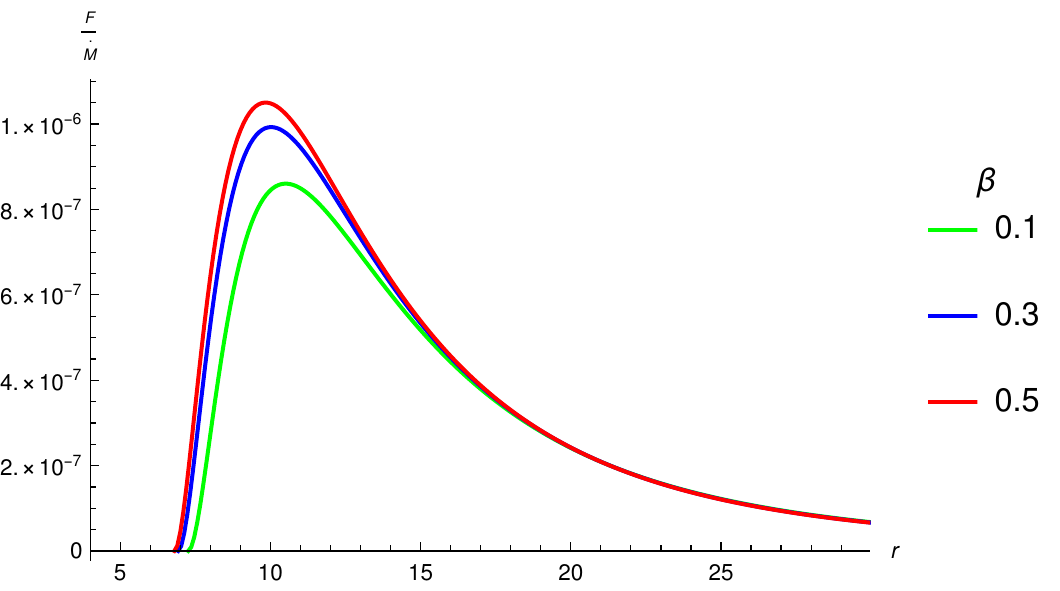}
    \includegraphics[width=0.48\textwidth]{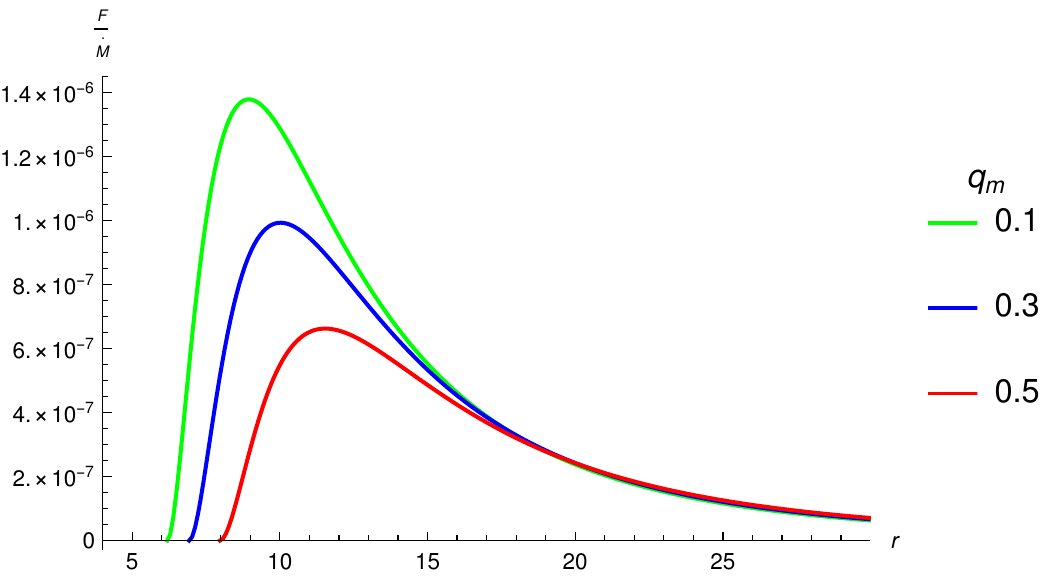}
    \caption{Radiation flux for different values of NLE parameter $\beta=0.1 (\text{green})$, $\beta=0.3 (\text{blue})$, and $\beta=0.5 (\text{red})$ when $q_m=0.3$ in the left side plot while for magnetic charge $q_m=0.1 (\text{green})$, $q_m=0.3 (\text{blue})$ and $q_m=0.5 (\text{red})$ when $\beta=0.3$ in the right.} \label{fig:m}
\end{figure*}

\begin{figure*}
    \centering
    \includegraphics[width=0.48\textwidth]{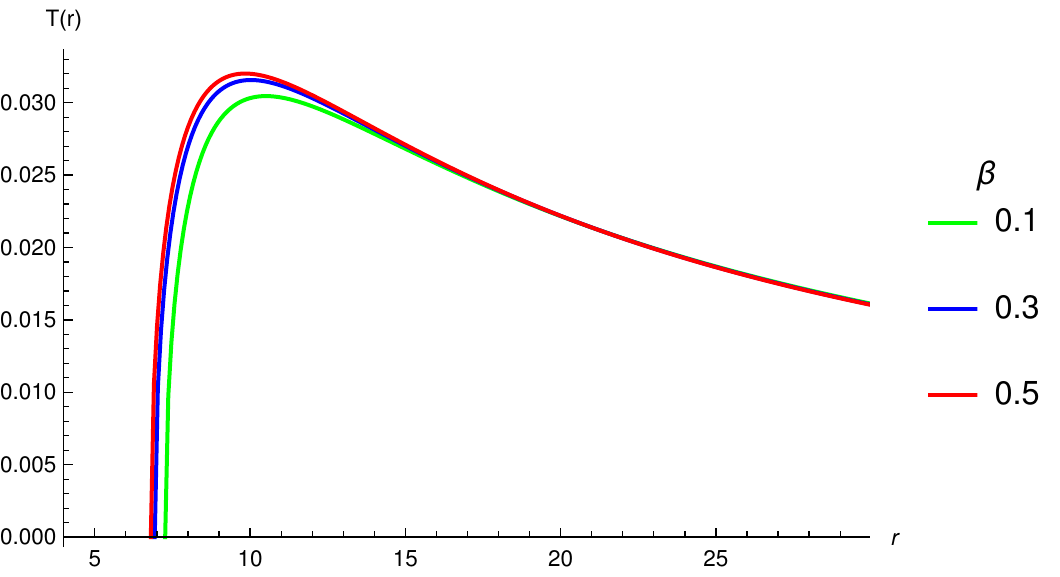}
    \includegraphics[width=0.48\textwidth]{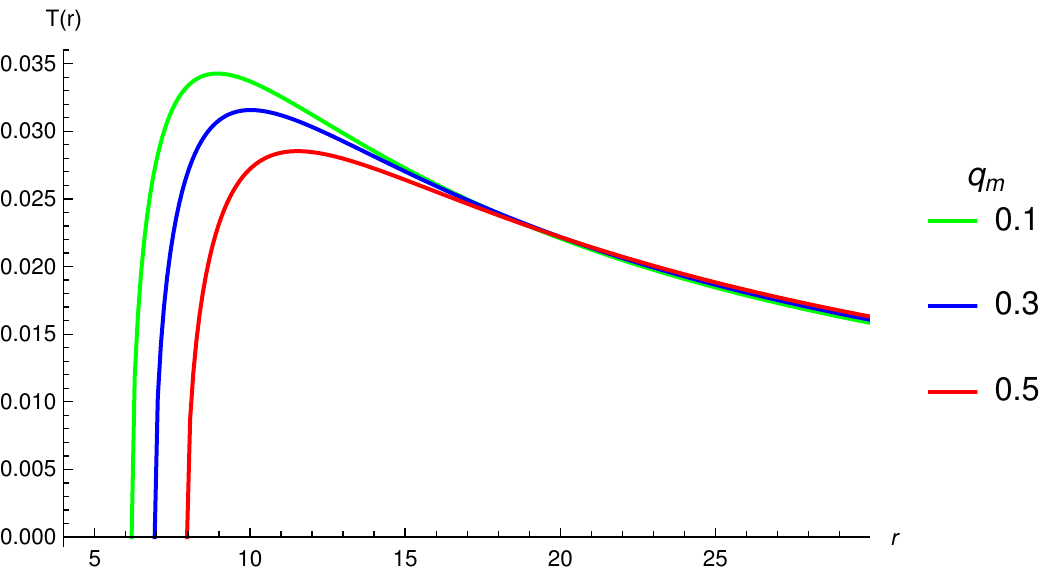}
    \caption{Temperature for different values of NLE parameter $\beta=0.1 (\text{green})$, $\beta=0.3 (\text{blue})$, and $\beta=0.5 (\text{red})$ when $q_m=0.3$ in the left side plot while for magnetic charge $q_m=0.1 (\text{green})$, $q_m=0.3 (\text{blue})$ and $q_m=0.5 (\text{red})$ when $\beta=1$ in the right.} \label{fig:n}
\end{figure*}

\begin{figure*}
    \centering
    \includegraphics[width=0.48\textwidth]{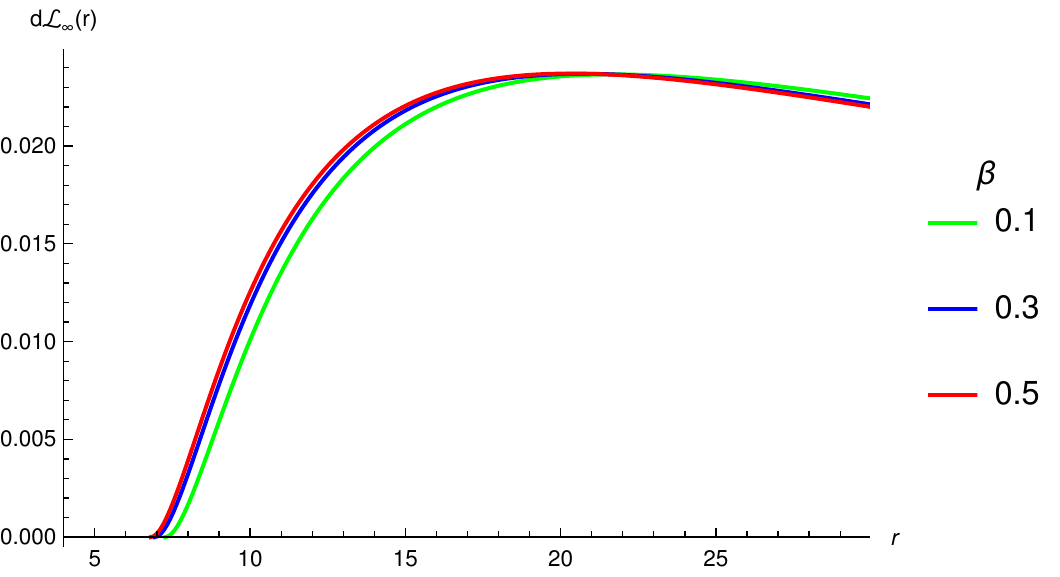}
    \includegraphics[width=0.48\textwidth]{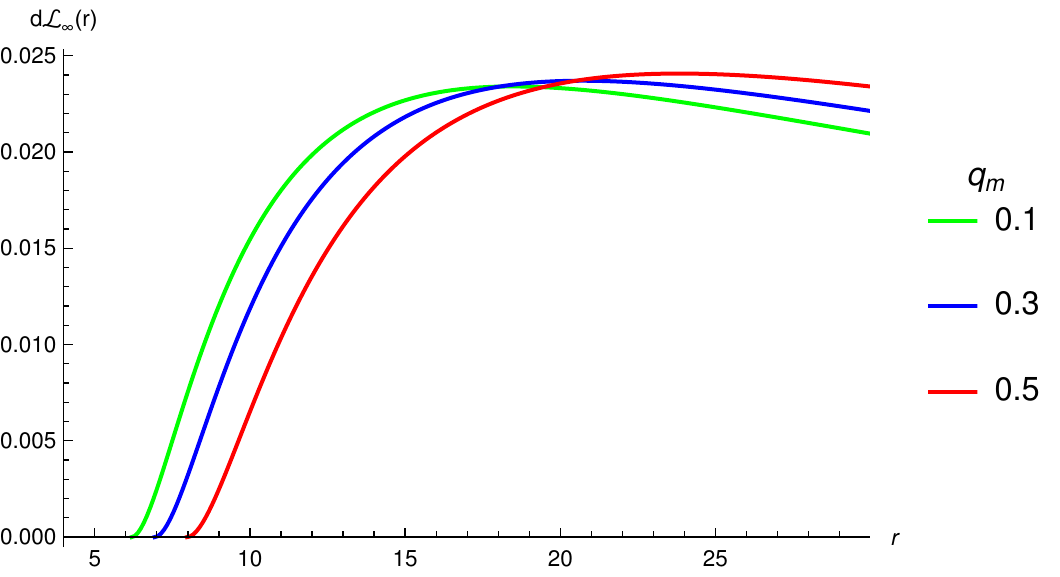}
    \caption{Differential luminosity for different values of NLE parameter $\beta=0.1 (\text{green})$, $\beta=0.3 (\text{blue})$, and $\beta=0.5 (\text{red})$ when $q_m=0.3$ in the left side plot while for magnetic charge $q_m=0.1 (\text{green})$, $q_m=0.3 (\text{blue})$ and $q_m=0.5 (\text{red})$ when $\beta=0.3$ in the right.} \label{fig:o}
\end{figure*}

\begin{figure*}
    \centering
    \includegraphics[width=0.48\textwidth]{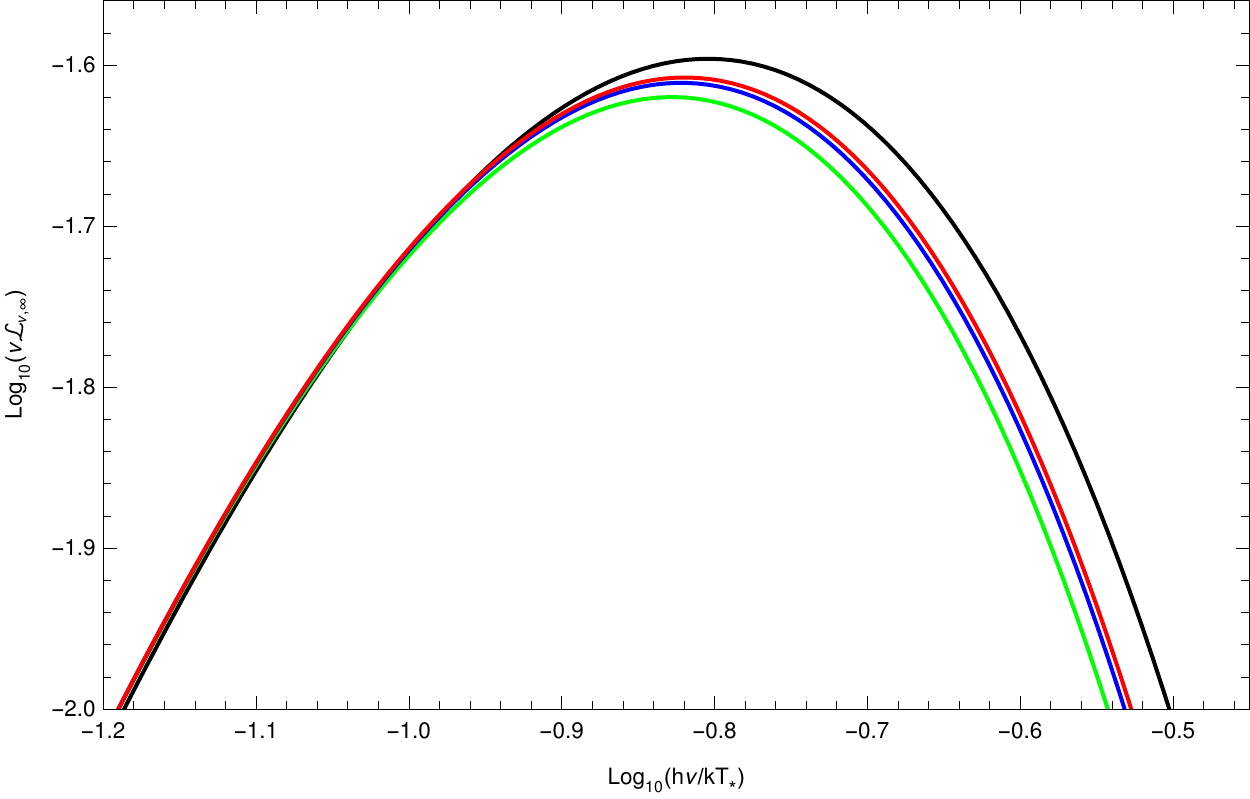}
    \includegraphics[width=0.48\textwidth]{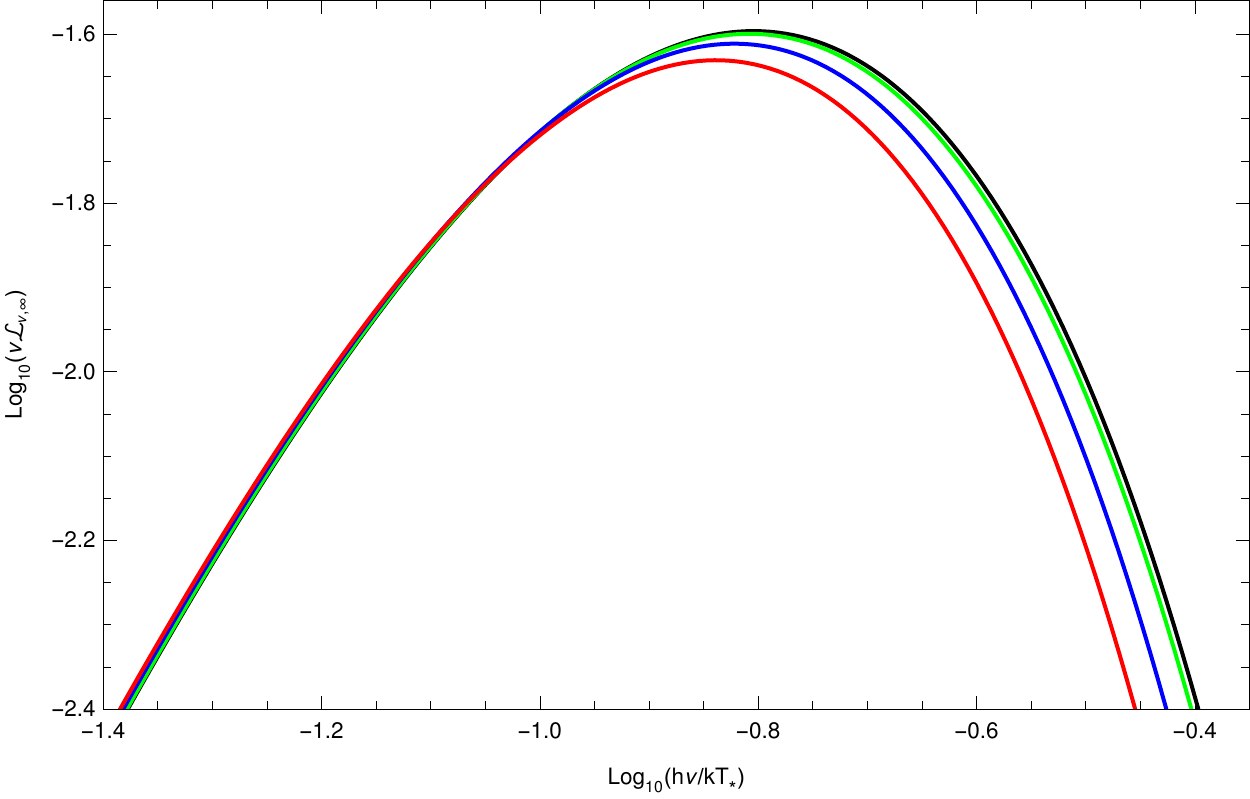}
    \caption{Spectral luminosity for different values of NLE parameter $\beta=0.1 (\text{green})$, $\beta=0.3 (\text{blue})$, and $\beta=0.5 (\text{red})$ when $q_m=0.3$ in the left side plot while for magnetic charge $q_m=0.1 (\text{green})$, $q_m=0.3 (\text{blue})$ and $q_m=0.5 (\text{red})$ when $\beta=0.3$ in the right. The black curve is for the Schwarzchild black hole.} \label{fig:o1}
\end{figure*}

By applying the energy-angular momentum relation, $E_{,r}=\Omega L_{,r}$, and eliminating $W^r_\phi$ from equations \eqref{e77} and \eqref{e78}, we can derive an expression for the time-averaged energy flux $F(r)$ that is emitted from the surface of an accretion disk surrounding a compact object
\begin{equation}
    F(r)=-\frac{\Omega_{,r}\dot{M}}{4\pi \sqrt{-g}(E-\Omega L)^2} \int_{r_\text{ISCO}}^{r}(E-\Omega L)L_{,r} dr.
\end{equation}
The relationship between the temperature of the accretion disk and the energy flux can be expressed as:
\begin{equation}
    F(r)=\sigma T^4(r).
\end{equation}
In the same vein, the Stefan-Boltzmann constant is denoted by $\sigma$. By taking into account both energy and angular momentum conservation laws, we can obtain the derivative of the luminosity $L_\infty$ at infinity as \cite{1974ApJ...191..499P,Joshi:2013dva}
\begin{equation}
    \frac{d\mathcal{L_\infty}}{d\ln r}=4 \pi r \sqrt{-g} E {F}(r).
\end{equation}
Now, considering the disk emission as a blackbody emission and calculating the spectral luminosity by converting the differential luminosity as a function of $(r)$ into the spectral luminosity with the dependency on frequency $(\nu)$ and get the following expression,
\begin{equation} \label{e79}
    \nu \mathcal{L}_{\nu,\infty}=\frac{60}{\pi^3} \int_{r_{ISCO}}^\infty \frac{\sqrt{g} E}{M^2} \frac{(u^t y)^4}{\exp{[u^t y/F^{1/4}]}-1} dr
\end{equation}
where, $y=h \nu /kT_*$ is a dimensionless parameter with $h$ Planck constant, $k$ Boltzmann constant, and $T$ characteristic temperature of the disk. The figures in Figs.\ref{fig:m}, \ref{fig:n}, \ref{fig:o} and \ref{fig:o1} depict the energy flux, temperature, differential luminosity, and spectral radiation, respectively. The results show that when the NLE parameter $\beta$ (left) increases, the radiative flux, temperature, differential luminosity, and spectral profile of the disk also increase whereas it decreases when the magnetic charge increases $q_m$ (right). This implies that for higher values of the $\beta$ and lower values of the $q_m$, the radiation emitted from the disk is greater, and the temperature of the disk will be higher compared to the other values of $\beta$ and $q_m$. The spectral luminosity is particularly interesting from an observational standpoint, as both the parameters $\beta$ and $q_m$ have shifted the curve downward compared to the Schwarzschild space-time case ($\beta=q_m=0$). The effect of $q_m$ is much more prominent in comparison to the effect of $\beta$ in the lower branch of the spectrum with respect to the Schwarzschild case. Therefore, magnetically charged black holes exhibit higher spectral luminosity than uncharged black holes, and in the future researchers may want to explore the possibility of detecting such black holes through observations.

\subsection{Direct Emission, Lensing Ring, And Photon Ring}

In this section, we investigate the impact of photon trajectories on the black hole shadow and the observed emission profile for an observer within the framework magnetically charged NLED black hole. 
Moreover, we aim to examine the trajectories of photons around a magnetically charged NLED black hole and their impact on the observed emission profile and black hole shadow for an observer located at a finite distance from the black hole. To accomplish this, we adopt a method introduced by Wald et al. \cite{Gralla:2019xty} to categorize null rays emanating from the observer's sky based on the number of orbits $n(\gamma)$, which represents the final azimuthal coordinate $\phi$ of the ray after it has completely escaped the effective gravitational field of the black hole. This parameter indicates how many times the geodesic $\gamma$ has crossed the accretion disk's equatorial plane and is defined as follows:
(1) Null rays with $n<3/4$ correspond to direct emission and have only crossed the equatorial plane once.

(2) Null rays with $3/4<n<5/4$ correspond to the lensing ring and have crossed the equatorial plane twice.

(3) Null rays with $n>5/4$ correspond to the photon ring and have crossed the equatorial plane more than twice.

\begin{figure*}[htbp]
    \centering
    \includegraphics[width=.3\textwidth]{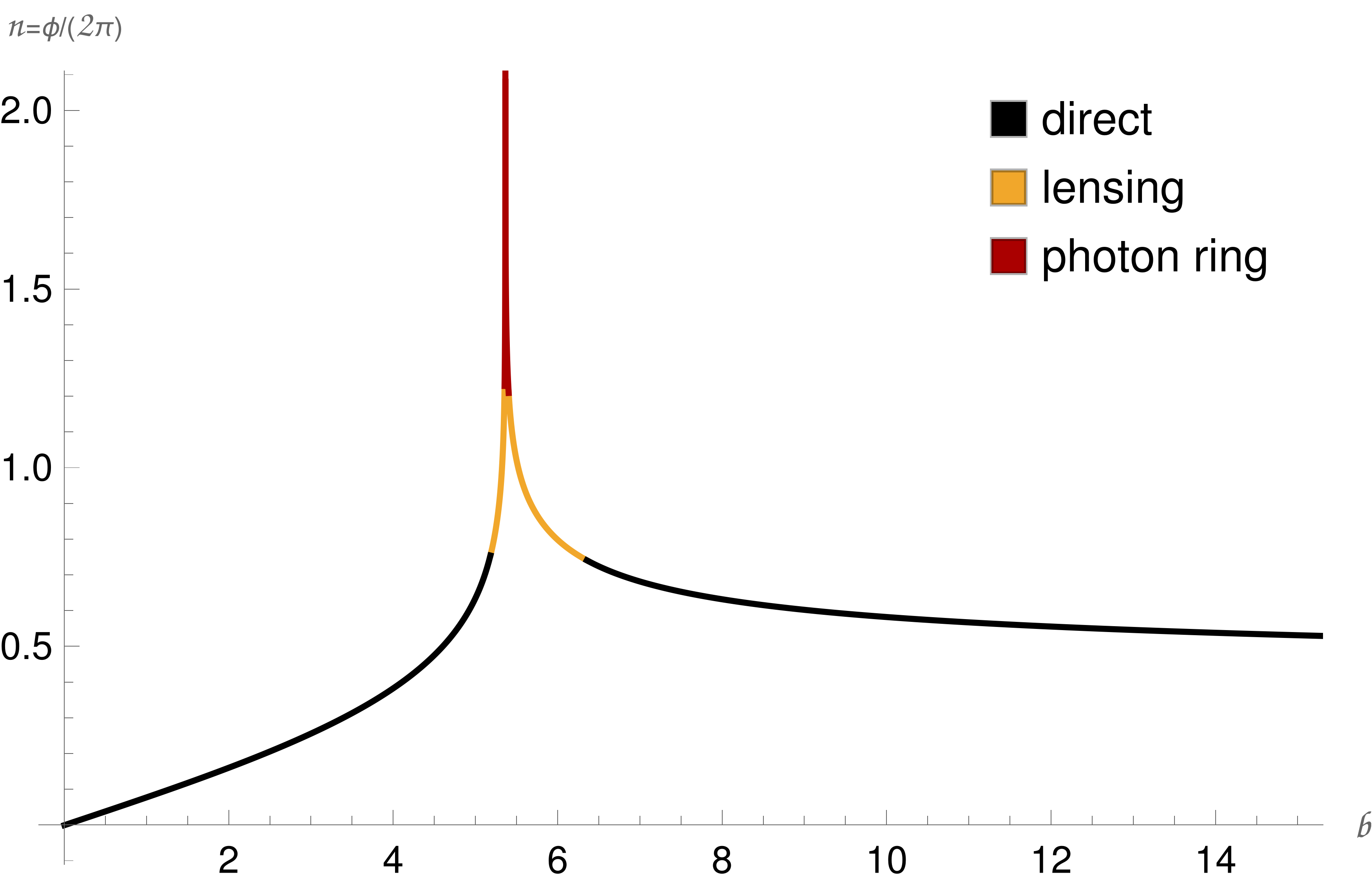} \hfill
    \includegraphics[width=.3\textwidth]{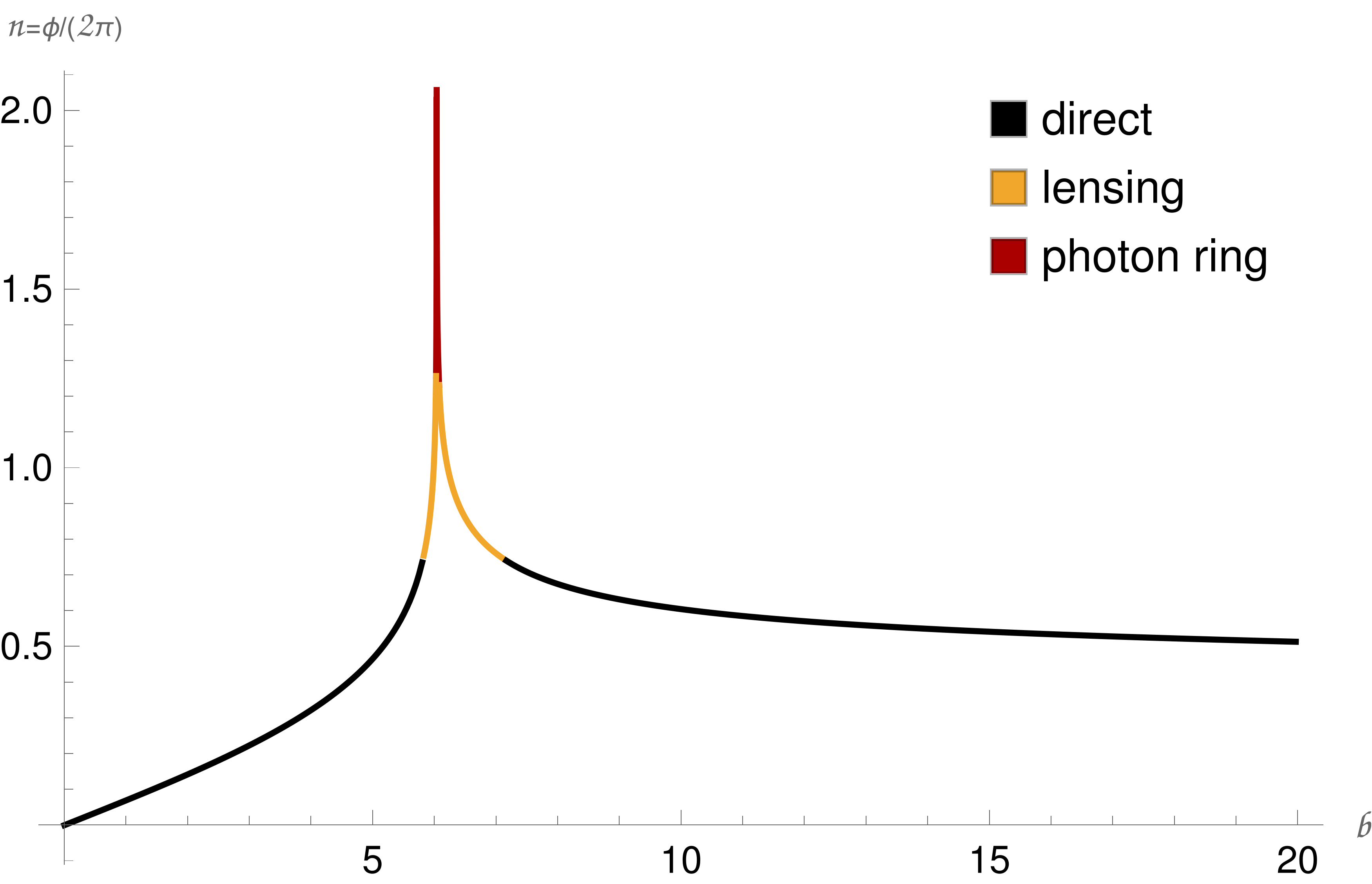} \hfill
    \includegraphics[width=.3\textwidth]{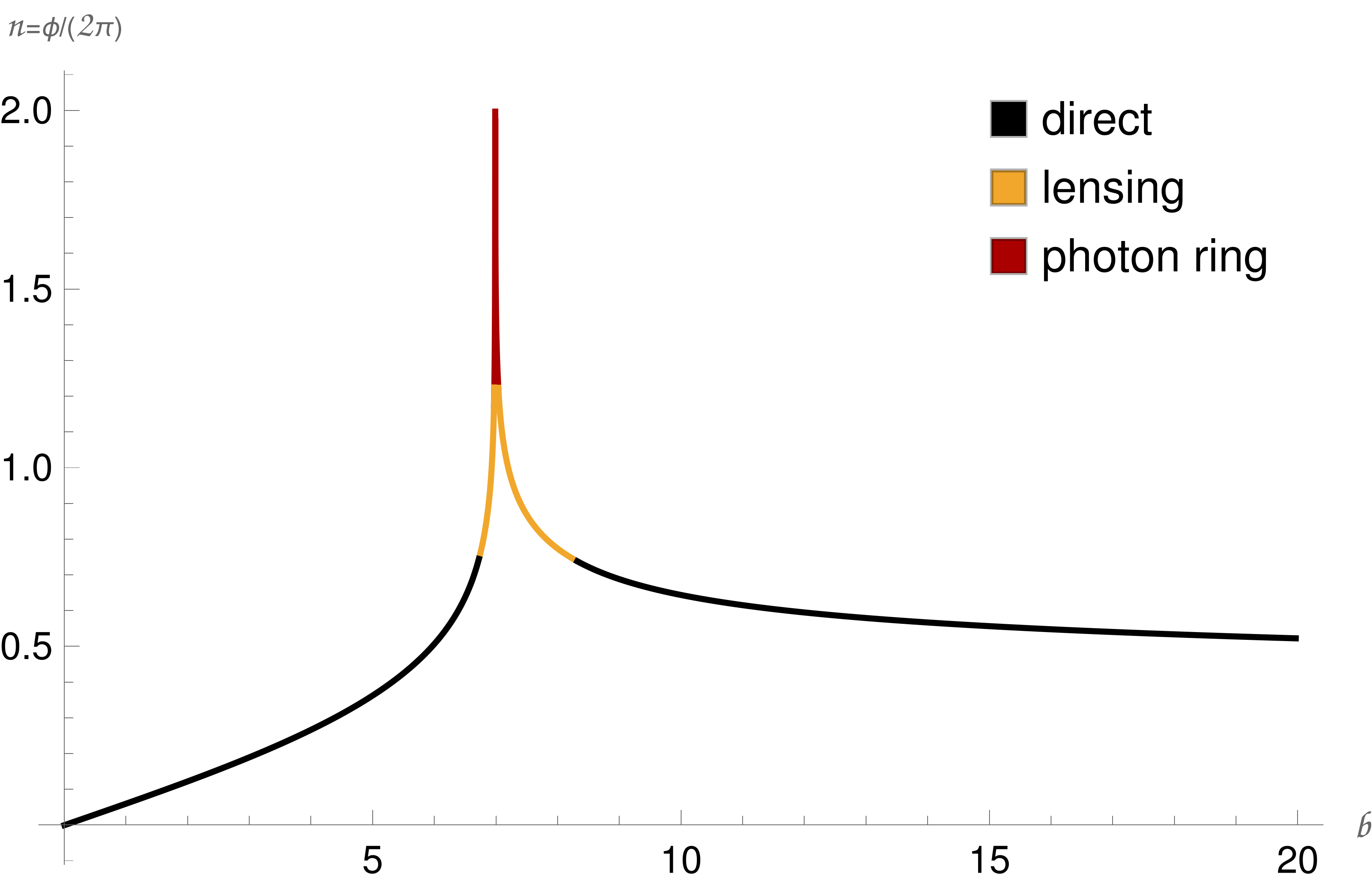} \hfill
    \vfill
    \centering
    \includegraphics[width=.3\textwidth]{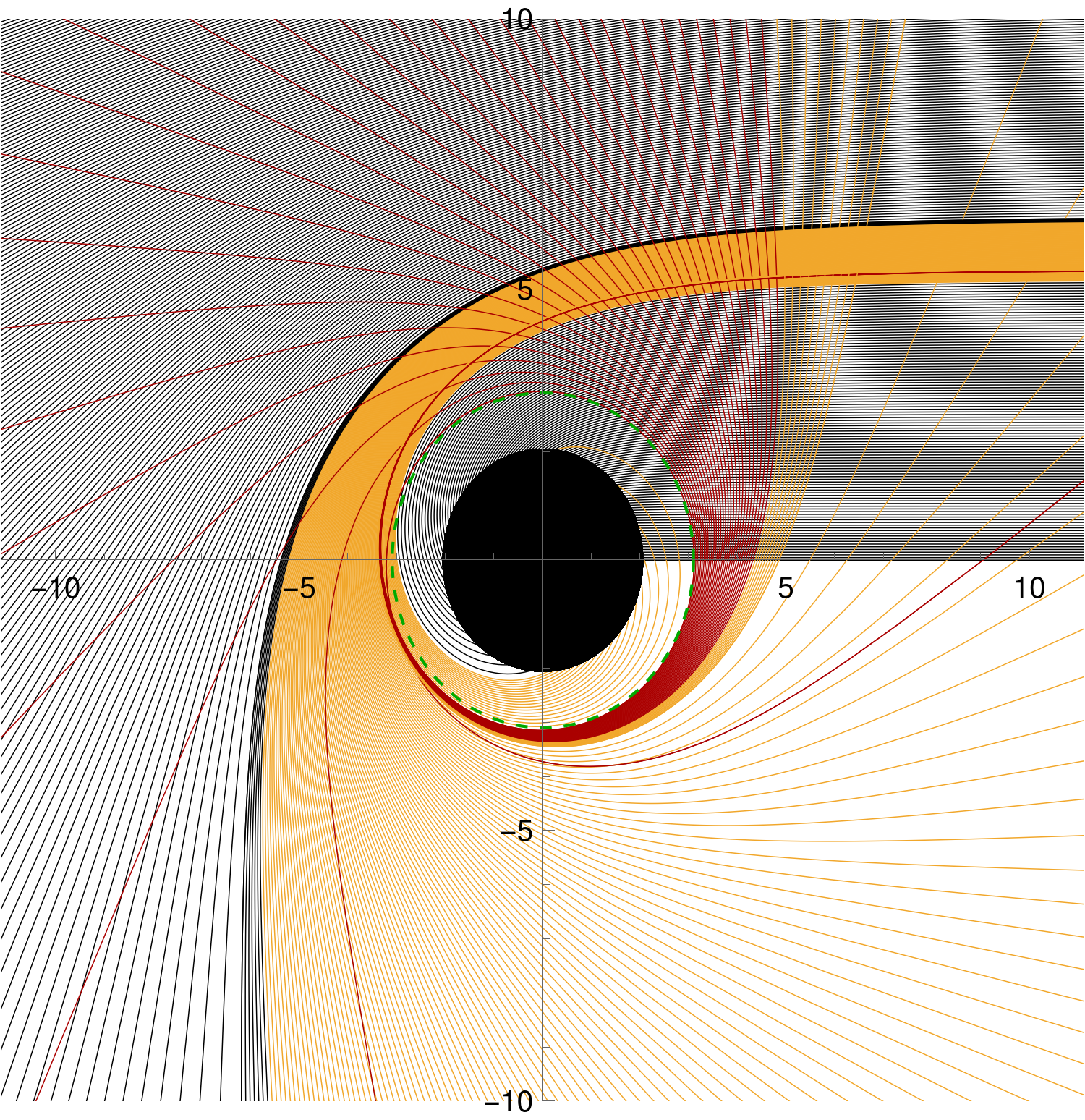} \hfill
    \includegraphics[width=.3\textwidth]{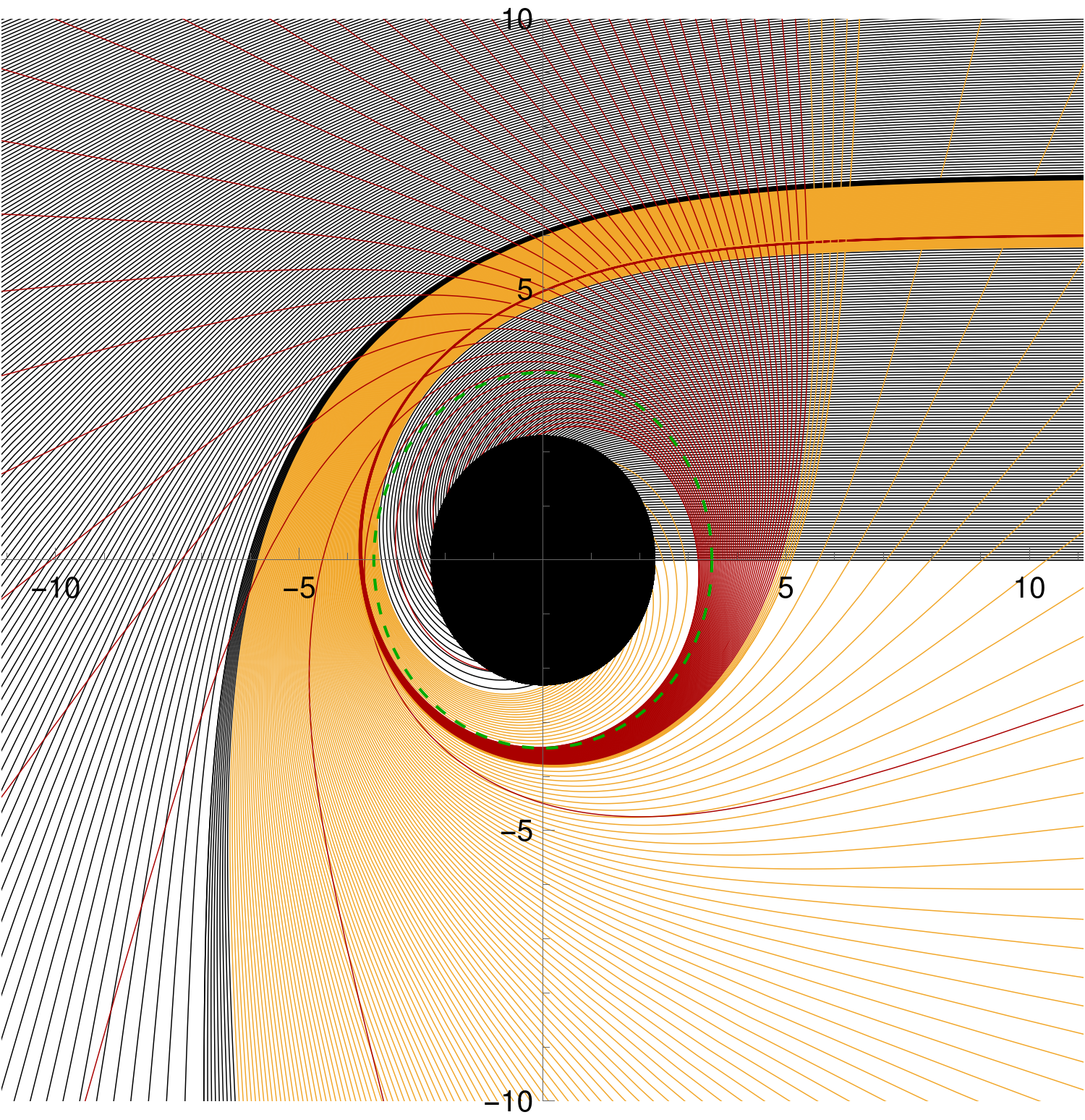} \hfill
    \includegraphics[width=.3\textwidth]{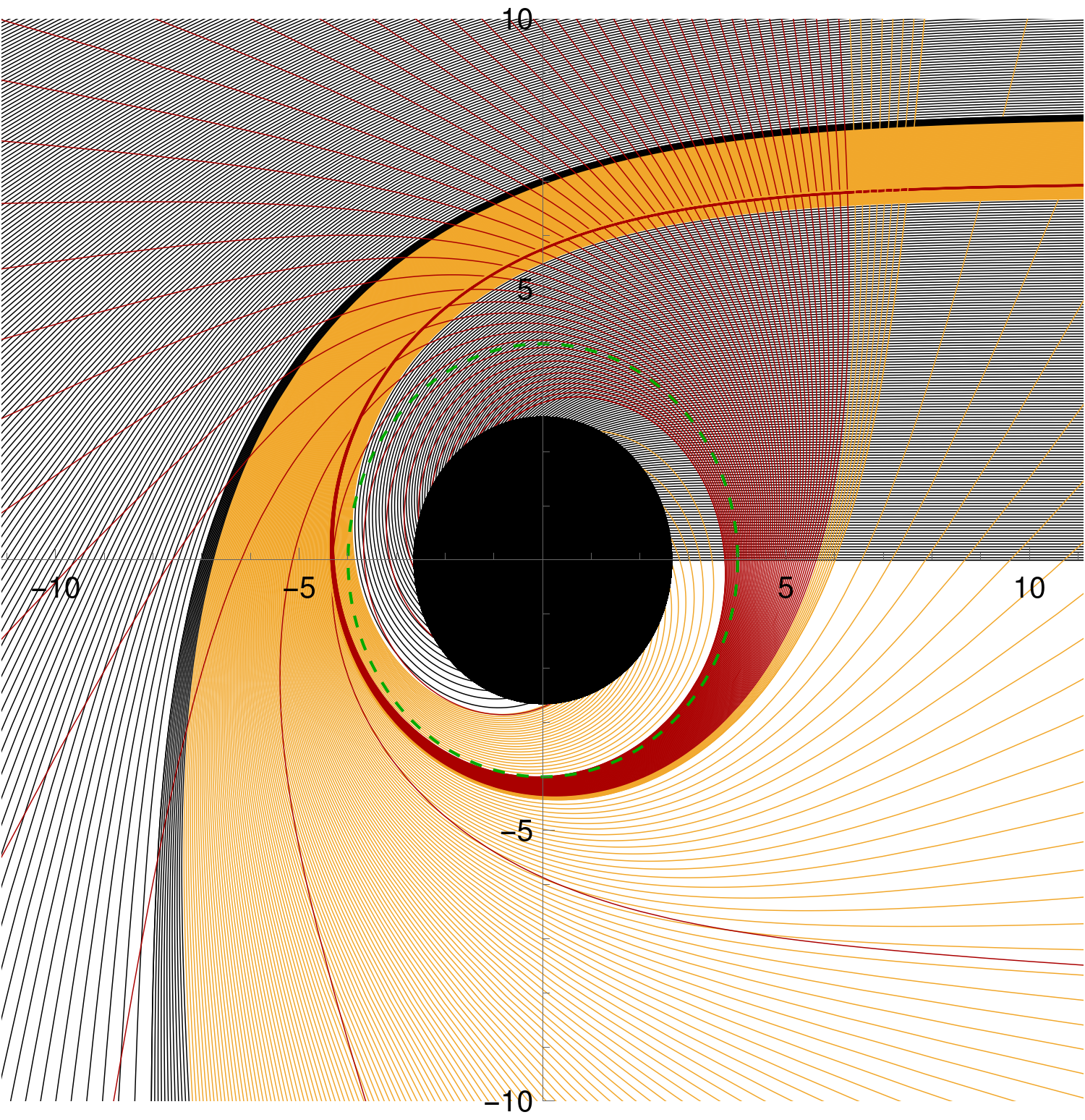} \hfill
    \caption{The behaviour of photons was observed at various $q_m$ values ($0.1$, $0.3$, and $0.5$), with the red, orange, and black lines indicating the photon ring, lensing ring, and direct emission, respectively with $\beta=0.3$. The green dotted line represents the photon orbit, while the black disk shows the event horizon of the black hole. The top panel displays the fractional number of orbits ($n = \phi/(2\pi)$), where $\phi$ is the total change in azimuthal angle outside the horizon. In the bottom panel, selected photon trajectories are displayed, treating $r$ and $\phi$ as Euclidean polar coordinates.}
    \label{fig:15}
\end{figure*}

In this figure \ref{fig:15}, we have presented a plot of the number of orbits $n$ versus the impact parameter $b$. The direct emission rays are represented by the color black, the lensing ray by yellow, and the photon ring rays by red. The green dashed circle in the ray tracing figure indicates the photon orbit. It should be noted that we utilized the Okyay-\"Ovg\"un \textit{Mathematica} notebook package \cite{Okyay:2021nnh} to calculate the thin-accretion disk, null geodesics, and shadow cast, which was also employed in \cite{Chakhchi:2022fls}.
\begin{table*}
\centering
\begin{tabular}{ p{3cm} p{4cm} p{4cm} p{4cm}}
 \hline
 \hline
 Parameter & $q_m=0.1$ & $q_m=0.3$ & $q_m=0.5$ \\
 \hline
 Direct Emission \newline $n<3/4$   & $b<5.18215$ \newline $b>6.35215$    & $b<5.82532$ \newline $b>7.16532$ &   $b<6.73498$ \newline $b>8.31498$ \\
 \\ Lensing Ring \newline $3/4<n<5/4$ &   $5.18215<b<5.36215$ \newline $5.40215<b<6.35215$  & $5.82532<b<6.02532$ \newline $6.07532<b<7.716532$   & $6.73498<b<6.97498$ \newline $7.03498<b<8.31498$ \\
\\ Photon Ring \newline $n>5/4$ & $5.36215<b<5.40215$ & $6.02532<b<6.07532$ &  $6.97498<b<7.03498$\\
\hline
\end{tabular}
\caption{Figure shows the region where the direct emission, lensing ring, and photon ring occur depending on the value of the magnetic charge $q_m$ with $\beta=0.3$.}
\label{Tab:II}
\end{table*}


In Table \ref{Tab:II}, we present the changes in the black hole shadow as the magnetic charge $q_m$ increases. Our main objective was to investigate the impact of the magnetic charge on photon trajectories around the NLED black hole. Fig.~\ref{fig:15} is a 2D plot showing the same. The range of lensing and photon rings increases with $q_m$, and the shadow angular diameter also increases as the horizons shift outwards with the magnetic charge. An increase in $q_m$ leads to an increase in the contribution of brightness from lensing and photon rings. Furthermore, when the impact parameter is near the critical impact parameter $b \pm b_\text{c}$ for the photon trajectories, the photon orbit indicates a narrow but sharp peak in the $(b,\phi)$ plane. For all the cases, the photon trajectories show that as we continually increase the impact parameter, all trajectories will fall into the category of direct emission.

\subsection{Transfer functions and observed specific intensities}
\begin{figure*}[htbp]
    \centering
    \includegraphics[width=.32\textwidth]{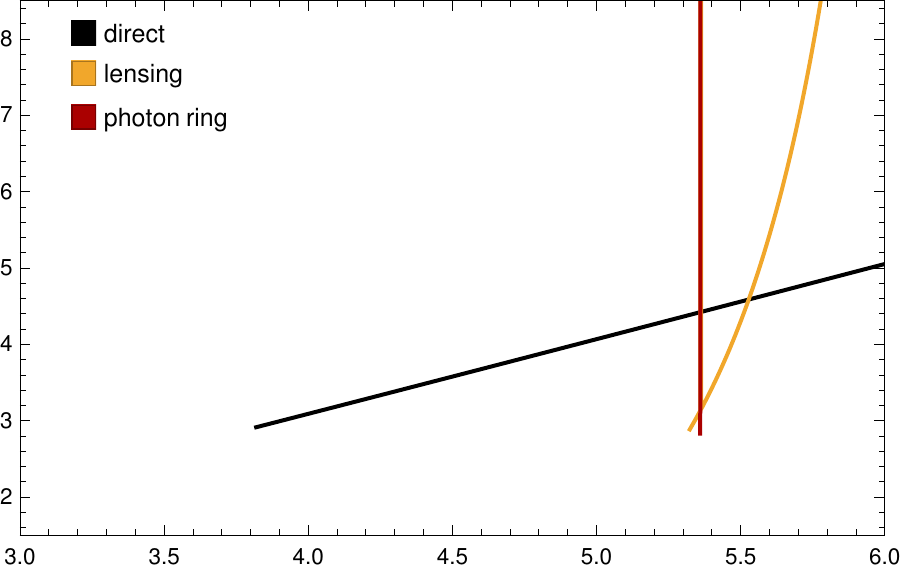}
    \includegraphics[width=.32\textwidth]{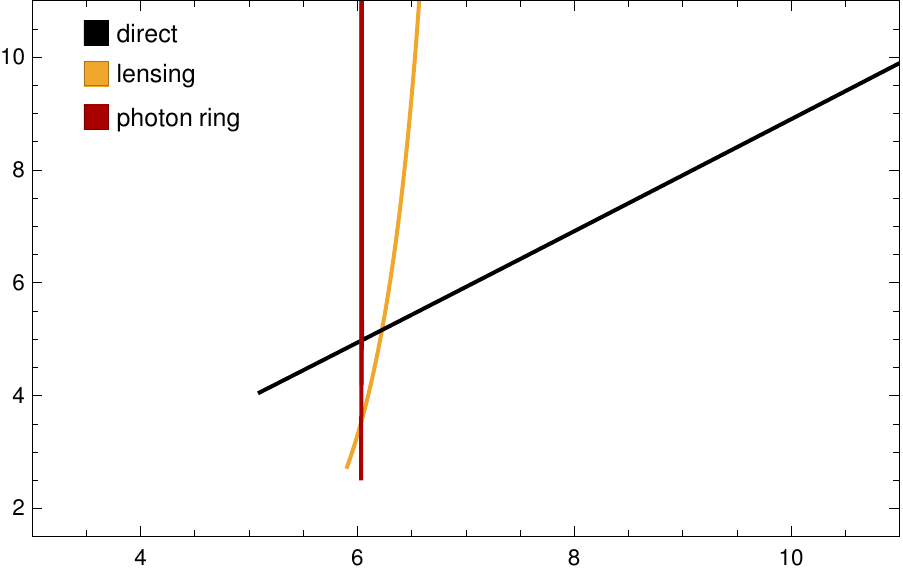}
    \includegraphics[width=.32\textwidth]{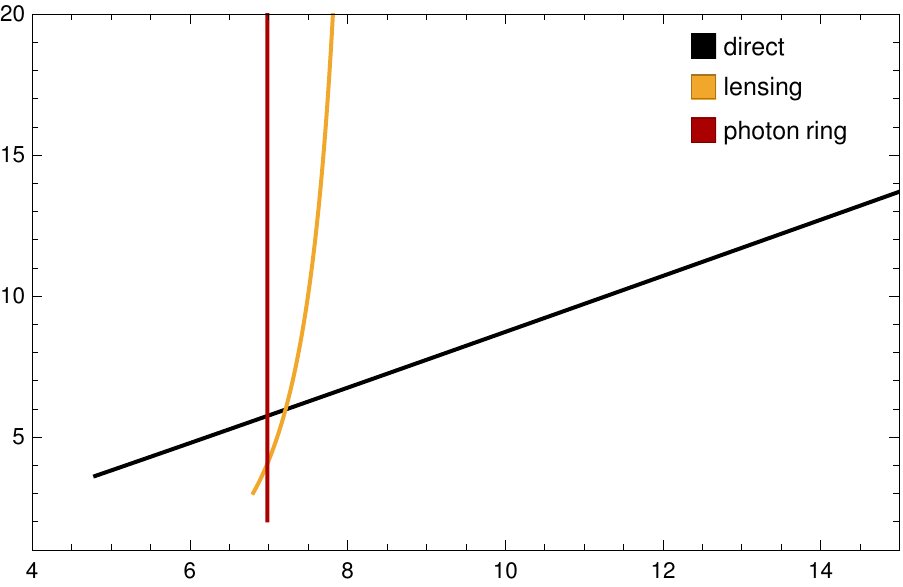}
    \caption{Different values of $q_m$ (including $0.1$, $0.3$, and $0.5$) were used to calculate the first three transfer functions of an NLE black hole with $\beta=0.3$. The transfer functions were plotted on the $y$-axis as $r_m(b)$, and the impact parameter ($b$) was plotted on the $x$-axis.} \label{fig:16}
\end{figure*}

Here, we aim to explore the intensity emanating from the black hole as detected by an observer located a finite distance away. We make the assumption that the disk releases radiation isotropically within the stationary worldline's rest frame, and therefore, Liouville's theorem can be applied to the ratio of the emitted intensity ($I^{em}_{\nu}$) and its frequency ($\nu_e^3$) along the path of the light ray. Consequently, the formula for the observed intensity can be derived
\begin{equation}
    I_{\nu'}^\text{obs}=g^3 I_\nu^\text{em}.
\end{equation}
We define $g=\sqrt{f(r)}$, use it to calculate the observed intensity $I_{\nu'}^{obs}$ of a frequency ($\nu'$) coming from a black hole. We integrate all the frequencies using the equation,  $\nu'=g d\nu$, with the total emitted specific intensity being given by $I_\text{em}=\int I_\nu^{em} d\nu$. Finally, we can calculate the observed frequency using this information
\begin{equation}
    I^\text{obs}=g^4 I_\text{em}.
\end{equation}
Therefore, the sum of the received intensity for an observer will be given as
\begin{equation}
    I(r)=\sum_{n} I^\text{obs}(r)|_{r=r_m(b)},
\end{equation}
The $m^{th}$ intersection in the equatorial plane outside the horizon is called the transfer function, denoted as $r_m(b)$ in the text. It gives us the relationship between the impact parameter of the photon with the radial coordinates. The slope of this transfer function is known as the demagnification factor, which tells us about the demagnified scale of the photon transfer function \cite{Gralla:2019xty,Zeng:2020vsj}. Accordingly, the transfer functions have been represented for different values of the magnetic charge "$q_m$" (see Fig. \ref{fig:16}). 

In Figure \ref{fig:16}, we plotted the transfer function $(r_m(b))$ with impact parameter $(b)$ for $q_m=0.1,0.3$, and $0.5$ with $\beta=0.3$. In all the plots, black dots depict the transfer function for the direct emission which has a nearly constant slope and corresponds to $m=1$. The slope of the direct emission represents the redshifted source profile for photons received by the observer. The transfer function for the lensing ring has been shown by yellow dots which corresponds to $m=2$. It is observed that for the lensing ring as $b$ approaches the critical impact parameter $b_\text{c}$, the slope for $m=2$ is small, but it increases rapidly as $b$ increases. Since the slope is much greater than one, the back side image of the accretion disk will be demagnified for the observer receiving the photons from the back side of the disk. And the photon ring is represented by the red dots which correspond to $m=3$ having an infinite slope for the transfer function. This indicates that the front side of the accretion disk will be highly demagnified for the observer. Therefore, the main contribution to the observed flux is coming from the direct emission of photons from the accretion disk. Here, we are not considering the emissions which correspond to $(m>4)$ because of their minimal contribution to the observed flux due to their infinite slopes in the transfer function. Hence, we are only concerned with the transfer function corresponding to $m=1,2$, and $3$.

\subsection{Observed characteristics of direct emission, photon, and lensing rings}
The specific brightness is solely dependent on the radial coordinate $r$ for the observer. Thus, we will now discuss three theoretical models that are ideal for the emission intensity profile from the accretion disk, denoted as $I_\text{em}$
\begin{itemize}
  \item Model 1:\[
    I_\text{em}^1(r)= 
\begin{cases}
    \left( \frac{1}{r-(r_\text{ISCO}-1)} \right)^2   ,&  r\geq r_\text{ISCO}\\
    0,              & r \leq r_\text{ISCO}
\end{cases}
\]
  \item Model 2:\[   I_\text{em}^2(r)= 
\begin{cases}
    \left( \frac{1}{r-(r_\text{ph}-1)} \right)^3   ,&  r\geq r_\text{ph}\\
    0,              & r \leq r_\text{ph}
\end{cases}
\]
   \item Model 3:\[   I_\text{em}^3(r)= 
\begin{cases}
    \frac{1-\arctan(r-(r_\text{ISCO}-1))}{1-\arctan(r_\text{ph})}    ,&  r\geq r_\text{h}\\
    0.              & r \leq r_\text{h}
\end{cases}
\]
\end{itemize}
The models consist of three distinct radii: $r_\text{ISCO}$, $r_\text{ph}$, and $r_\text{h}$, representing the innermost stable circular orbit, photon sphere around the black hole, and the event horizon of the black hole respectively. Each of the models exhibits unique features. For example, the second model has a rapid decay rate, while the third model has a very slow decay rate compared to the others. Also, the third model considered the disk to be distributed till the horizon and the emission starts directly from the horizon, while the second model considered that emission starts from the photon sphere and the first model considered emission from the ISCO.

\begin{figure*}[htbp]
    \centering
    \includegraphics[width=.3\textwidth]{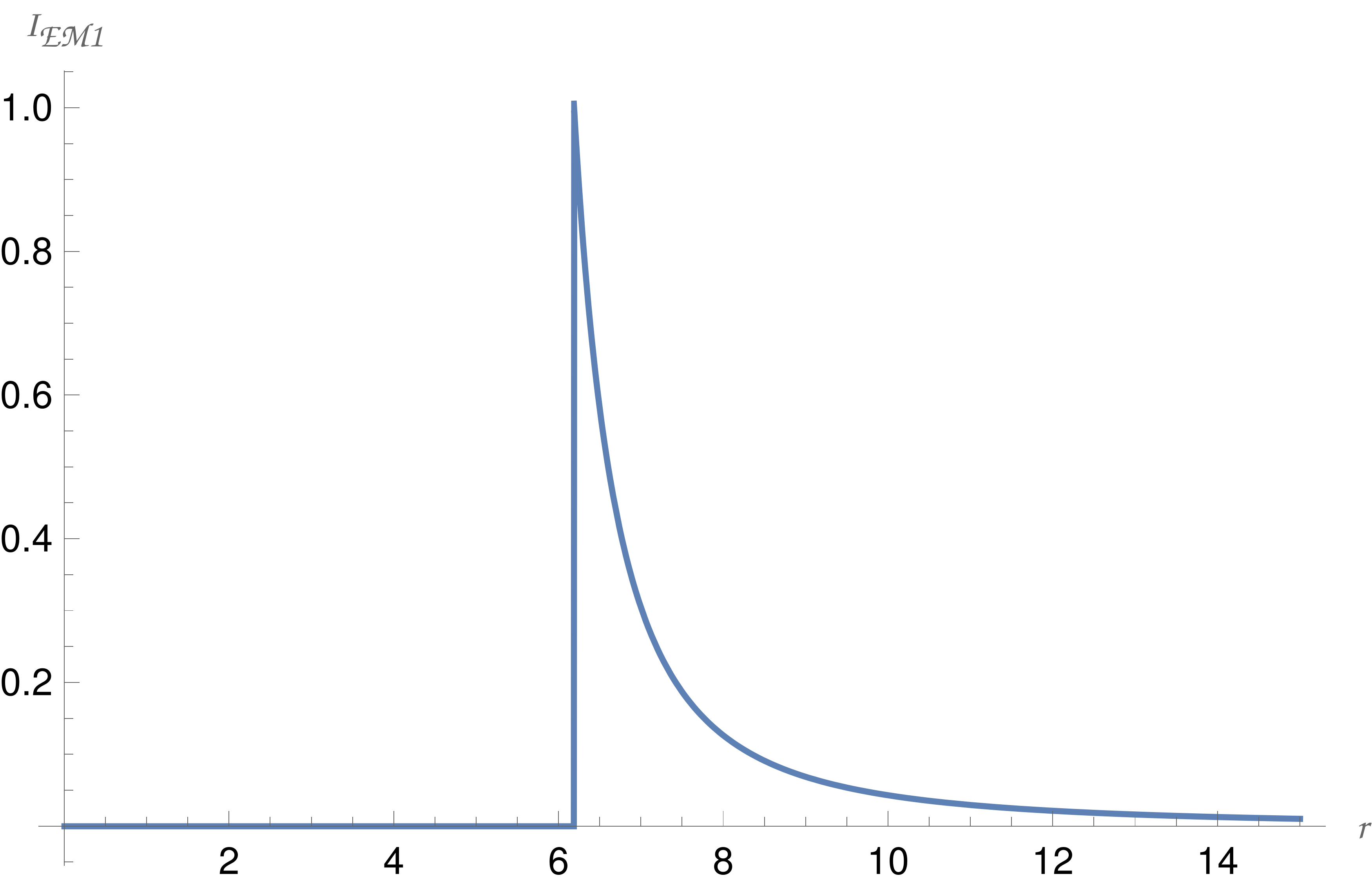}
    \includegraphics[width=.3\textwidth]{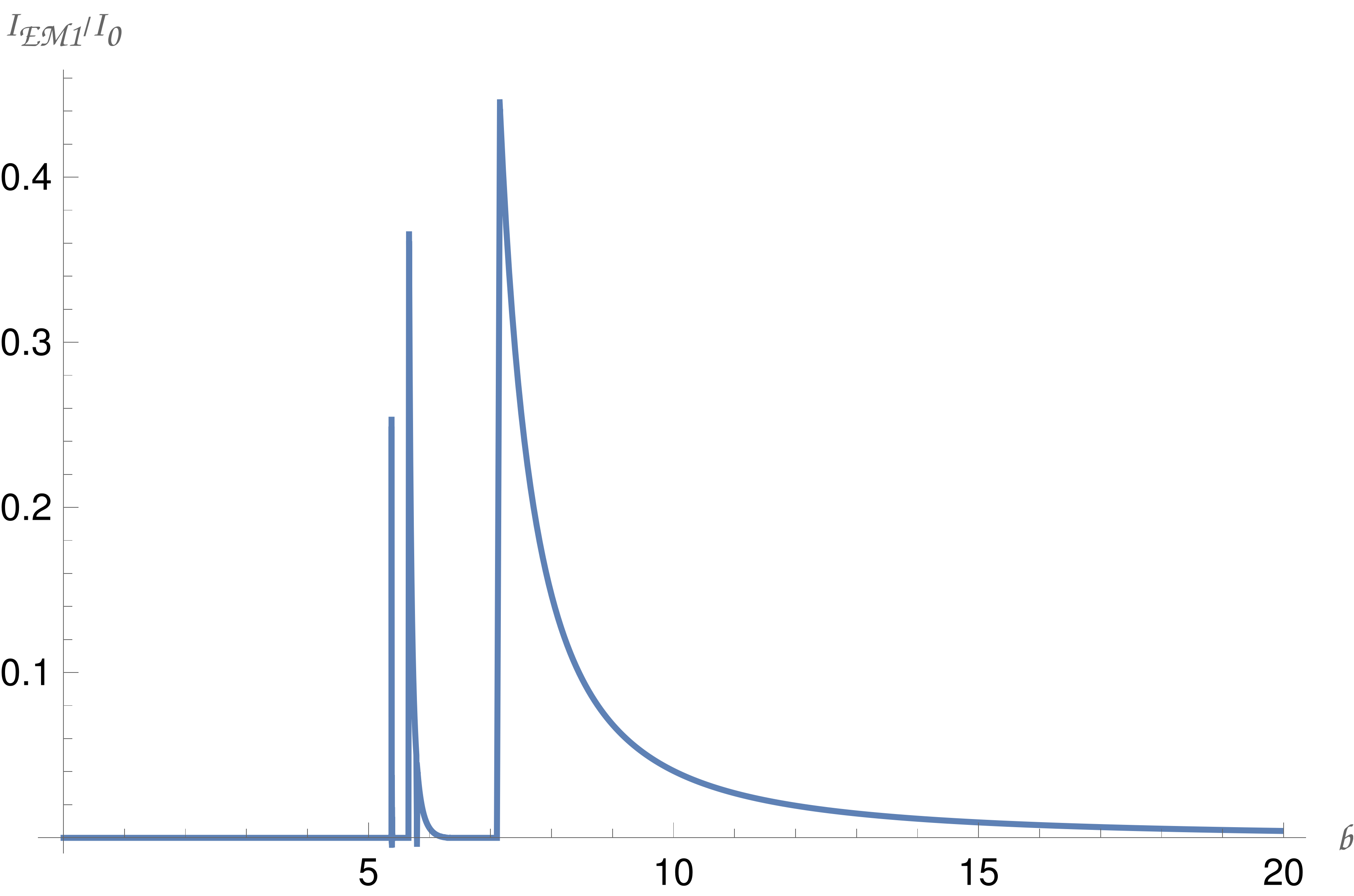}
    \includegraphics[width=.3\textwidth]{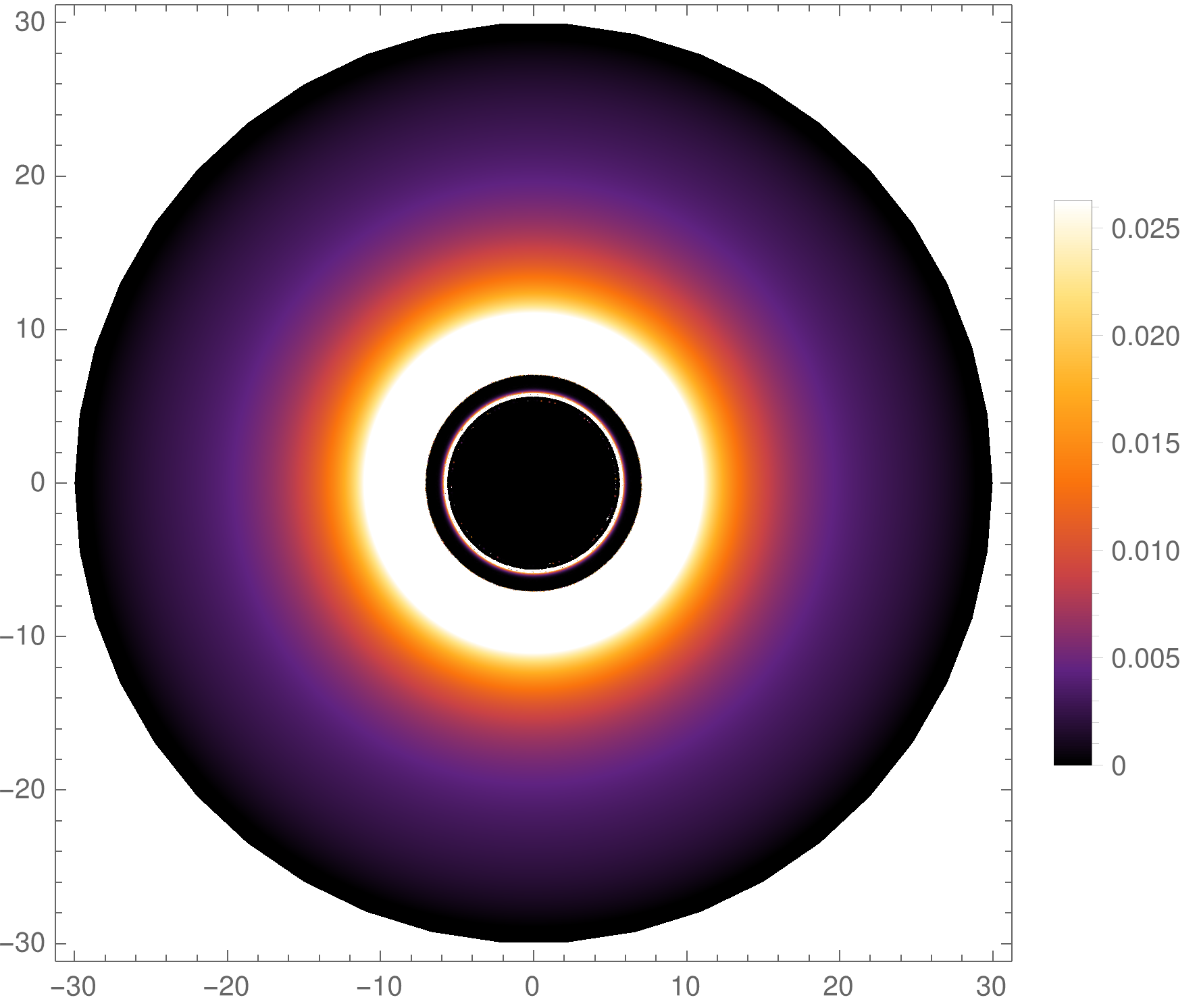}
    \vfill
    \centering
    \includegraphics[width=.3\textwidth]{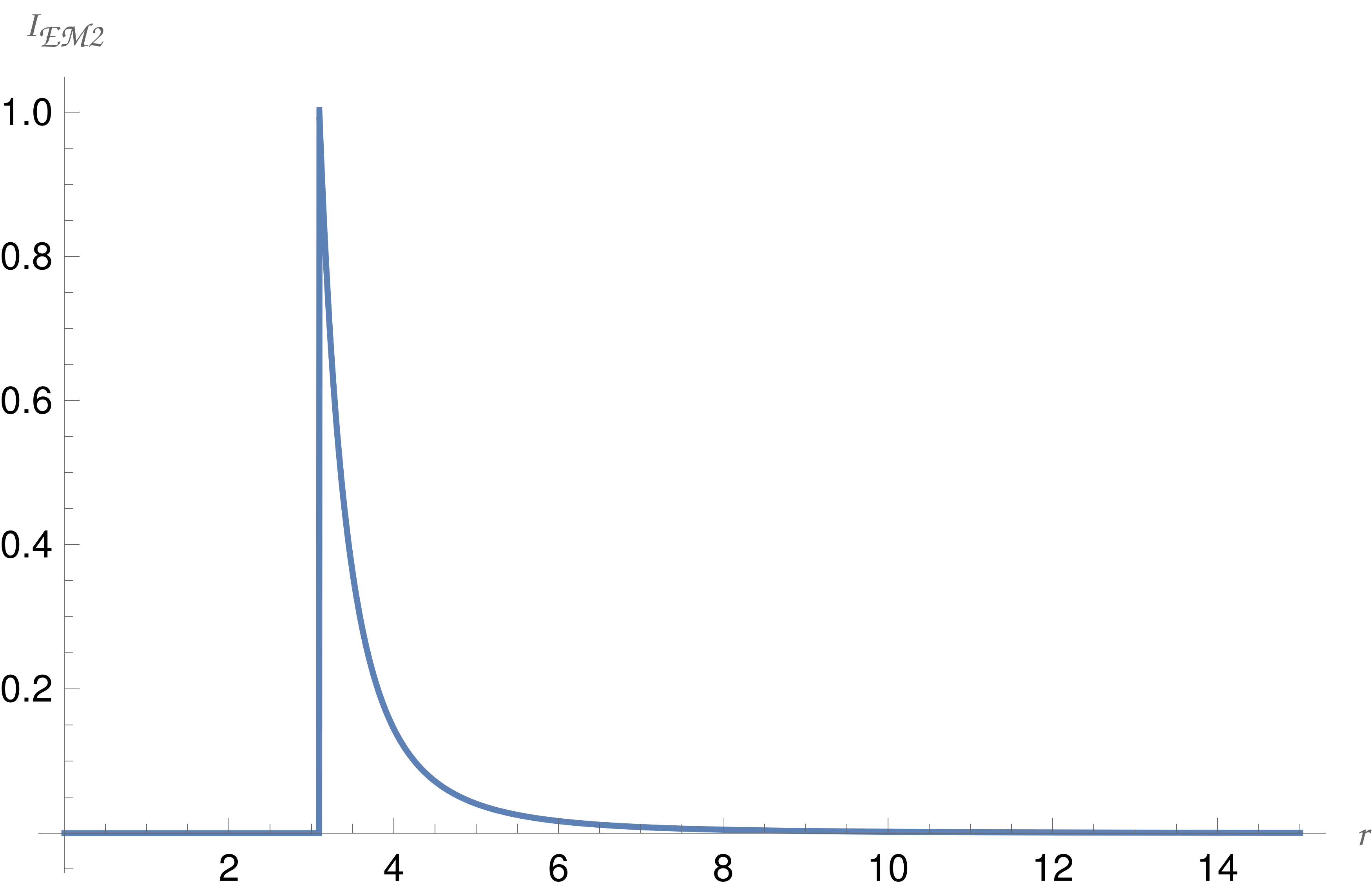}
    \includegraphics[width=.3\textwidth]{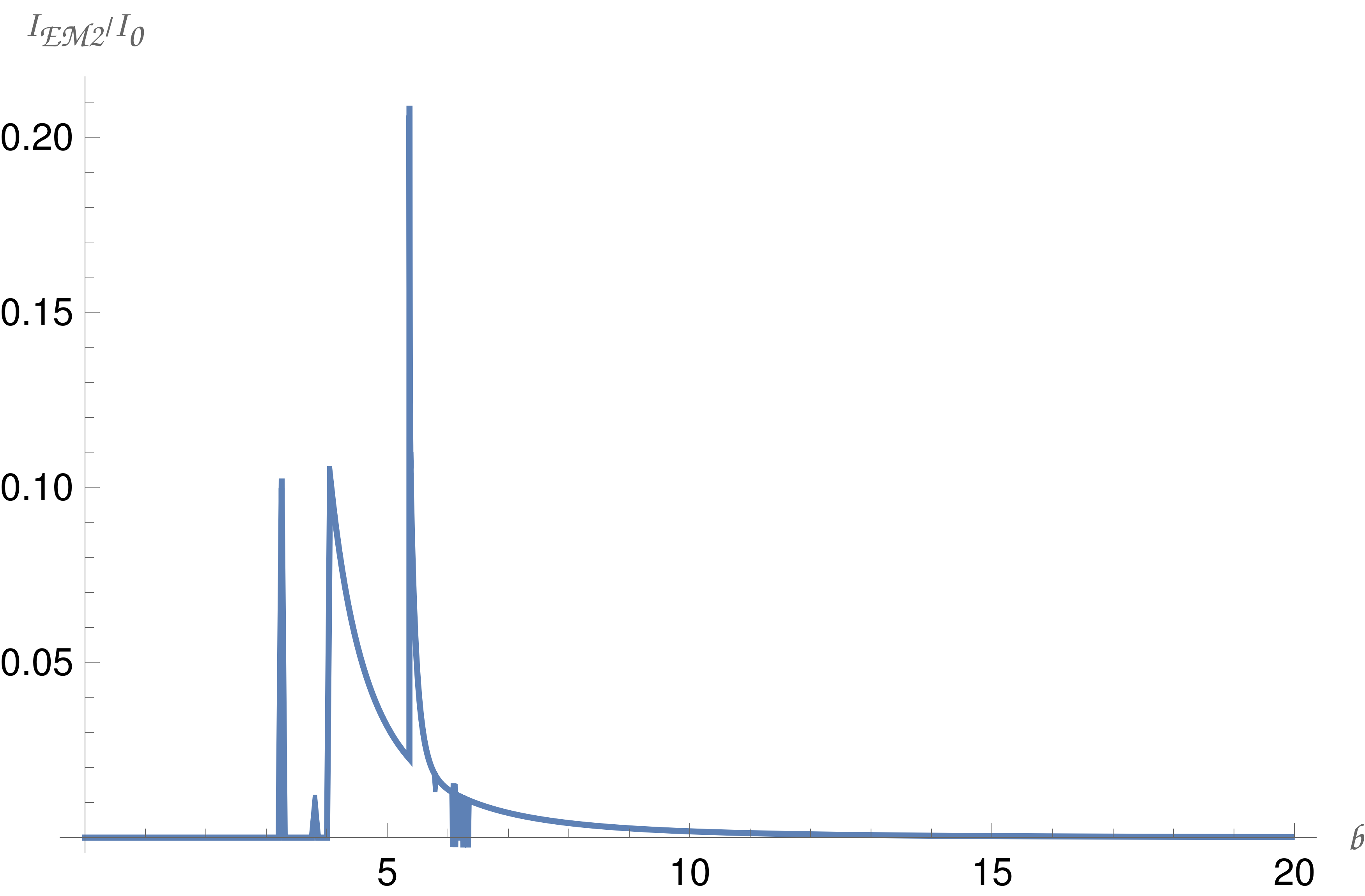}
    \includegraphics[width=.3\textwidth]{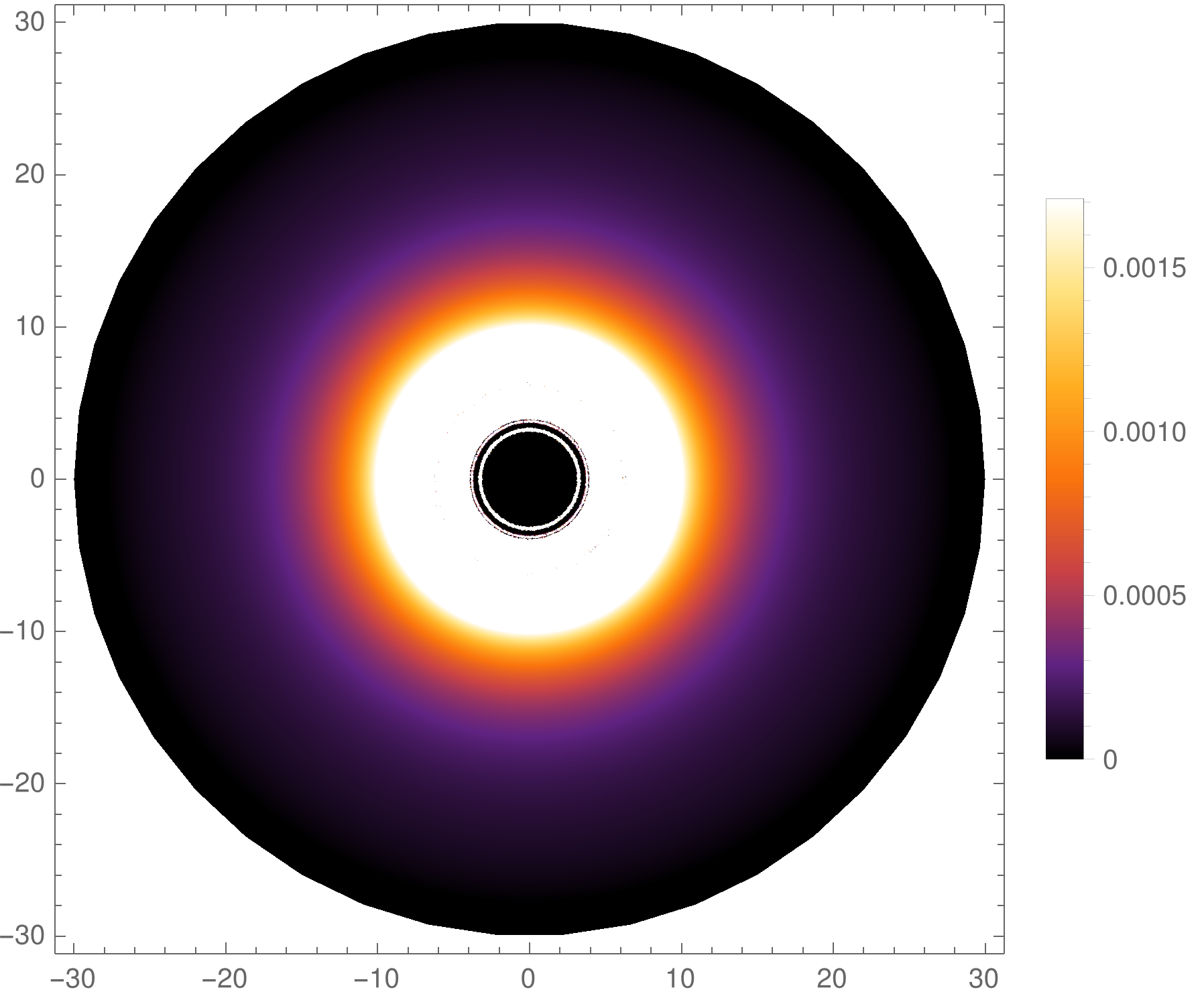}
    \vfill
    \centering
    \includegraphics[width=.3\textwidth]{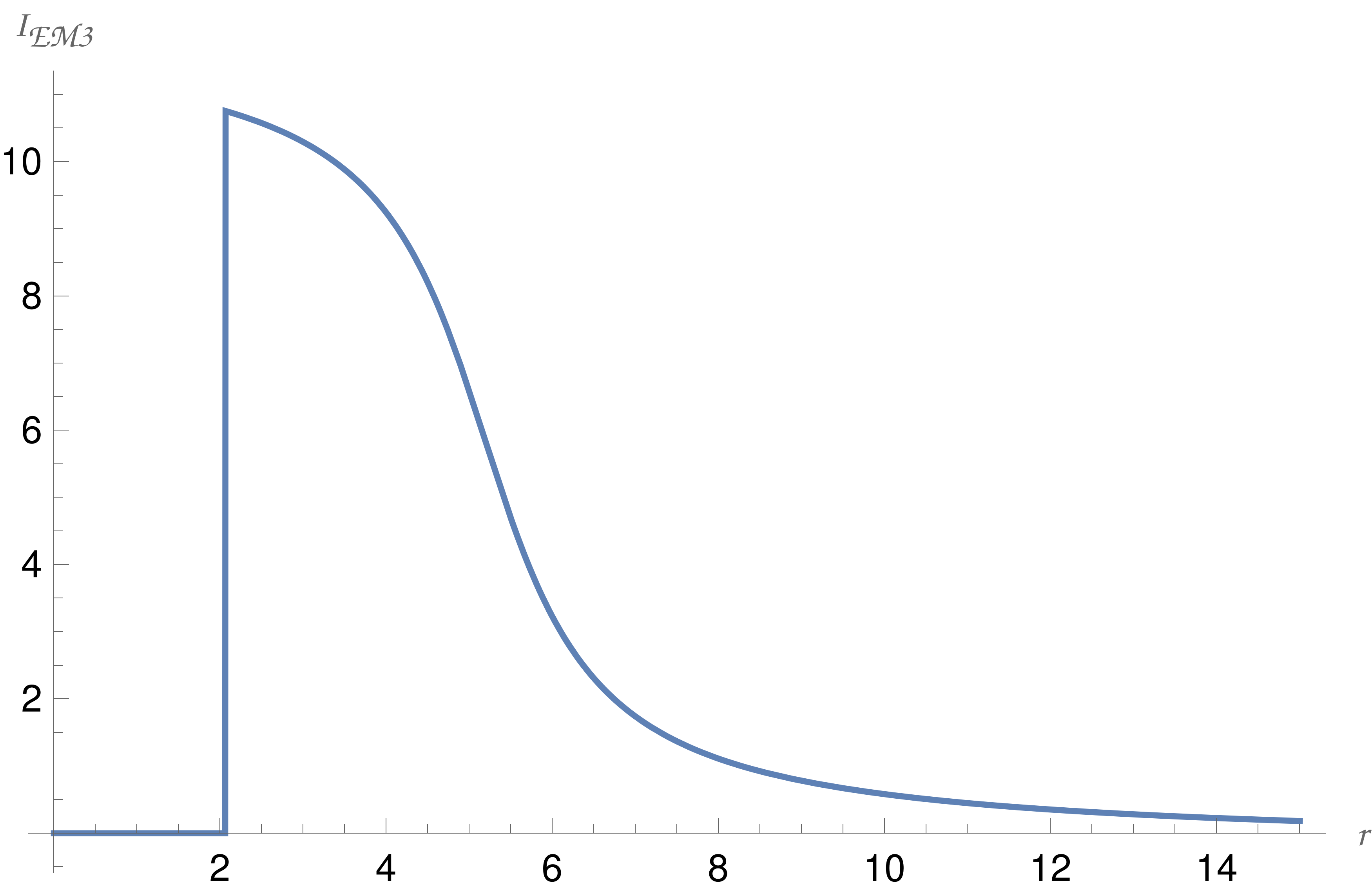}
    \includegraphics[width=.3\textwidth]{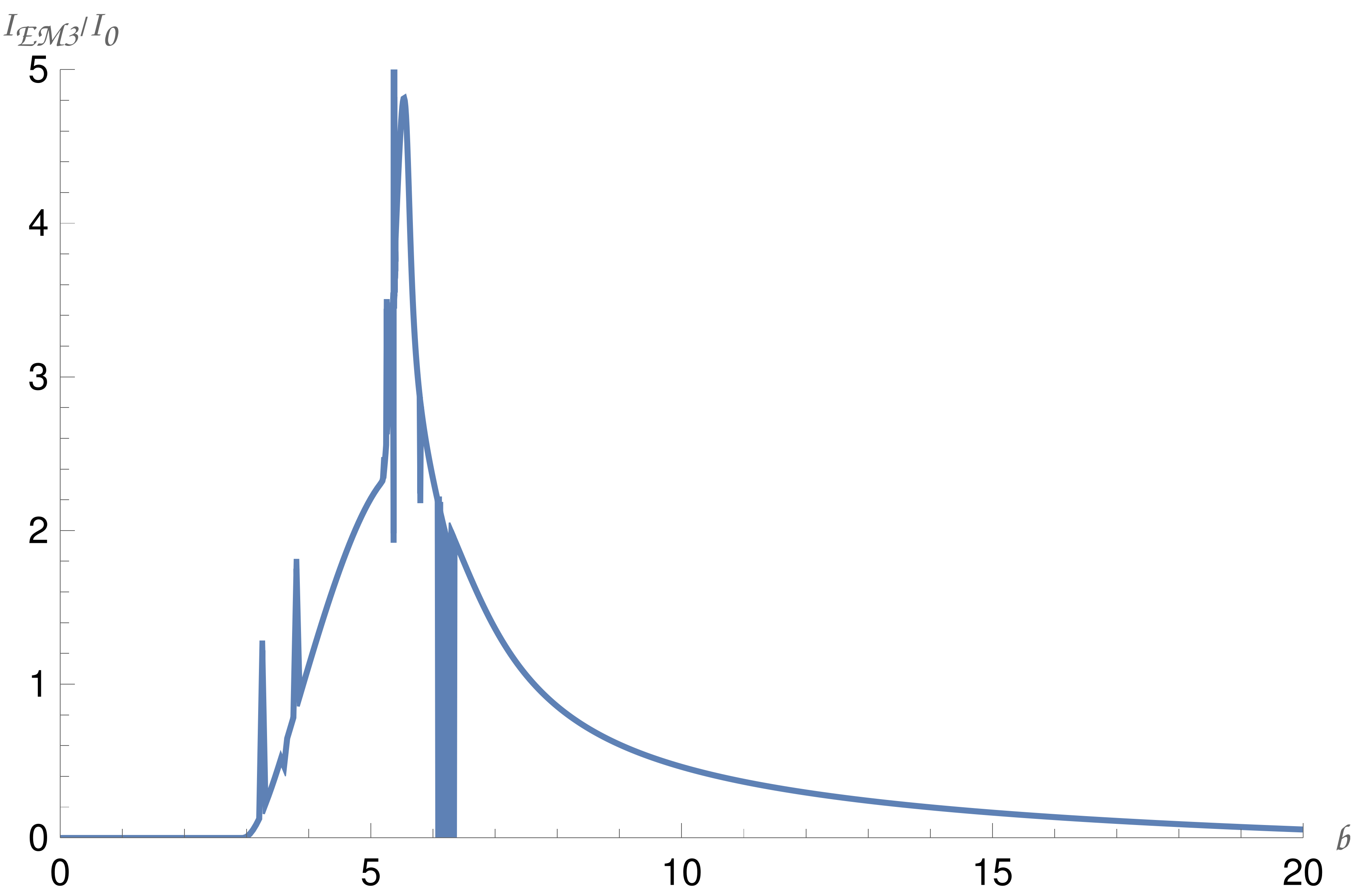}
    \includegraphics[width=.3\textwidth]{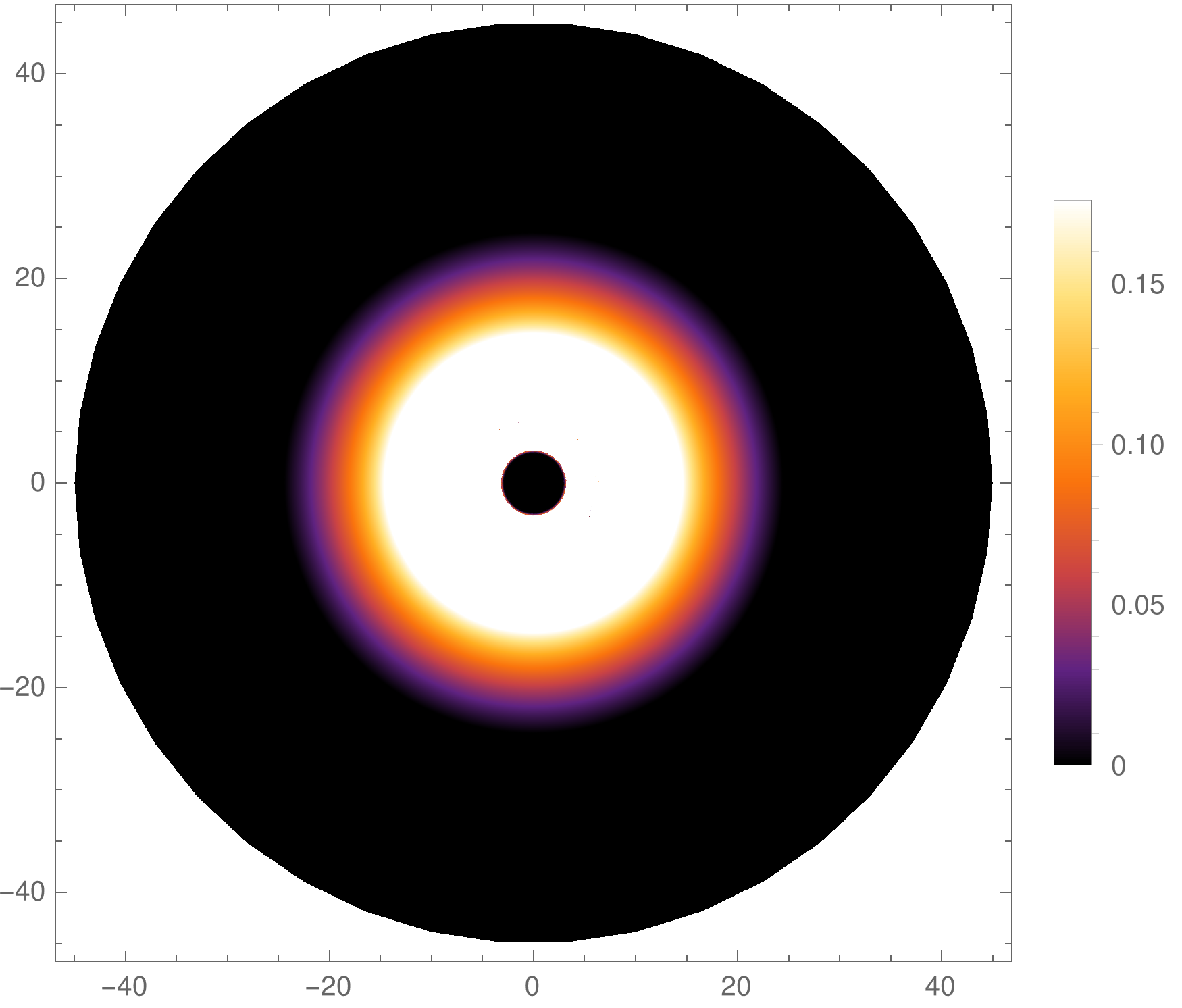}
    \caption{The thin disk's appearance was observed from a face-on perspective, with varying emission profiles for $q_m=0.1$ and $\beta=0.3$. The top row shows model $1$, the second row shows model $2$, and the third row shows model $3$, as described in section C. The emitted and observed intensities ($I_{EM}$ and $I_{obs}$) in the plots were normalized to the emitted intensity's maximum value outside the horizon ($I_0$).}
    \label{fig:17}
\end{figure*}

\begin{figure*}[htbp]
    \centering
    \includegraphics[width=.3\textwidth]{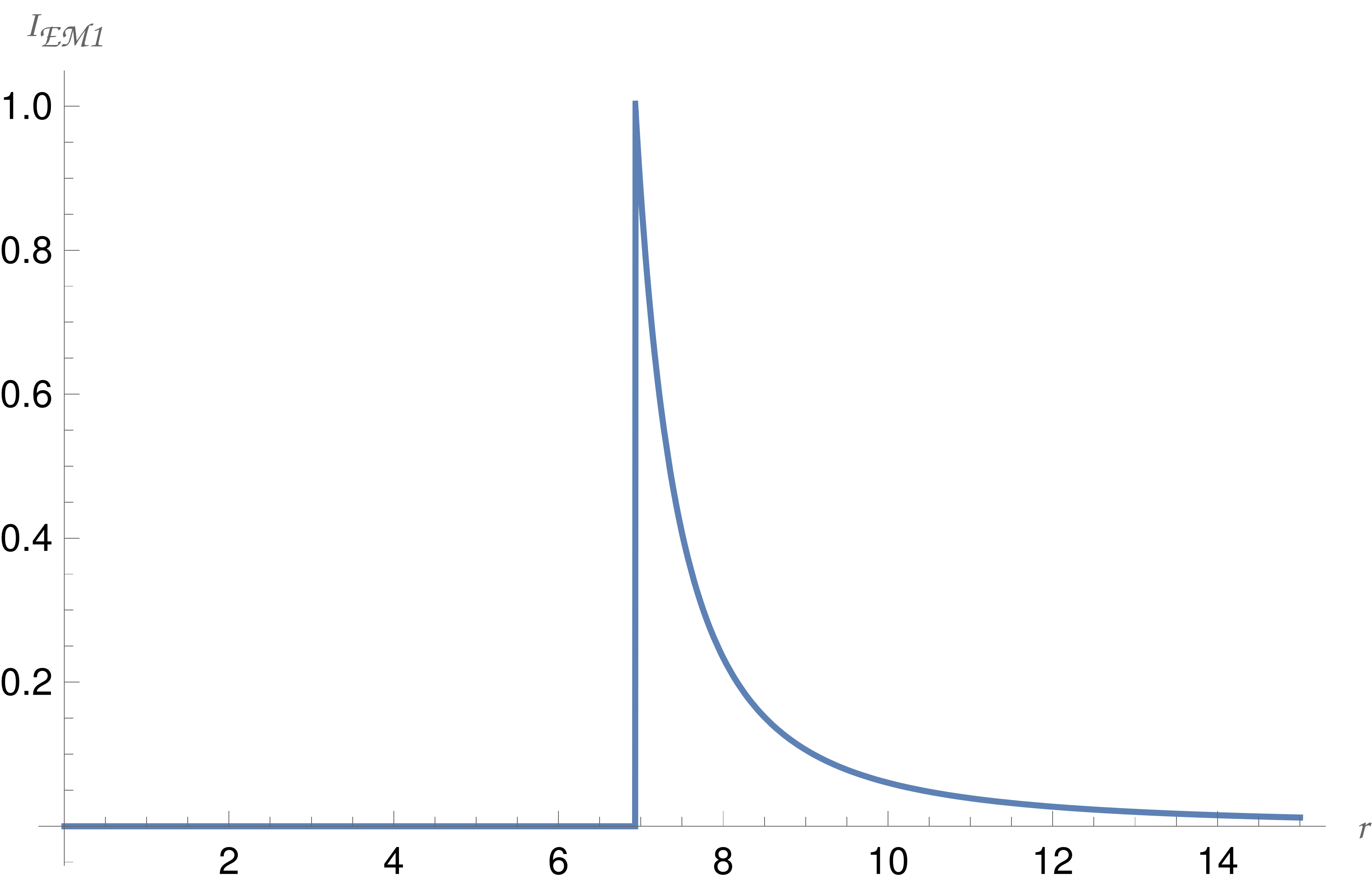}
    \includegraphics[width=.3\textwidth]{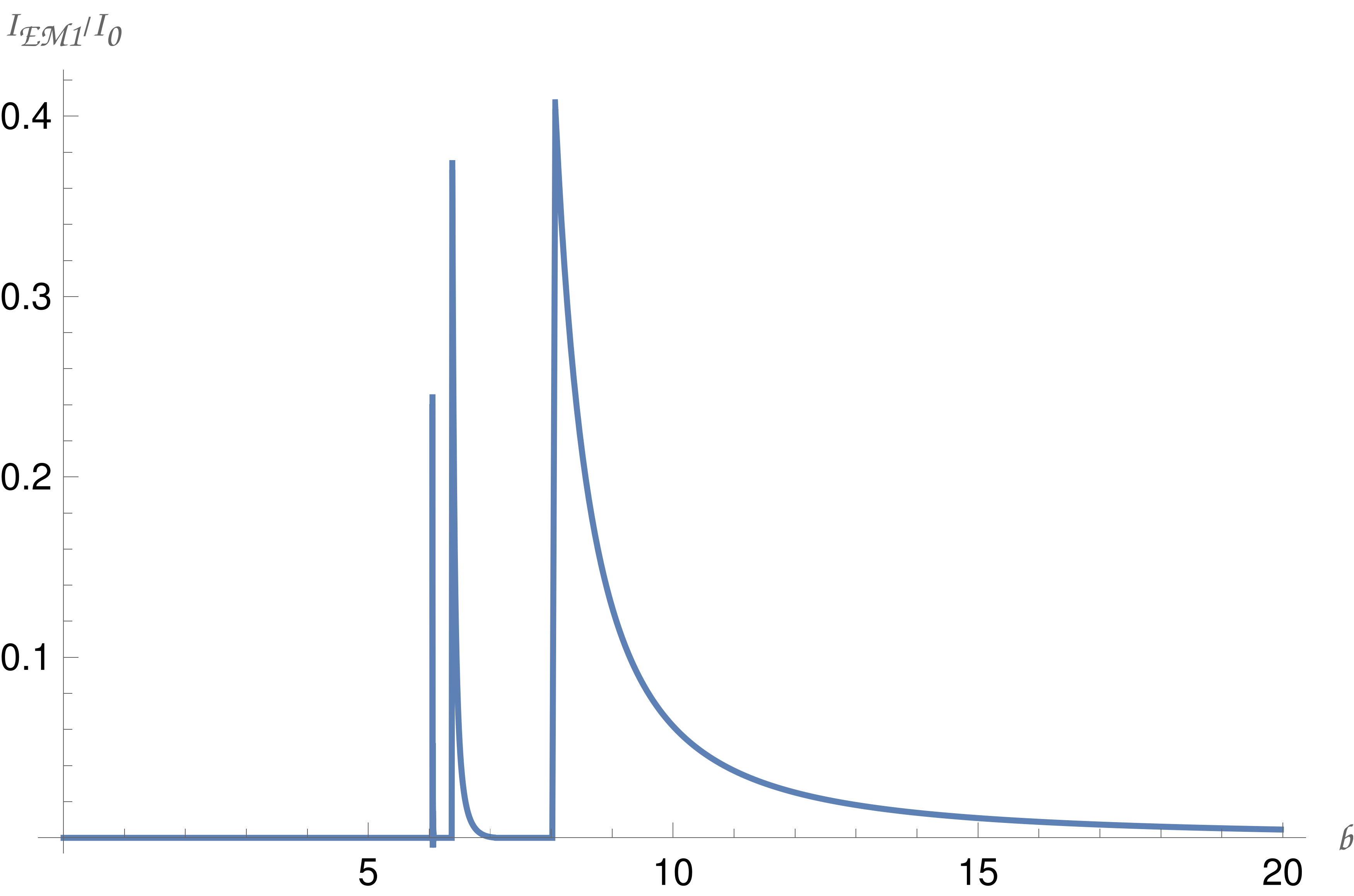}
    \includegraphics[width=.3\textwidth]{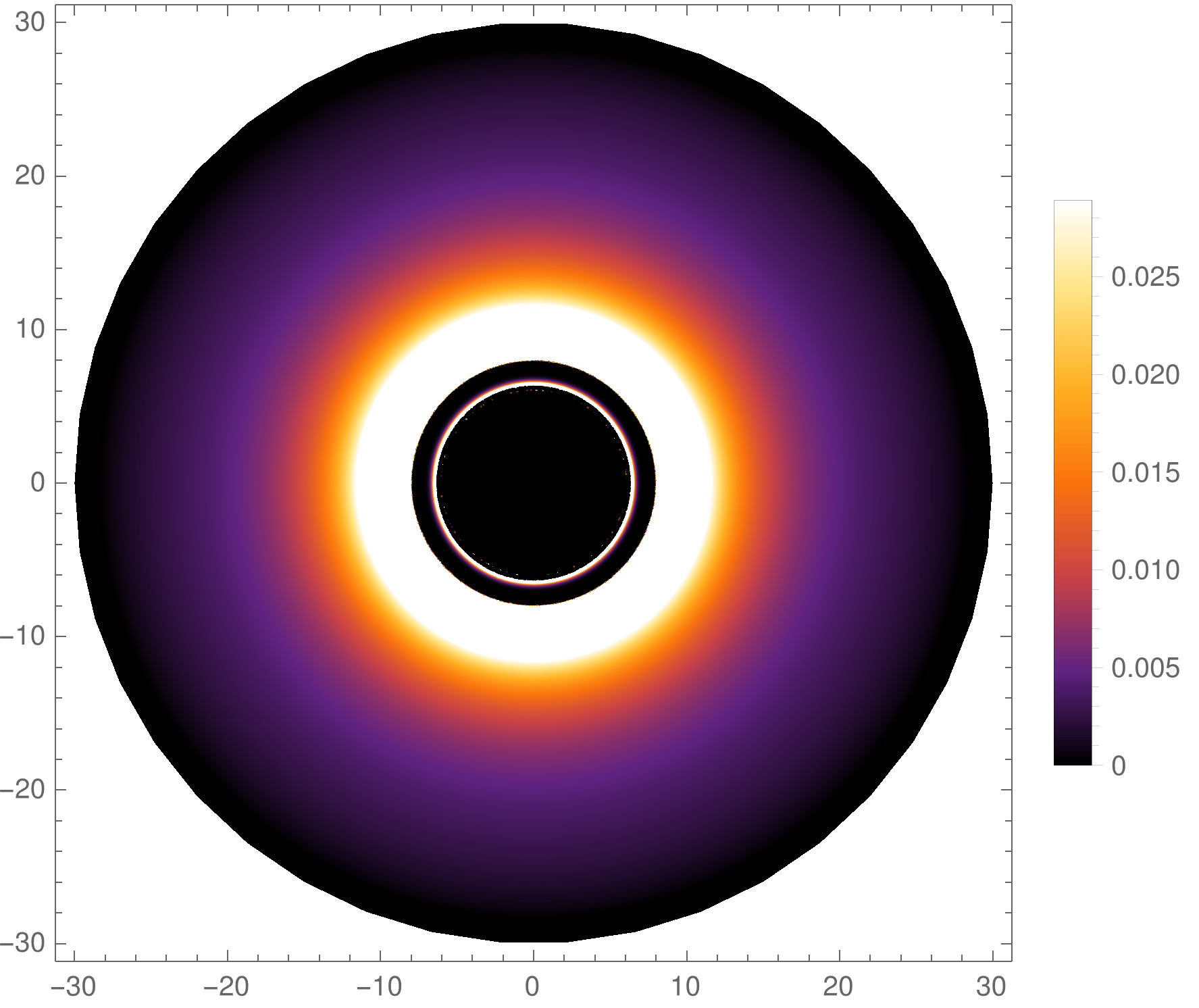}
    \vfill
    \centering
    \includegraphics[width=.3\textwidth]{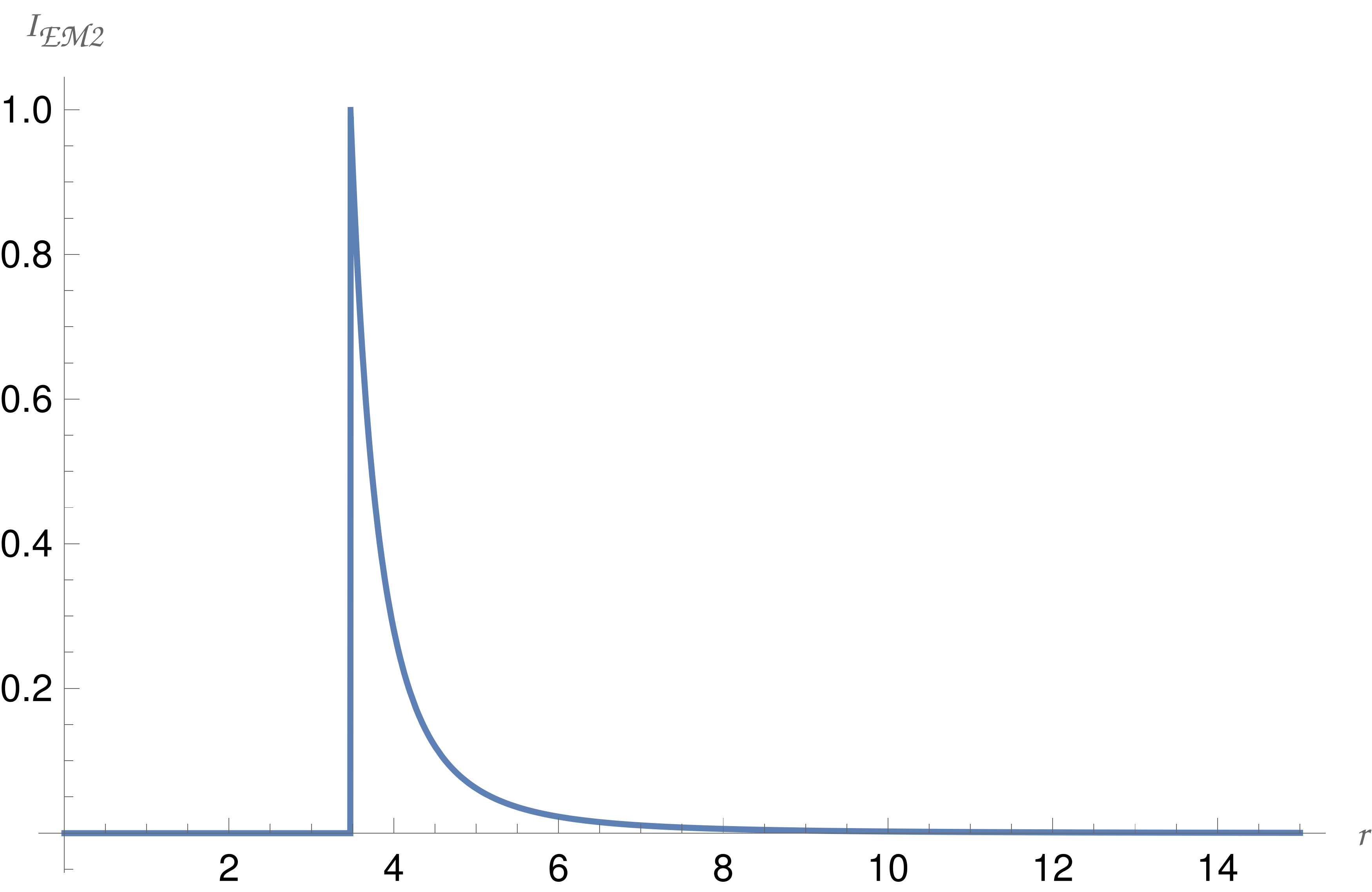}
    \includegraphics[width=.3\textwidth]{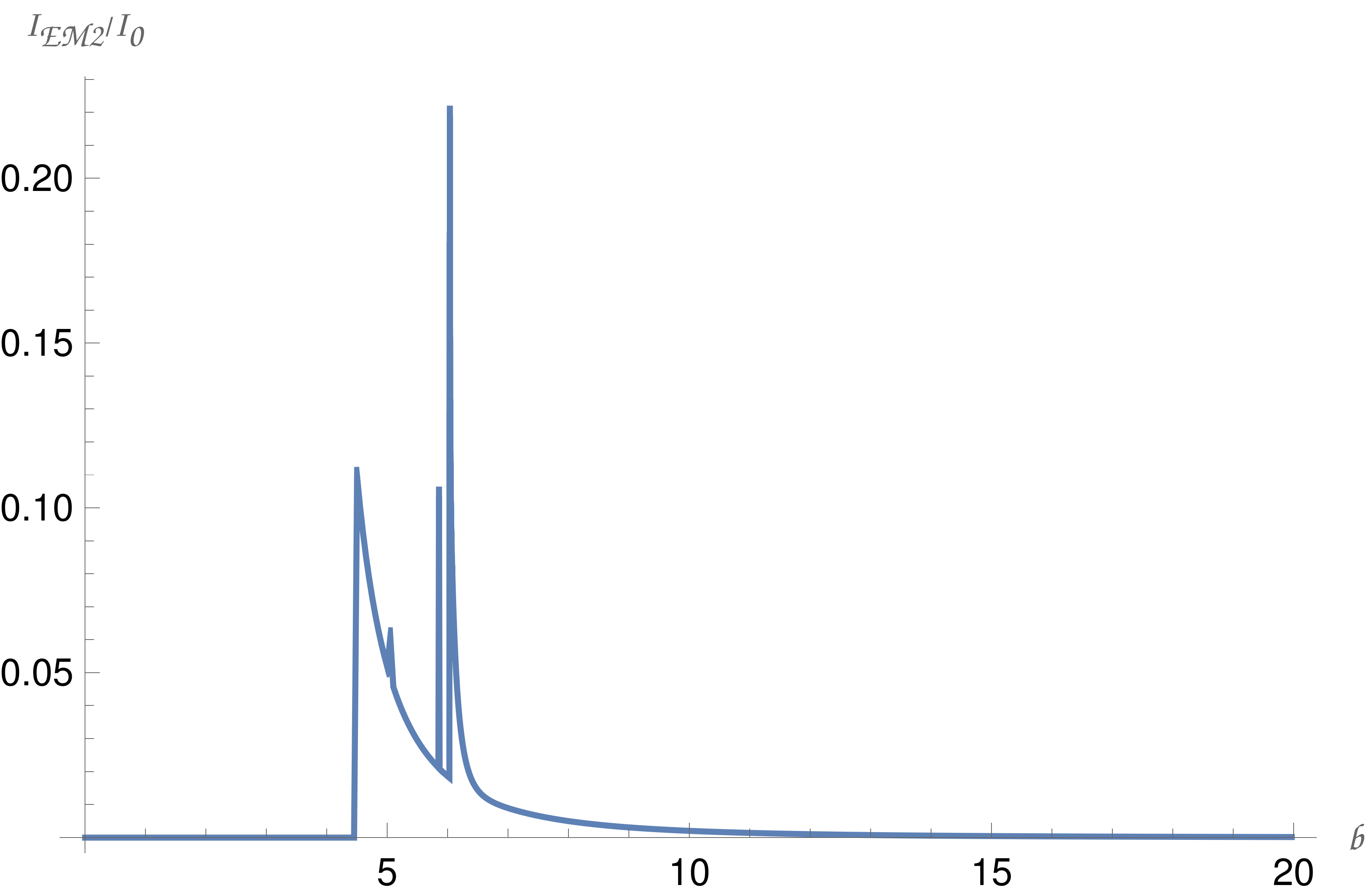}
    \includegraphics[width=.3\textwidth]{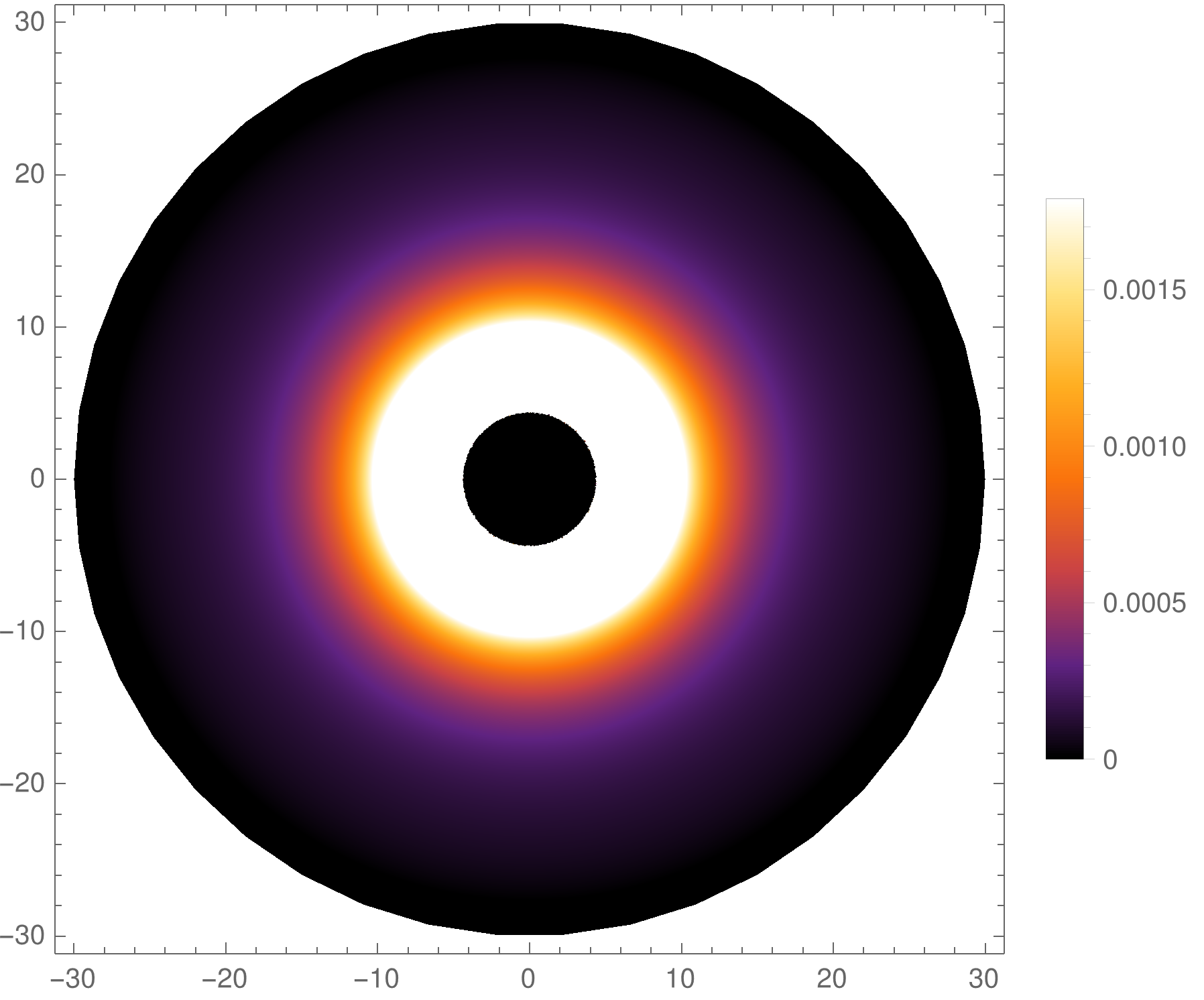}
    \vfill
    \centering
    \includegraphics[width=.3\textwidth]{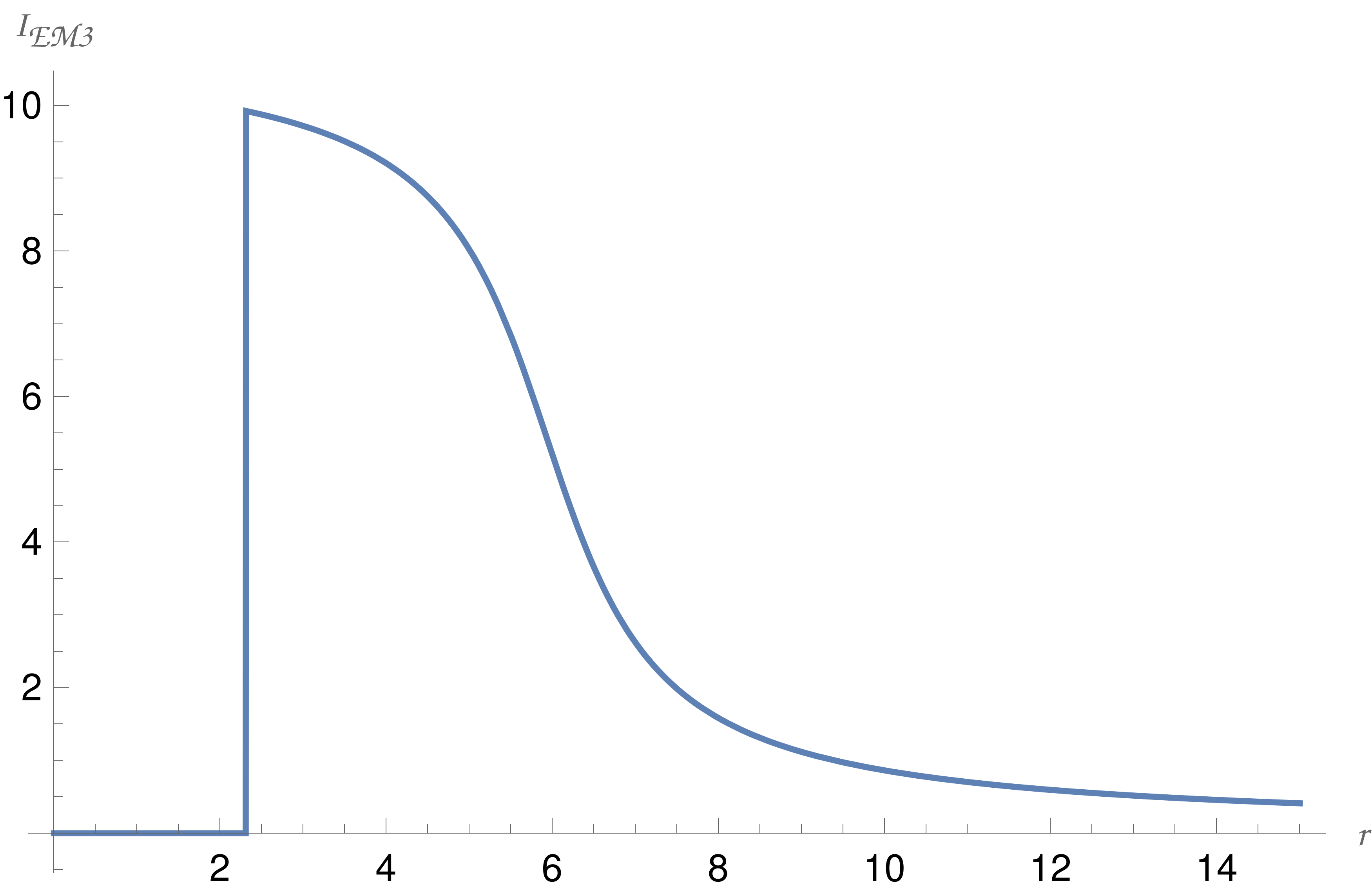}
    \includegraphics[width=.3\textwidth]{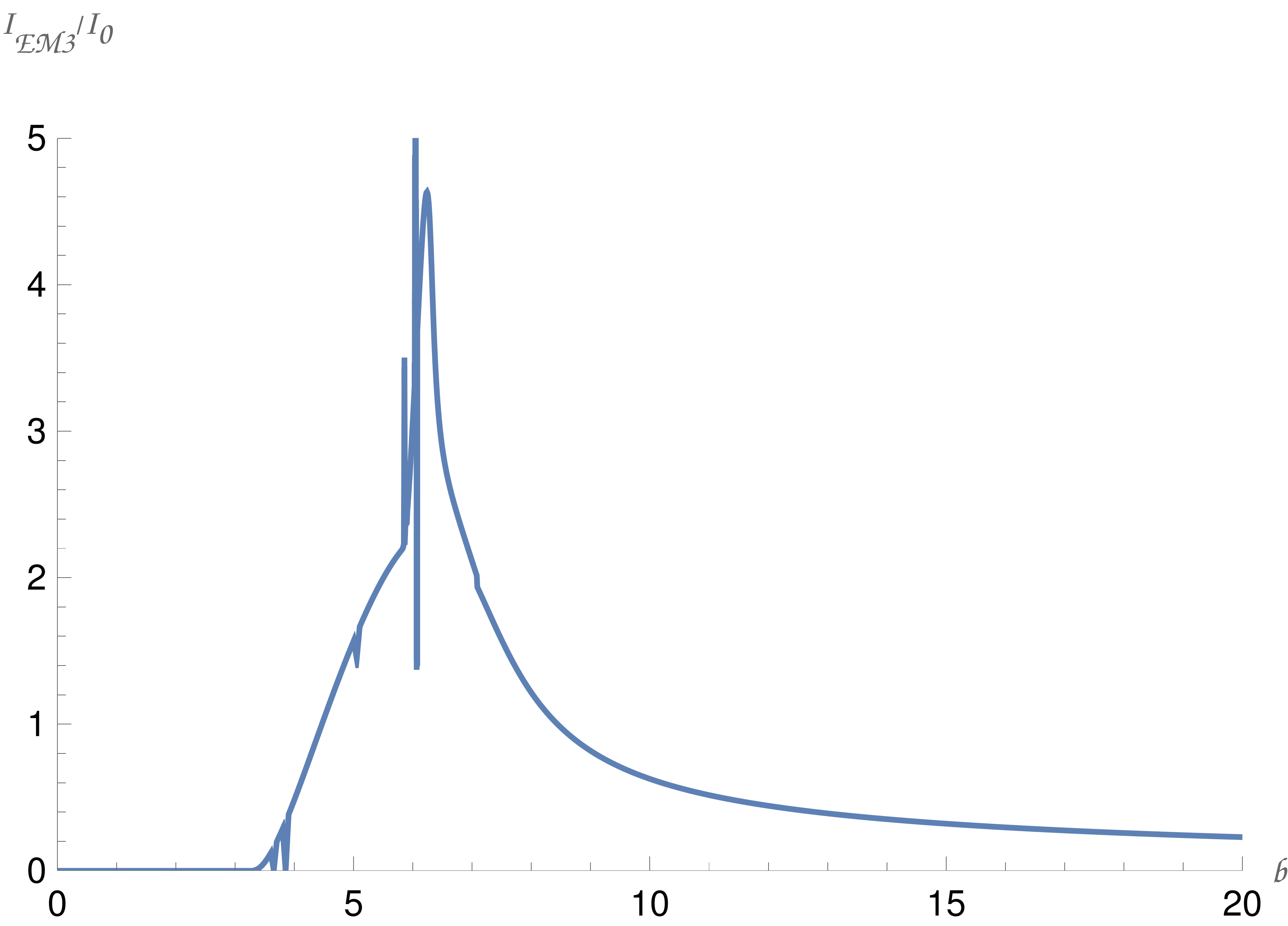}
    \includegraphics[width=.3\textwidth]{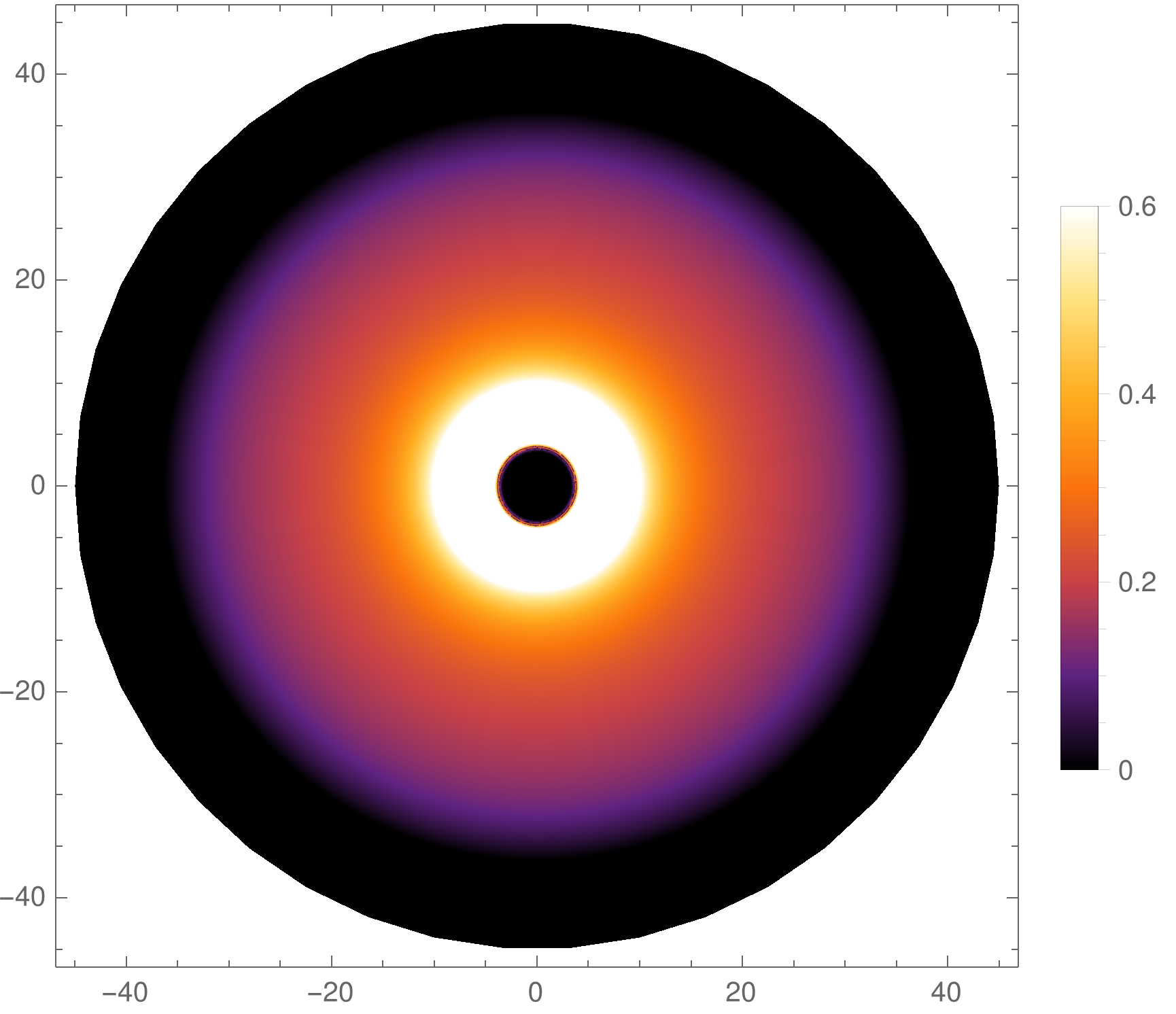}
    \caption{The thin disk's appearance was observed from a face-on orientation with varying emission profiles for $q_m=0.3$ and $\beta=0.3$. The plots show three models, where the top row is for model $1$, the second row is for model $2$, and the third row is for model $3$ which are described in section C. The intensities of emitted and observed light ($I_{EM}$ and $I_{obs}$) are normalized to the maximum emitted intensity outside the horizon ($I_0$).}
    \label{fig:18}
\end{figure*}

\begin{figure*}[htbp]
    \centering
    \includegraphics[width=.3\textwidth]{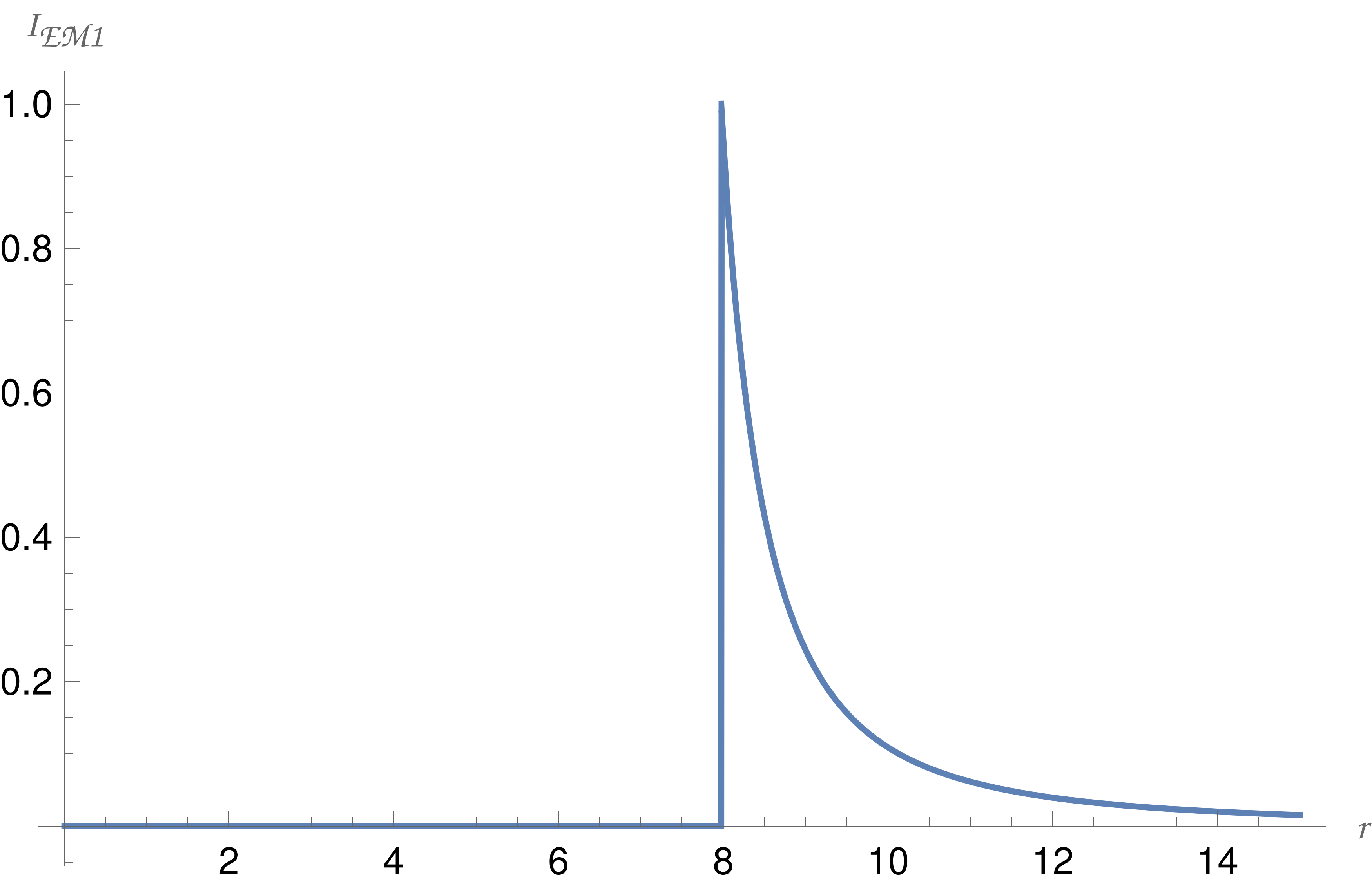}
    \includegraphics[width=.3\textwidth]{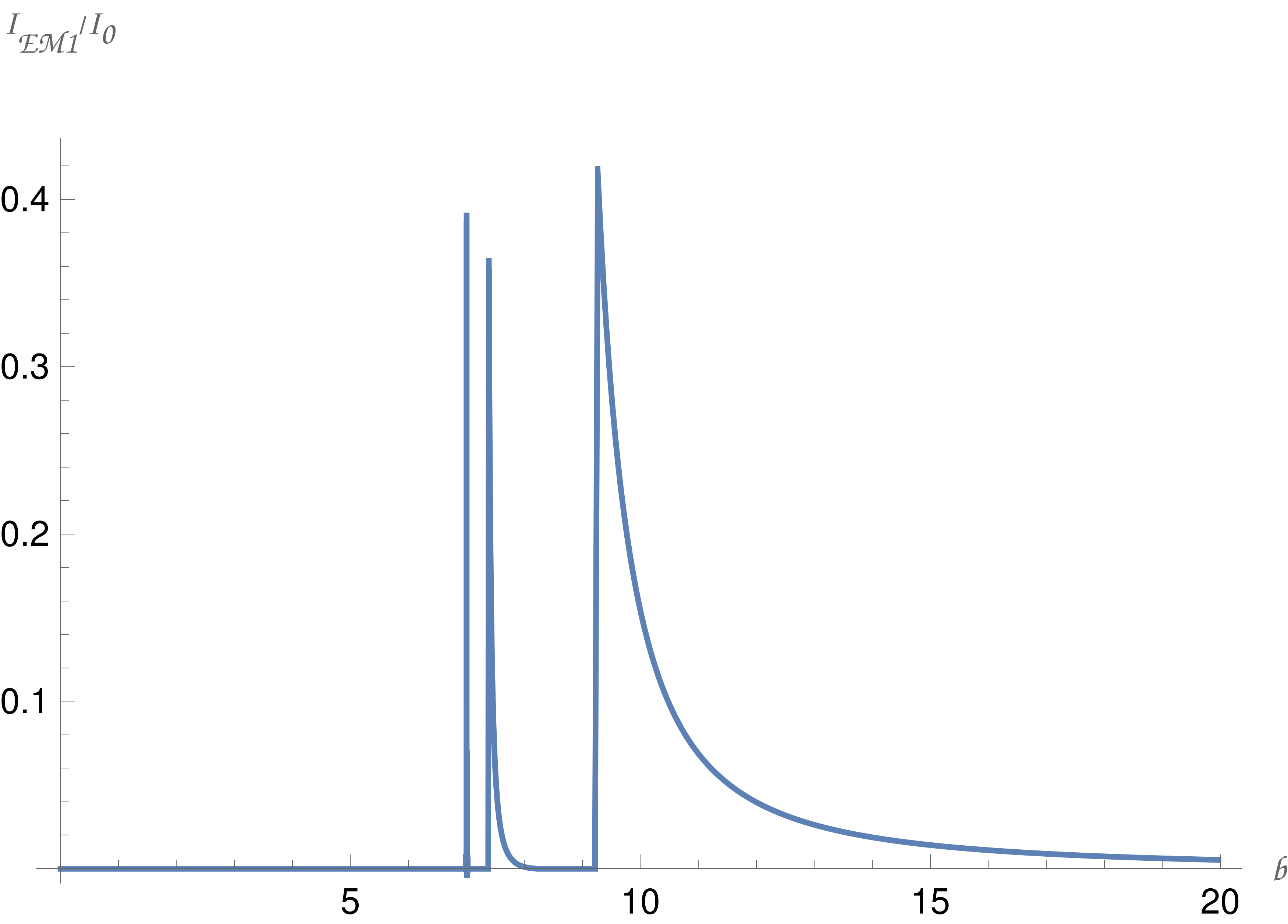}
    \includegraphics[width=.3\textwidth]{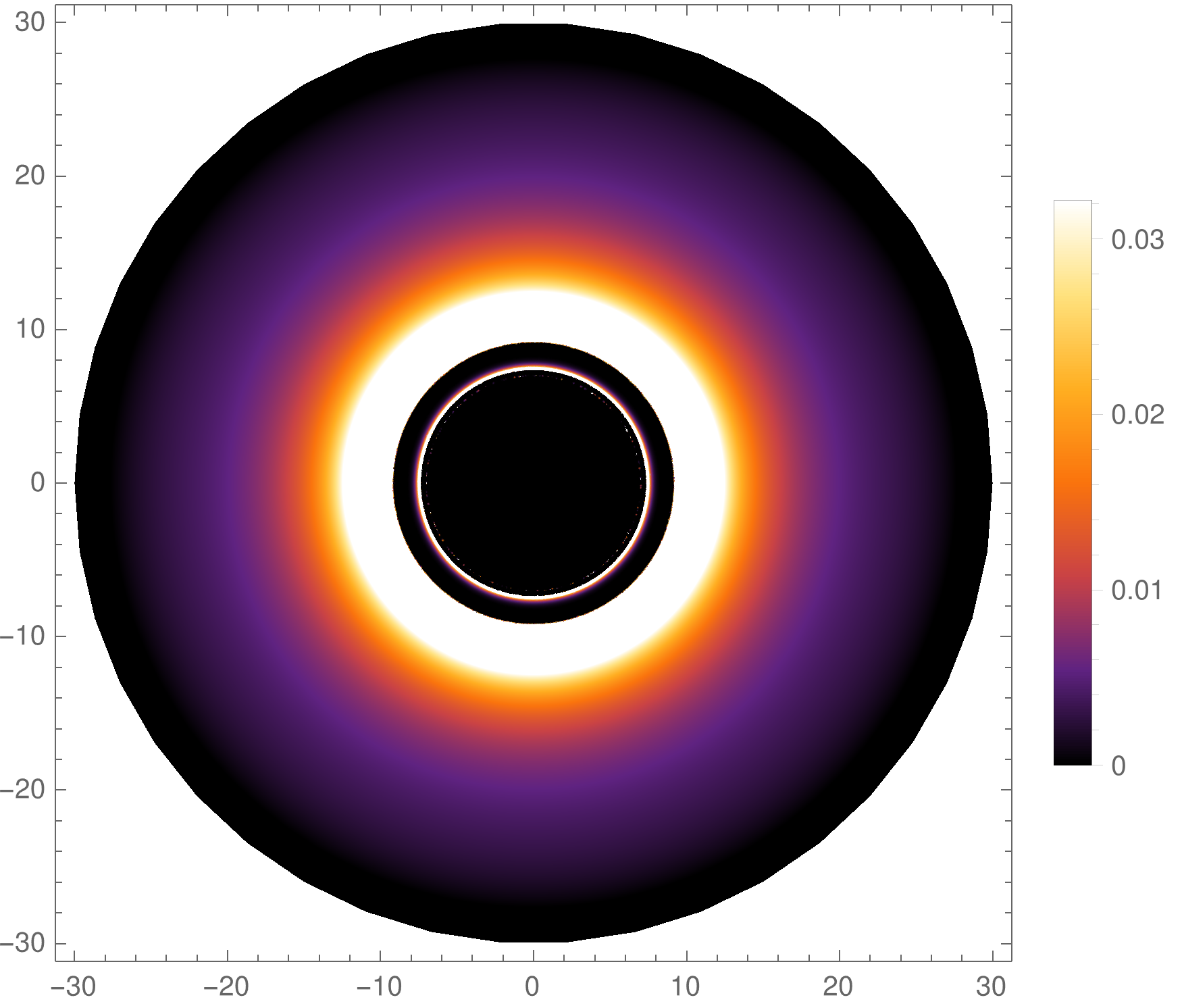}
    \vfill
    \centering
    \includegraphics[width=.3\textwidth]{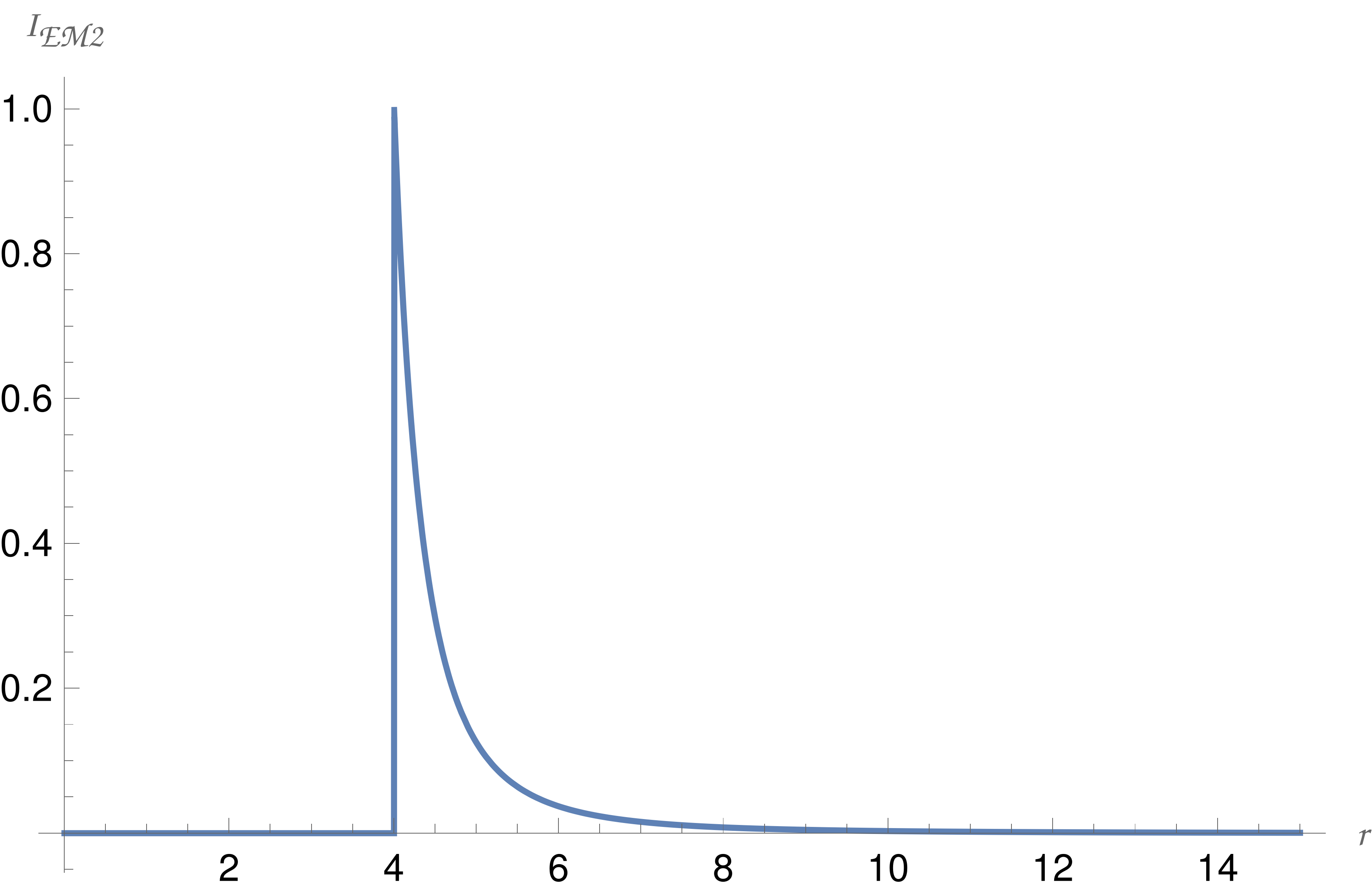}
    \includegraphics[width=.3\textwidth]{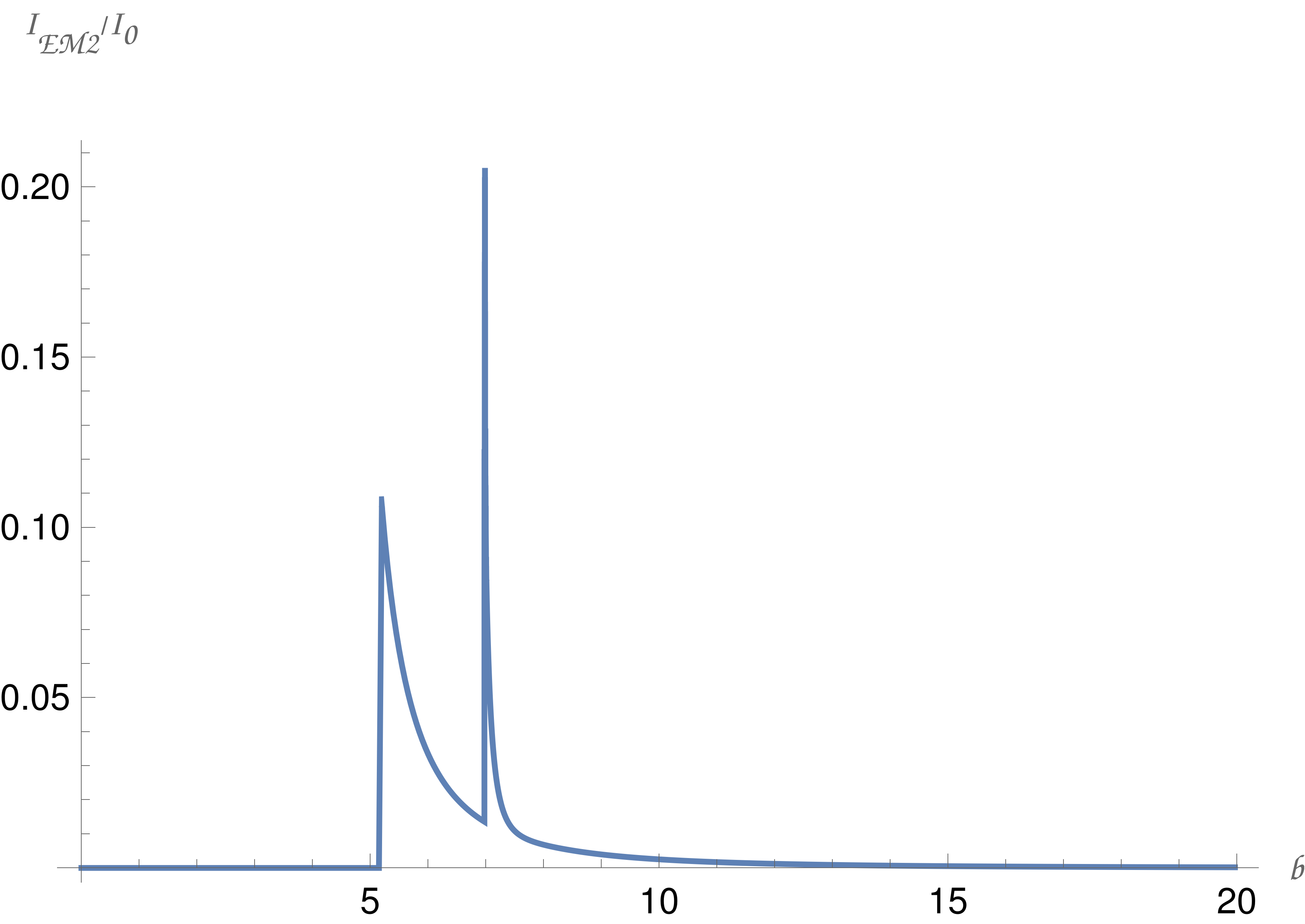}
    \includegraphics[width=.3\textwidth]{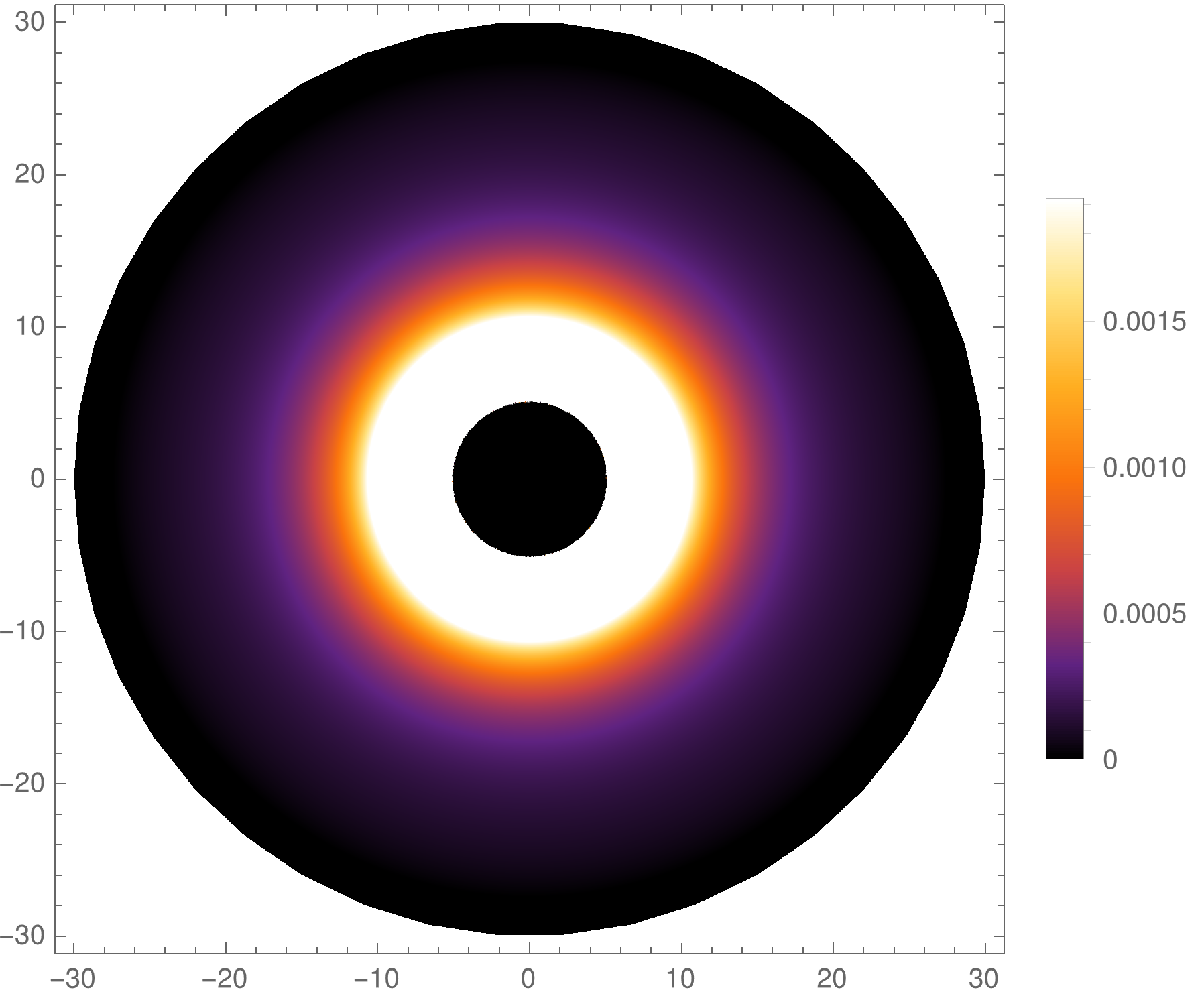}
    \vfill
    \centering
    \includegraphics[width=.3\textwidth]{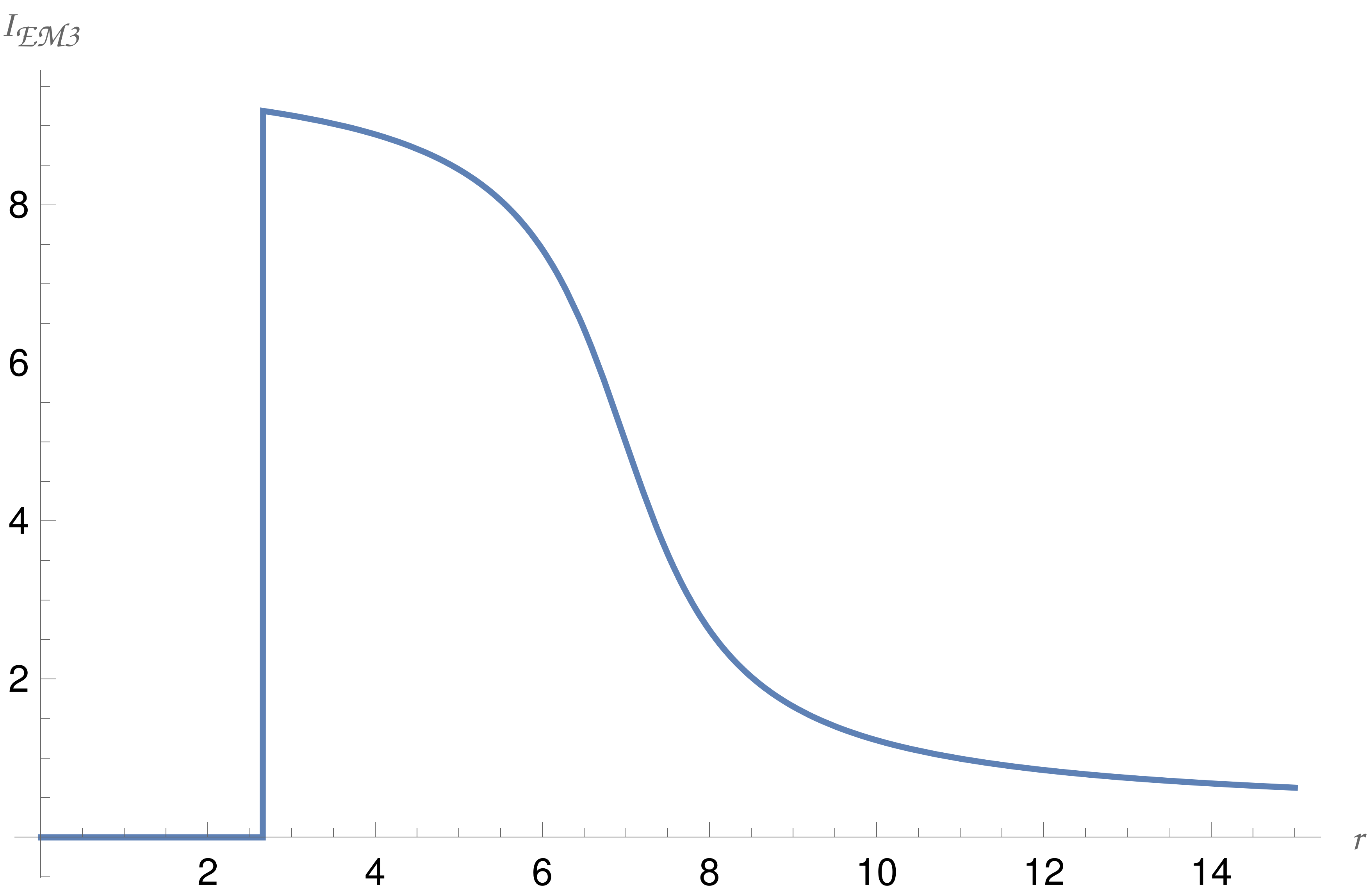}
    \includegraphics[width=.3\textwidth]{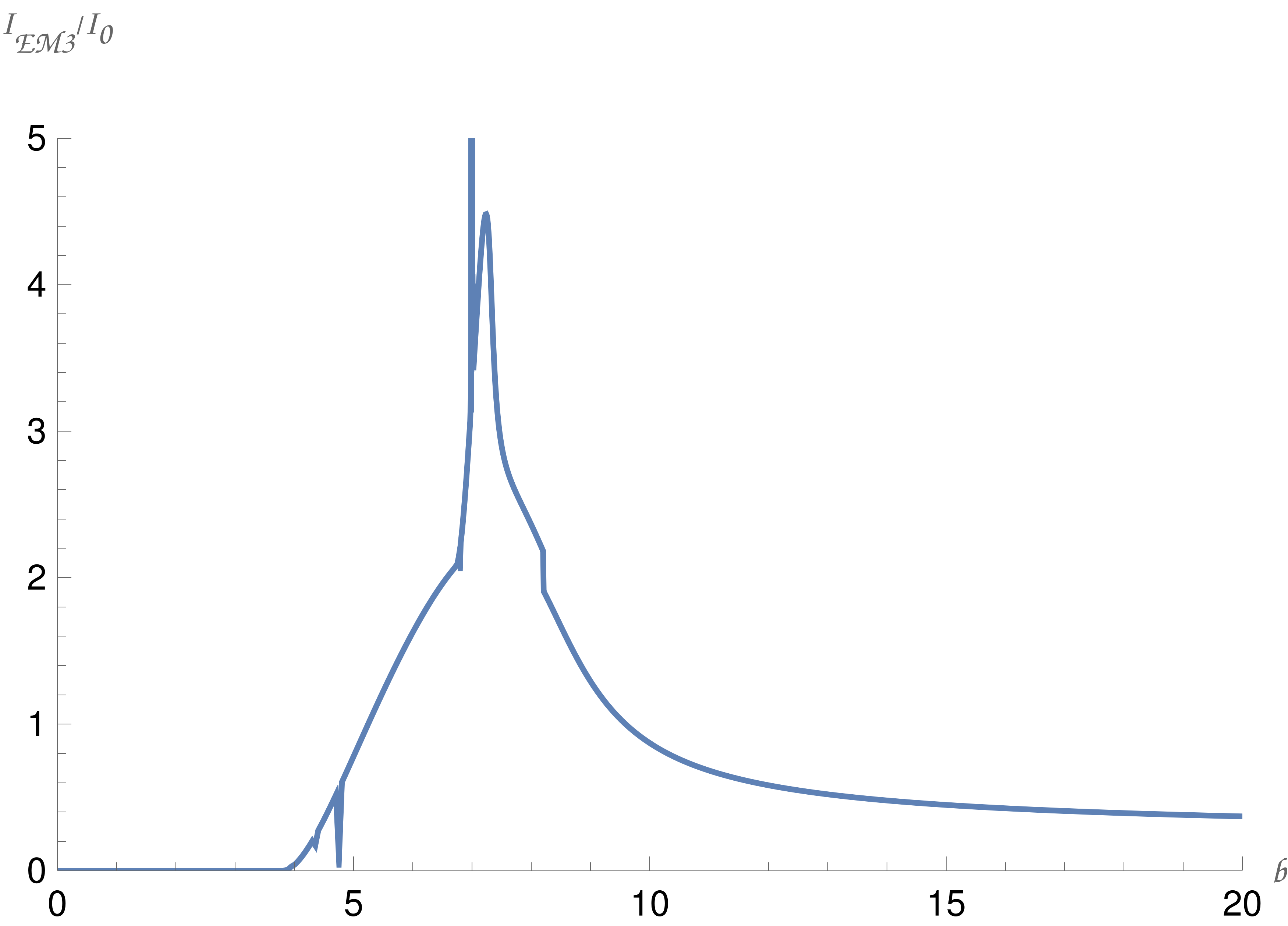}
    \includegraphics[width=.3\textwidth]{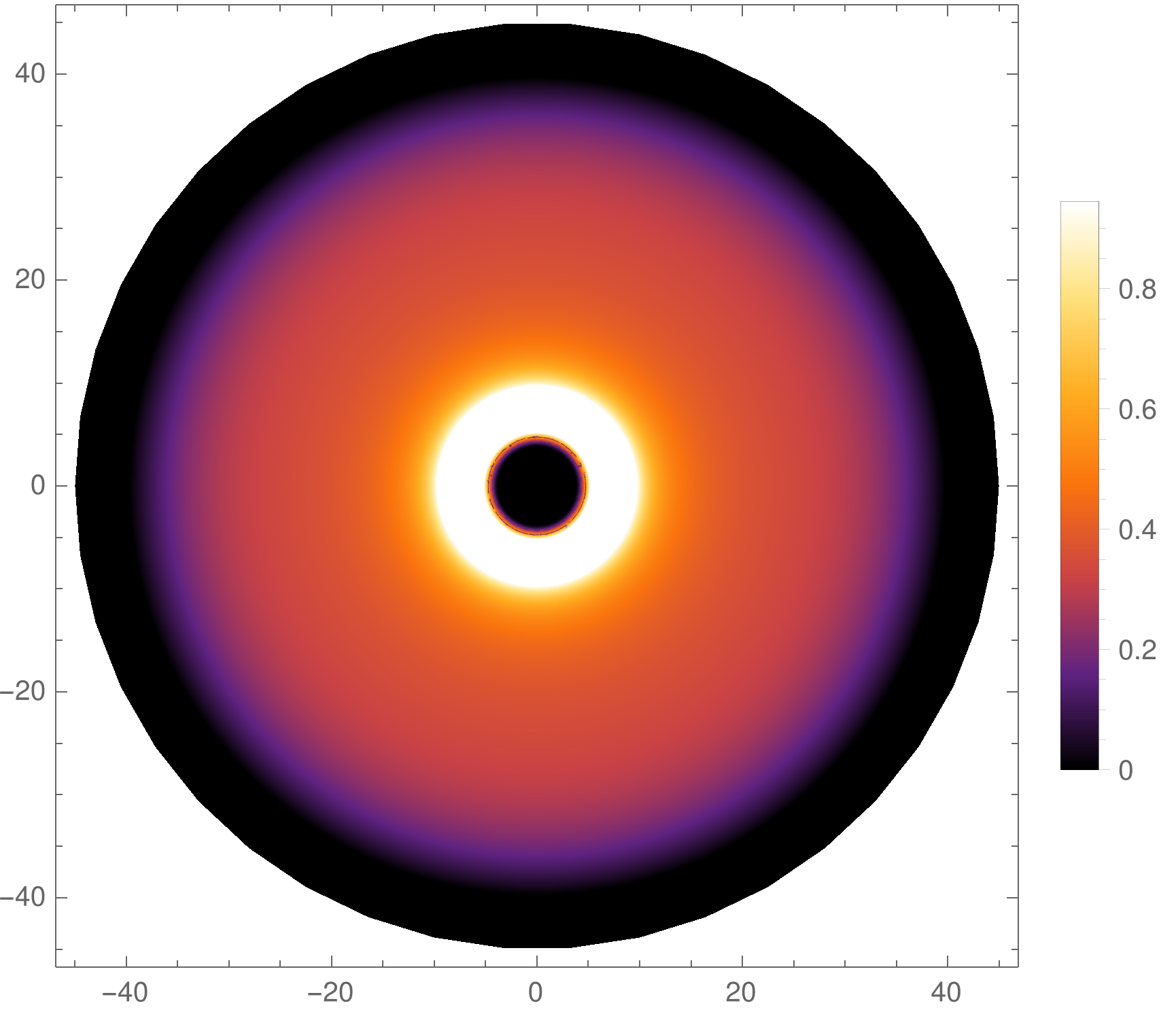}
    \caption{The thin disk's visual appearance was observed in a face-on orientation with various emission profiles for $q_m=0.5$ and $\beta=0.3$. Model $1$ "model $2$ and "model $3$ from section C were used for the upper, second, and third rows, respectively. The emitted and observed intensities ($I_{EM}$ and $I_{obs}$) in the plots are standardized to the highest emitted intensity outside the horizon ($I_0$).}
    \label{fig:19}
\end{figure*}

Figures \ref{fig:17}, \ref{fig:18}, and \ref{fig:19} display the observed features of the thin accretion disk, including the observed intensity and impact factor, corresponding to the above described three models for different magnetic charges of $q_m=0.1,0.3,0.5$, respectively with $\beta=0.3$. Across all figures, we observe that the emission intensity coming from the accretion disk (first row, first column) for the first model as described earlier (top row), has a peak near the critical impact parameter at $r\sim b_\text{c}$, which then decreases with the radial distance and ultimately reaches zero. Additionally, we note that the location of the photon sphere is inside the interior region of the disk's emission profile. Furthermore, due to the effect of the gravitational lensing, we observe two independently separated peaks corresponding to the lensing and photon rings (first row, second column) within the region of the observed emission profile. However, we find that the peaks of the photon and lensing rings have a narrower observational area with respect to the direct emission and are also smaller in size. Thus, we conclude that in the observed intensity, the main contribution is coming from the direct emission, with smaller contributions from the lensing rings and an even smaller contribution from the photon rings. Furthermore, it is evident from the $2D$ shadow image (first row, third column) that the direct emission and photon ring are dominating the optical appearance of the accretion disk in this model.

In the second toy model (second row of the figures), the emission intensity from the disk reaches its peak value at the photon sphere ($r\sim r_\text{ph}$) and then decreases as we move away radially (first column of the second row). The observed intensity profile (second column of the second row) also has a peak due to the direct emission, but it then attenuates as $r$ increases. The lensing and photon rings in this model are overlapping, which results in an increase in the total intensity of that area, leading to a new peak with multiple small peaks due to the effect of the photon ring, lensing ring, and direct emission. However, as we discussed earlier, the photon and lensing rings have a narrow area and are highly demagnified. Therefore, the observed intensity is still dominated by the direct emission, as can be observed in the $2D$ shadow image of the accretion disk (third column of the second row).

In the last third model (shown in the third row of the figures), the intensity peaks at the black hole's horizon and decreases as the radial distance from the black hole increases (as seen in the first column of the third row). In this particular case, the photon ring, lensing ring, and direct emission regions overlap for a significant range of radial distance, resulting in a peak in the observed intensity (shown in the second column of the third row). The intensity increases gradually just above the horizon and then shows a sharp increment to a peak value in the photon ring region. Also, due to the existing contribution of the lensing ring, the observed intensity of the disk shows a higher peak value, followed by a sudden and slow decrease, and then eventually decreases to zero as shown in the Figure. Consequently, the $2D$ shadow image exhibits multiple rings resulting from the observed contribution of the photon ring, lensing ring, and direct emission (shown in the third column of the third row) It is clear that for $q_m=0.1$ to $q_m=0.5$, the variation in the observed intensity is very prominent hence with some relativistic emission intensity profile, one can explore more about the effect of the charge on the observed shadow.

In this section, we compared the emission profiles of the three models emitted from the accretion disk with the observed intensity observed by an observer and the optical appearance of the disk for various magnetic charge values ($q_m=0.1,0.3,0.5$) when $\beta=0.3$. We noticed that as we increase the magnetic charge, the observed intensity decreases with the peak of the direct, lensing, and photon rings occurring at different locations. However, the magnitude of the peak intensity is much less than that for the Schwarzschild spacetime \cite{Gralla:2019xty}. This indicates that there are significant changes in the observed intensities of the disk when we consider the magnetically charged NLED effect in the theory, and we can differentiate between the NLED black hole with a magnetic charge and the Schwarzschild black hole. This could be more interesting to study in detail and we expect that in future with more precious observations, we will be able to make this difference more clear.

\subsection{Shadow With Infalling Spherical Accretion}
Let's dive into this section, where we'll explore the infalling spherical accretion around a magnetically charged NLED black hole. Specifically, we'll be using the optically thin disk model developed by Bambi \cite{Bambi:2013nla} to examine the behaviour of radiative gas as it moves around the black hole to form the spherical accretion disk. Our goal is to gain insight into the observable features of this system by studying the effect of the magnetic charge $q$ on the spherical accretion. To do this, we'll be analyzing the specific intensity observed by an observer located at a finite distance from the black hole
\begin{equation}
 I_\text{obs}= \int_\Gamma g^3 j(\nu_e) dl_\text{prop},
 \label{is}
\end{equation}
it is noted that $g$, $\nu_e$, and $\nu_{obs}$ represent the redshift factor, photon frequency, and observed photon frequency, respectively. We can shift our focus to the emitter's rest frame, where we have the emissivity per unit volume, $j(\nu_e) \propto \frac{\delta (\nu_e-\nu_f)}{r^2}$, which has a radial profile proportional to $1/r^2$. In the above equation, $\delta$ is the delta function, $\nu_f$ refers to the frequency of the radiative light, which is considered to be monochromatic, and $dl_\text{prop}$ is an infinitesimal proper length. The redshift factor for any black hole space-time can be expressed as:
\begin{equation}
g=\frac{\mathcal{K_{\rho}}u_{o}^{\rho}}{\mathcal{K_{\sigma}}u_{e}^{\sigma}}.
\end{equation}
The four-velocity of a photon is denoted as $\mathcal{K^{\mu}}$, which represents its velocity in four-dimensional spacetime. On the other hand, $u_o^{\mu}=(1,0,0,0)$ is the known four-velocity of a static observer located at infinity. Therefore, the four-velocity $(u_e^{\mu})$ of the matter which is in the infalling accretion can be written as,
\begin{equation}
u^t_e=f(r)^{-1},u^r_e=-\sqrt{1-f(r)},u^{\theta}_e=u^\phi_e=0.
\end{equation}
Therefore, we can express the four velocities of the photon utilizing the null geodesic,
\begin{equation}
\mathcal{K}_t=\frac{1}{b},\frac{\mathcal{K}_r}{\mathcal{K}_t}=\pm \frac{1}{f(r)}\sqrt{1-f(r)\frac{b^2}{r^2}},
\end{equation}
We use the symbols $+$ and $-$ to represent whether the photon moves towards or away from the black hole. With this in mind, we can express the redshift factor for the infalling accretion
\begin{equation}
g=\left[ u_e^t + \left( \frac{\mathcal{K}_r}{\mathcal{K}_e} \right) u^r_e \right]^{-1}.
\end{equation}
The proper distance will be changed as follows:
\begin{equation}
dl_\text{prop}=\mathcal{K}_{\mu}u^{\mu}_e d\lambda = \frac{\mathcal{K}_t}{g |\mathcal{K}_r |}dr.
\end{equation}
Therefore, we can now calculate the observed intensity of the disk with the infalling spherical accretion by integrating equation (\ref{is}) over all frequencies
\begin{equation}
I_\text{obs} \propto \int_{\Gamma} \frac{g^3 \mathcal{K}_t dr}{r^2 |\mathcal{K}_r |}.
\end{equation}
Using the above equations, we examine how the distribution of the brightness of the black hole shadow can be affected by the magnetically charged NLED black hole.

\begin{figure}[htbp]
    \centering
    \includegraphics[width=\linewidth]{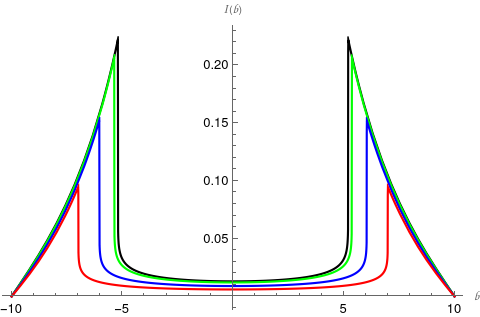}
    \caption{The intensity of infalling spherical accreting matter was observed at varying values of $q$, represented by different colors. The values include Schwarzchild in black, $q_m=0.1$ in green, $q_m=0.3$ in blue, and $q_m=0.5$ in red with $\beta=0.3$.} \label{fig:compp}
\end{figure}

\begin{figure}[htbp]
    \centering
    \includegraphics[width=.3\textwidth]{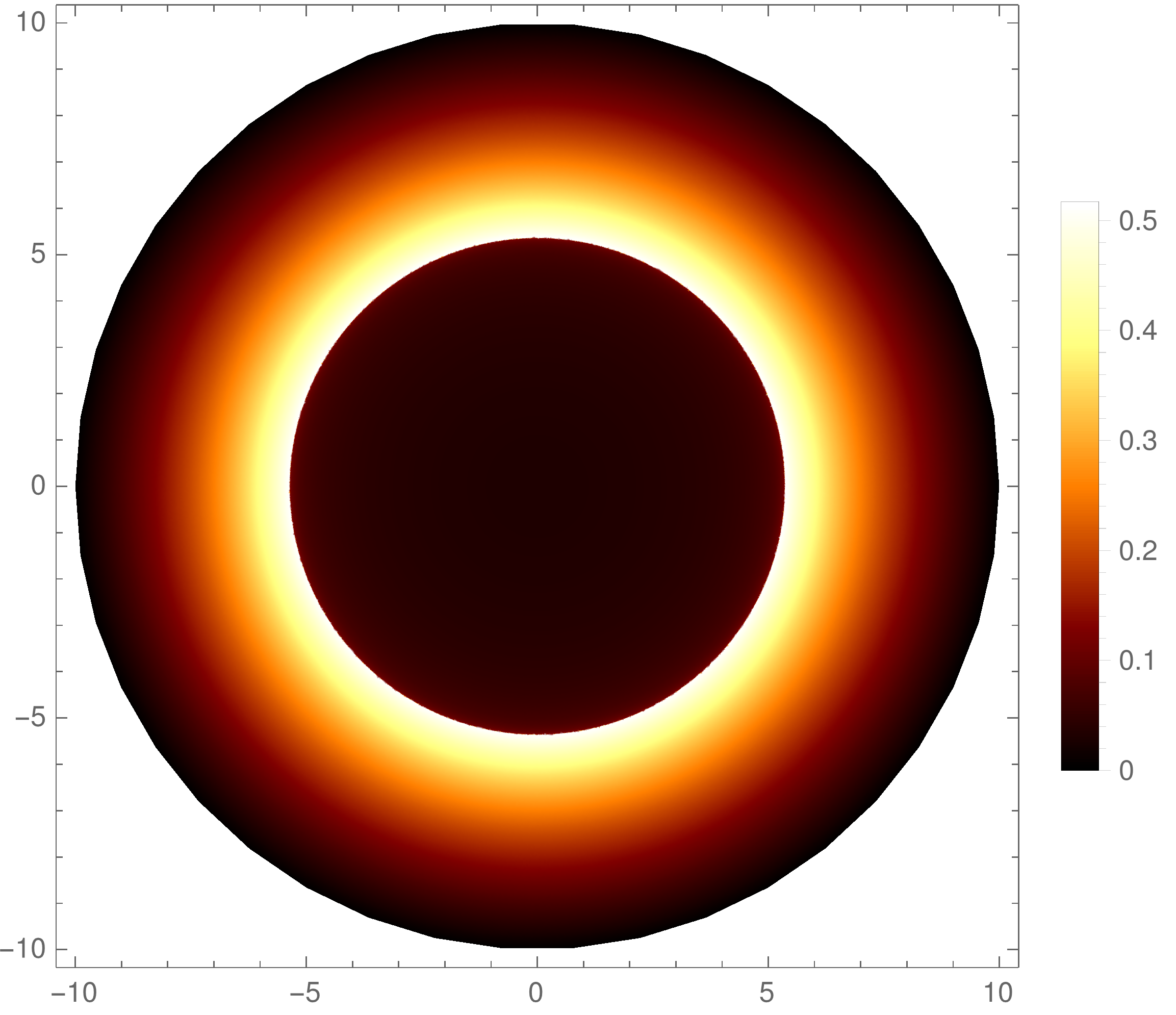}
    \includegraphics[width=.3\textwidth]{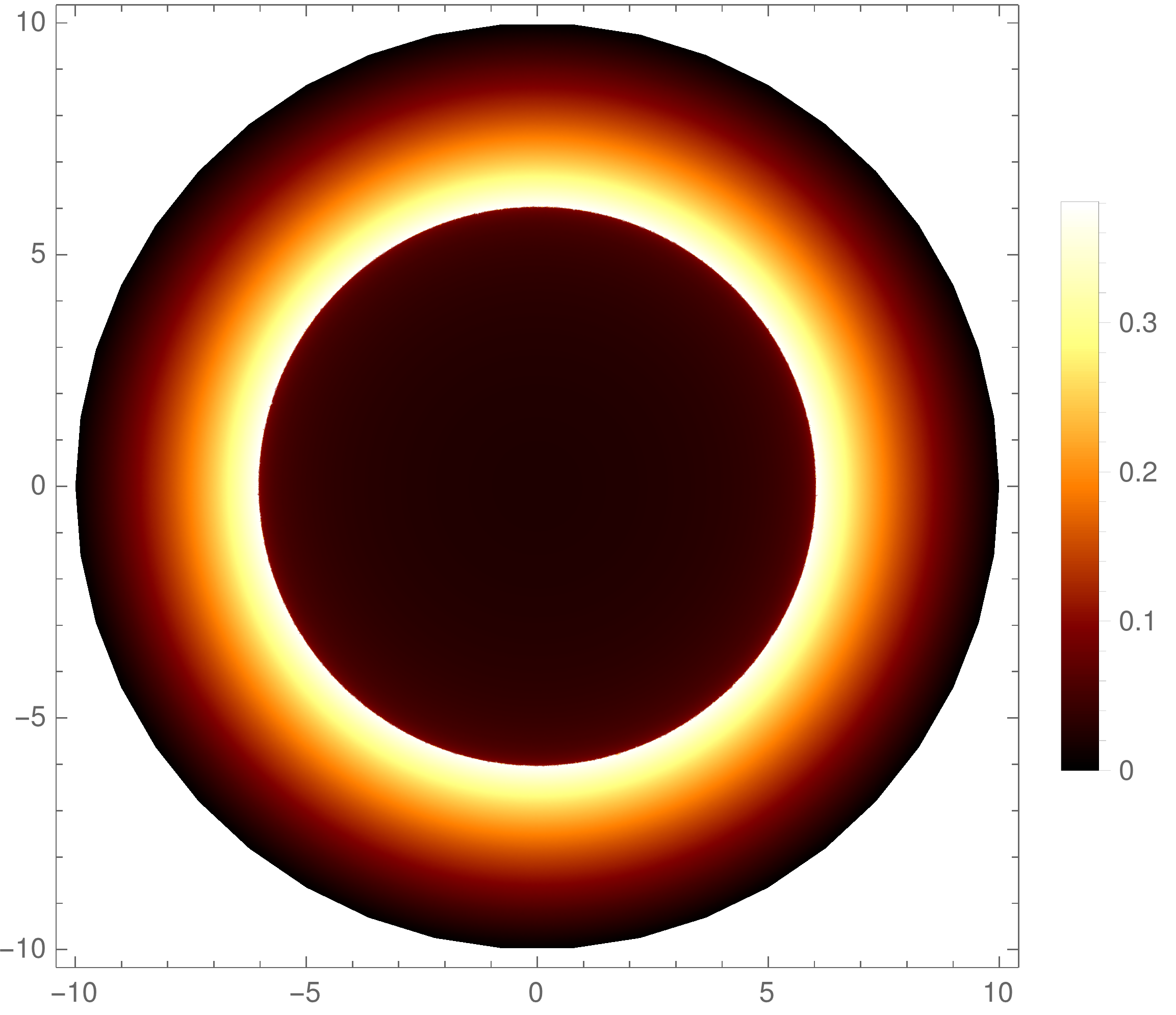}
    \includegraphics[width=.3\textwidth]{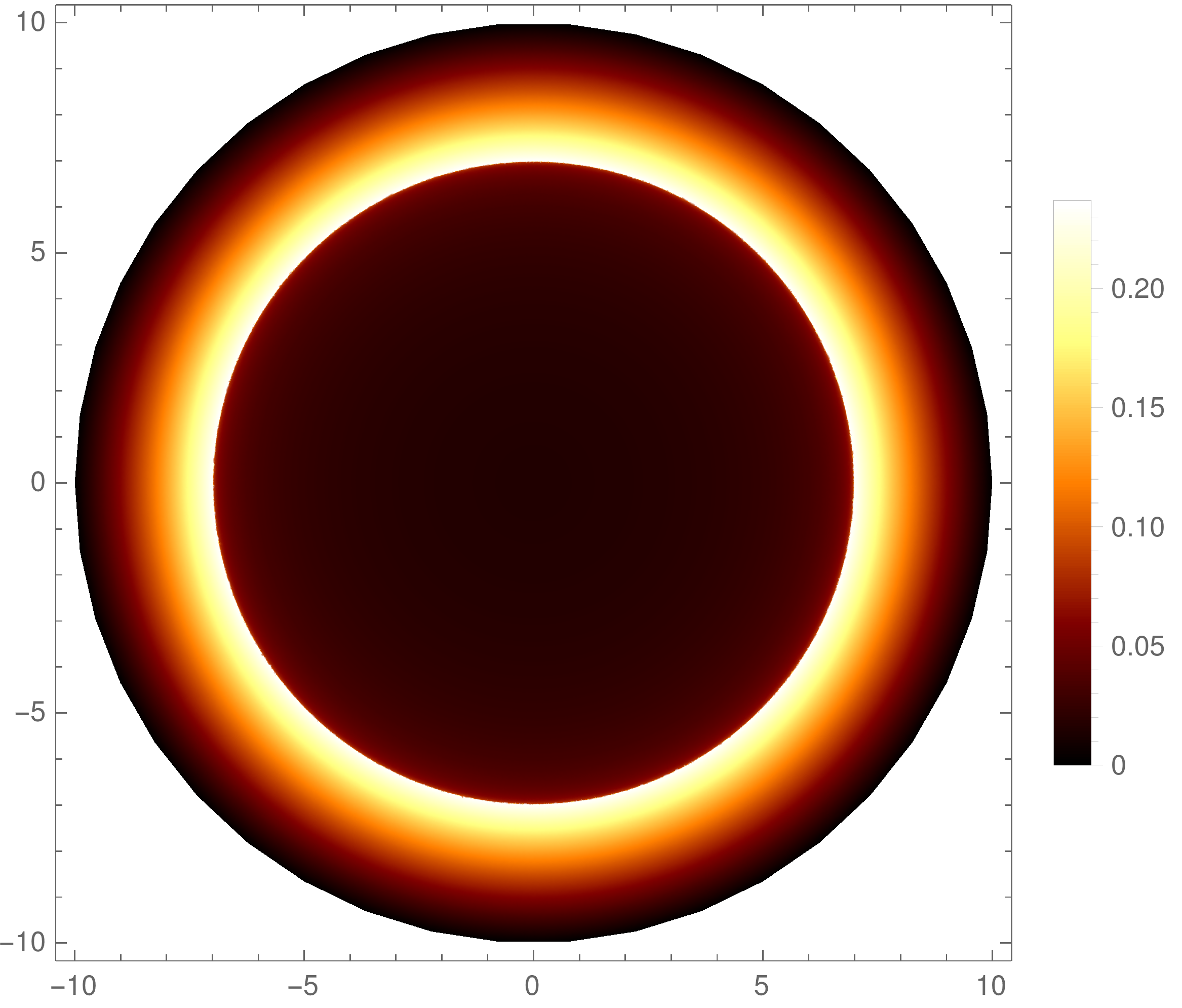}
    \caption{An image was captured of the black hole shadow with infalling spherical accretion, with different values of $q_m$ represented by $0.1$, $0.3$, and $0.5$, respectively with $\beta=0.3$.} \label{fig:21}
\end{figure}

The plot in Fig. \ref{fig:compp} indicates that the observed specific intensity $I_\text{obs}$ rises sharply as the impact parameter $b$ increases for all the values of the magnetic charge and reaches a peak value at $b \sim b_{ph}$. Beyond the point where $b>b_\text{c}$, the observed intensity ($I_\text{obs}$) diminishes and ultimately reaches zero as $b$ approaches infinity. An increase in magnetic charge $q_m$ has been observed to decrease the peak value and goes to zero at the same rate for all cases as the impact parameter increases. The graph includes Schwarzchild $\beta=q_m=0$ (in black), $q_m=0.1$ (in green), $q_m=0.3$ (in blue), and $q_m=0.5$ (in red) for comparison with $\beta=0.3$. A similar effect is observable in the $2D$ shadow image (refer to Fig. \ref{fig:21}) where it is clear that the brightness decreases with increasing the magnetic charge ($q_m$) and the angular diameter also increases. From the observation point of view, it is clear that for the higher value of the magnetic charge, the intensity of the shadow should decrease. 


\section{Conclusions}\label{sec:conclude}
In this paper, we aimed to investigate the impact of a magnetically charged NLED black hole on various physical parameters, such as the radiation emitted by a thin accretion disk. Our preliminary investigation examines the behavior of the photonsphere under the effect of the NLED parameter $\beta$ and magnetic charge $q_m$. For instance, the magnetic charge generally increases the photonsphere radius proportionally, in contrast to the effect of $\beta$. It turns out that as photons traverse the modified black hole geometry caused by the NLED parameter and $q_m$, an increase in shadow size is perceived by some remote observer. Using realistic values for the observer position and SMBH data from the EHT, M87* gives a wide range for $q_m$ than in Sgr. A*. In M87*, there are values of $q_m$ that match exactly the observed shadow radius. We observe that as the NLED parameter increases, the wider the range for $q_m$ would be.

Next, we analyzed the qualitative behaviour of the $\beta$ and $q_m$ on the time-averaged energy flux $(F)$, disk temperature $(T)$, differential luminosity $(dL_{\infty})$, and spectrum $\nu \mathcal{L}_{\nu \infty}$ produced by the thin accretion disk in the equatorial plane surrounding the black hole. Our findings indicate that an increase in the NLED parameter ($\beta$) of the black hole results in an increase in all the mentioned physical parameters, as well as some deviation from the radiation spectrum of a Schwarzschild black hole while it shows completely opposite behaviour as we increase the magnetic charge ($q_m$). We observed in the spectrum profile that the change in the spectrum for $q_m$ is more prominent than for $\beta$ with respect to the Schwarzchild black hole therefore we decided to study the effect of the magnetic charge on the shadow profile of the black hole.

In order to do that we focused on studying the emission profile of a thermally stabilised black hole and looked at the effect of the magnetic charge on some specific ideal emission profiles to get a qualitative idea about the observed emission profiles. We used the Okyay-\"Ovg\"un \textit{Mathematica} notebook package \cite{Okyay:2021nnh}, which has also been used in \cite{Chakhchi:2022fls}, to investigate the effect of the black hole's magnetic charge on the shadow and rings using three toy models of a thin accretion disk. The observed profiles displayed multiple peaks, but it was determined that direct emission was the main factor contributing to the shadow image for the first two models. However, for the third model of the emission profile, the lensing ring was observed to have a significant impact due to the presence of a magnetic charge in the NLE black hole.

The current study investigated the impact of the magnetic charge of the NLED black hole on the observed features of spherical accretion flows. Results showed that an increase in magnetic charge leads to a decrease in the peak of observed intensity. However, this effect is accompanied by a decrease in intensity at the same rate. The impact of increased intensity can be observed in the 2D shadow image Fig. \ref{fig:21}, where the plot of $q=0.1$ exhibits higher brightness in comparison to $q_m=0.3$ and $q_m=0.5$ with $\beta=0.3$. In the future, this research could be expanded to explore the impact of other variables on the observed features of spherical accretion flows. Additionally, more studies could be carried out to explore the implications of these findings for our understanding of black holes and the universe.

\acknowledgements 
A. {\"O}. and R. P. would like to acknowledge networking support by the COST Action CA18108 - Quantum gravity phenomenology in the multi-messenger approach (QG-MM). The work of SC is supported by the SERB-MATRICS grant MTR/2022/000318.

\bibliography{references}
\end{document}